\documentclass[12pt]{article}
\usepackage{latexsym}
\usepackage{amsfonts}

\newcommand{\mysection}[1]{\section{#1}\setcounter{equation}{0}}

\newtheorem{theorem}{Theorem}

\newcommand{\qed}{\hbox{${\vcenter{\vbox{
            \hrule height 0.4pt\hbox{\vrule width 0.4pt height 6pt
            \kern5pt\vrule width 0.4pt}\hrule height 0.4pt}}}$}}

\newtheorem{lemma}{Lemma}

\newtheorem{corollary}{Corollary}

\newtheorem{definition}{Definition}

\newcommand{\bea}{\begin{eqnarray}} 
\newcommand{\eea}{\end{eqnarray}}
\newcommand{\beann}{\begin{eqnarray*}} 
\newcommand{\eeann}{\end{eqnarray*}}
\newcommand{\beq}{\begin{equation}} 
\newcommand{\eeq}{\end{equation}}
\newcommand{\ba}{\begin{array}} 
\newcommand{\ea}{\end{array}}
\newcommand{\ben}{\begin{enumerate}} 
\newcommand{\een}{\end{enumerate}}

\newcommand{\G}{\Gamma}
\newcommand{\D}{\Delta}
\newcommand{\proof}[1]{{\bf Proof.} #1~$\qed$}

\begin{document}

\begin{titlepage}

\begin{flushright} 
ULB-TH/00-28\\
hep-th/0011120
\end{flushright}

\begin{centering}

\vspace{0.5cm}

\huge{Classical and quantum aspects of the 
extended antifield formalism} \\

\vspace{1.5cm}

\large{Glenn Barnich}

\vspace{.5cm}

Physique Th\'eorique et Math\'ematique,\\ Universit\'e Libre de
Bruxelles,\\
Campus Plaine C.P. 231, B--1050 Bruxelles, Belgium

\end{centering}

\vspace{1.5cm}

\begin{abstract}
Starting from a solution to the classical Batalin-Vilkovisky master 
equation, an
extended solution to an extended master equation is constructed by
coupling all the observables, the anomaly candidates and the
generators of global symmetries. The construction of the formalism and
its applications in the context of the renormalization of generic and
potentially anomalous gauge theories are reviewed.   
\end{abstract}

\vspace{.5cm}

\end{titlepage}

\newpage


The main aim of the standard antifield formalism is the
construction, for generic gauge theories, of the proper solution of
the master equation. The master equation is formulated in terms of the
antisymplectic structure for the fields and antifields, and its
solution is the generator of the BRST differential. The coupling
constants play a passive role in the usual discussions. The
most important feature of the 
extended antifield formalism is to promote the
coupling constants to active participants of the construction, by
allowing the BRST differential to act on them, and by introducing, in an
intermediate stage ``anticoupling constants'' which are their partners
in an extended antibracket. The understanding that the physically
relevant coupling constants are related to independent
local BRST cohomology classes in ghost number 0 naturally leads to
consider the constant ghosts coupled to the generators of generalized
global symmetries (local BRST cohomology classes in negative ghost
numbers) as generalized coupling constants and to include 
the couplings to anomaly and anomaly for anomaly
candidates (local BRST cohomology classes in positive ghost numbers).
 
The heart of the extended
antifield formalism is the construction of an extended master equation
for an extended action to which all these cohomology classes have been
coupled and a characterization of the cohomology of the associated
BRST differential, which takes into account in a systematic way higher
order cohomological restrictions through the Lie-Massey brackets. 
This construction is shown to be a particular case
of a general structure that is available as soon as one has a
differential graded Lie algebra, i.e., a graded vector space with an
even or odd Lie bracket and a
differential that is a graded derivation of the bracket. 

Applications of these ideas in the context of renormalized quantum
field theory are then discussed. More precisely:

\begin{itemize} 

\item The existence and relevance of higher order cohomological
  restrictions on anomalies and counterterms is demonstrated. As an
  illustration, the case of Yang-Mills theories with abelian factors
  is discussed.  

\item It is shown that the use of the extended formalism guarantees
  stability independently of power counting restrictions
  (``renormalizability in the modern sense'') for generic gauge
  theories. 

\item The anomalous Zinn-Justin equation for the renormalized
  effective action can be written to all orders as a functional differential
  equation. This allows to prove the existence of a quantum BRST
  differential and to extend the whole framework of algebraic
  renormalization to anomalous gauge theories. In particular, the
  existence of well defined quantum BRST cohomologies is proved.  

\item The dependence of the quantum theory on the parameters of the
  gauge fixing is considered and the general structure of the
  Callan-Symanzik and the renormalization group equations is
  discussed. 

\item A refined anomaly consistency condition for local BRST
  cohomology classes is derived. As an application, a new
  approach to the Adler-Bardeen theorem on the non renormalization of
  the non abelian gauge anomaly is proposed. 

\end{itemize}

This review is based on the papers 
\cite{Barnich:1998ph,Barnich:1999qz,Barnich:1998ke,Barnich:1999nv,%
Barnich:2000ue,Barnich:2000yx}, with the following
improvements: 

(i) The analysis of \cite{Barnich:1999qz,Barnich:1998ke} 
is done without the assumption that there are no anomalies.

(ii) The analysis of \cite{Barnich:2000yx} is rewritten in the context
of the extended antifield formalism, so that the assumptions of
stability and absence of anomalies are now not necessary. As a
consequence, the dependence of the anomaly coefficients on the gauge
parameters and the renormalization scale is obtained.  

(iii) A new section on the quantum BRST cohomology groups has
been added. 

\newpage

\tableofcontents

\newpage

\mysection{Introduction}
\label{Intro}

\subsubsection*{Yang-Mills theories}
The best known example of renormalization of a theory
invariant under a non linear symmetry is probably non abelian
Yang-Mills theory: on the level of the gauge fixed Faddeev-Popov
action \cite{Faddeev:1967fc}, gauge invariance is expressed through
invariance under the non linear global BRST symmetry
\cite{Becchi:1974xu,Becchi:1974md,Becchi:1975nq,BecchiTira,Tyutin:1975qk}. 

Some of the crucial points in the analysis are:
(i)~the importance of the BRST cohomology as a constraint on the
anomalies and the counterterms of the theory, 
(ii)~the anticanonical structure of the theory in terms of
the fields and the sources, to which the BRST variations are coupled,
together with the compact reformulation of all the 
Ward identities in terms of
the Zinn-Justin equation \cite{Zinn-Justin:1974mc,Zinn-Justin:1989mi}, 
and (iii)~the insight that BRST exact counterterms can be
absorbed by anticanonical fields and sources redefinitions 
\cite{Dixon:1975si}.

The question whether the remaining
counterterms can be absorbed by a redefinition of the coupling
constants of the theory could be settled to the affirmative, in the
power counting renormalizable case based on a semi-simple gauge group, 
through an exhaustive enumeration
of all possible renormalizable interactions \cite{Becchi:1975nq}. 
In the case where one
includes higher dimensional gauge invariant operators, such a property
depends crucially on a conjecture by Kluberg-Stern and Zuber
\cite{Kluberg-Stern:1975hc} on the BRST cohomology
in ghost number $0$,
which states that it should be describable by off-shell gauge invariant
operators not involving the ghosts or the sources. This conjecture can be
proved \cite{Barnich:1994ve,Barnich:1995mt} in the semi-simple case
for which it has been
originally formulated, but it is not valid in the presence of abelian
factors, not even for power counting renormalizable
theories 
\cite{Bandelloni:1978ke,Bandelloni:1978kf,Barnich:1994ve,Barnich:1995mt}.
In this last case,
the counterterms violating the generalized Kluberg-Stern and Zuber conjecture 
have been shown to be absent by more involved
arguments from renormalization theory 
\cite{Bandelloni:1978ke,Bandelloni:1978kf}, so that
renormalizability still holds, even if the conjecture does not. 

\subsubsection*{Extended antifield formalism}
The classical Batalin-Vilkovisky formalism 
\cite{Batalin:1981jr,Batalin:1983wj,Batalin:1983jr,Batalin:1984ss,%
Batalin:1985qj} 
(for reviews, see e.g. \cite{Henneaux:1992ig,Gomis:1995he}) extends the above
techniques to the case of general gauge theories with open gauge algebras
and structure functions, the invariance of the action being expressed
through the central master equation. 

These cohomological techniques can also be used in order to control 
the renormalization of non linear global symmetries with a closed
algebra by coupling their generators with constant ghosts 
\cite{Voronov:1981du,Becchi:1988vc}. The generalization of the
formalism to the case of systems including both gauge and generalized global
symmetries with generic algebra has been achieved only recently
\cite{BHW1,Brandt:1998cz}. It is based on the observation that the generators
of generalized global symmetries correspond to local BRST
cohomological classes in negative ghost numbers \cite{Barnich:1995db}
(for a review, see \cite{Barnich:2000zw}).

\subsubsection*{Stability in the BV formalism}
A detailed analysis of the
compatibility of the renormalization procedure with invariance
expressed through the master equation has been performed in
\cite{Voronov:1982cp,Voronov:1982ur,Voronov:1982ph,Lavrov:1985hr}, 
where it has been shown 
that the renormalized action
is a deformation of the starting point solution to the master
equation. Independently of this result, the fundamental problem of
locality of the construction is raised and 
a locality hypothesis is stated \cite{Voronov:1982cp}. 
This hypothesis can be
reinterpretated in a more general framework as the assumption that 
the cohomology of the Koszul-Tate differential 
\cite{Fisch:1990rp,Henneaux:1991rx}
vanishes in the space of local functionals. While the assumption holds
under certain conditions, which are in particular fulfilled for the
construction of the solution of the master equation, thus guaranteeing
its locality \cite{Henneaux:1991rx}, it does not hold in general:
the obstructions are related to the non
trivial global currents of the theory \cite{Barnich:1995db}, and give rise to
BRST cohomology classes with a non trivial antifield dependence. 

A consequence is the possibility of existence of observables 
that cannot be made off-shell gauge invariant, even in the case of
closed gauge theories, so that the associated 
deformed solutions of the master equation cannot be
related by a field, antifield and
coupling constant renormalization 
to the starting point solution extended by coupling 
all possible off-shell observables.
 
In \cite{Anselmi:1994ry,Anselmi:1995zx}, renormalization in the
context of the
Batalin-Vilkovisky formalism is
reconsidered precisely under the assumption that there are no such
deformations, i.e., 
in the closed case under the analog of the Kluberg-Stern and Zuber
conjecture, and in the open case under the conjecture that all the
BRST cohomology is already contained in the solution to the master equation
coupled with independent coupling
constants\footnote{Note that
it is not true that the 
antifield independent part of the 
cohomology of the differential $(S,\cdot)$ is off-shell
gauge invariant, 
it is in general only weakly
gauge invariant.}, with the conclusion, that the infinities can then
be absorbed by renormalizations.

Finally, in \cite{Gomis:1996jp} the problem of renormalization under non
linear symmetries is readressed in the context of
effective field theories: it is for instance shown that 
semi-simple Yang-Mills theory and gravity, to which are coupled all possible
(power counting non renormalizable) off-shell observables, 
are such that all the local counterterms needed to
cancel the infinities, can be absorbed through coupling constants,
field and antifield renormalizations, while preserving the symmetry (in
the form of the Batalin-Vilkovisky master equation). 
Theories possessing this last
property, even if an infinite number of coupling constants is needed,
are defined to be renormalizable in the modern sense. 
The difficulty, that is also discussed, is that the non trivial 
infinities are a priori only constrained to belong to 
the BRST cohomology in ghost number $0$, which, because of the non
validity of the generalized Kluberg-Stern and Zuber conjecture 
(taken as an example of a so called structural constraint), does not
guarantee that they can be absorbed by redefinitions of coupling
constants of an action extended by all possible off-shell
observables. What good structural
constraints might be in the general case and if they can be chosen in
such a way as to guarantee renormalizability in the modern sense for
all theories is left as an open question in 
\cite{Gomis:1996jp,Weinberg:1996kw}.

In order to relate the terminology in \cite{Gomis:1996jp,Weinberg:1996kw}
to the one coming from the algebraic approach to renormalization under
symmetries pioneered in 
\cite{Lowenstein:1973qt,Becchi:1974xu,Becchi:1974md,Becchi:1975nq,BecchiTira}
(see for instance 
\cite{Piguet:1981nr,Bonneau:1990xu,Piguet:1995er} for reviews), 
we note that what is called 
(extended) master equation in the former corresponds to a generalized
Slavnov-Taylor identity in the latter. By generalized, we mean 
the definition of this identity in theories with arbitrary gauge 
structure as proposed by Batalin and Vilkovisky, then
the extension to include the case of (a closed subset of) global
symmetries.
What is called (local) BRST
cohomology corresponds to the cohomology of
the generalized linearized Slavnov-Taylor operator $S_L$
acting in the space of (integrated) polynomials in the fields, the
sources and their derivatives.  
The question of {\em renormalizability in the modern sense} in 
\cite{Gomis:1996jp,Weinberg:1996kw} is then the
question of {\em stability independently of power counting
  restrictions} in
the language of  algebraic renormalization\footnote{Stability is
defined for instance in \cite{Bonneau:1990xu} as ``the dimension of the
cohomology space of the $S_L$ operator in the Faddeev-Popov neutral
sector should be equal to the number of physical parameters of the
classical action''.}.

\subsubsection*{Higher order cohomological restrictions}
A clue to the answer to these questions can be found in
\cite{Voronov:1982cp,Voronov:1982ur,Voronov:1982ph,Lavrov:1985hr}. 
Indeed, the fact that the divergences 
are such that they
always provide a deformation of the solution of the master equation,
implies in general that the non trivial first order deformations 
satisfy additional cohomological restrictions \cite{Barnich:1993vg}
besides 
belonging to the BRST cohomology. The problem with these
additional restrictions is that they are non linear.
The starting point for the extended antifield formalism 
to be discussed in this review is to show that the non trivial
counterterms and anomalies
satisfy linear higher order cohomological constraints, 
by coupling arbitrary BRST cocycles to the solution of the master equation. 

As an (academical) example of how these higher order cohomological 
restrictions work, consider Yang-Mills theories with 
free\footnote{By free, we mean
that the abelian gauge fields have no couplings to 
matter fields,
hence, they have no interactions at all. Their quantization is of
course trivial and we know a priori that no counterterms are
needed.} abelian
gauge fields $A^a_\mu$ as discussed for instance in \cite{Gomis:1996jp}.
The BRST cohomology in ghost number zero 
contains the term \cite{Barnich:1994ve,Barnich:1995mt} 
$$
K=f_{abc}\int d^nx\ F^{a\nu\mu}A^b_\mu A^c_\nu+2A^{*a\mu}A^b_\mu C^c
+C^{*a}C^bC^c,
$$
for completely antisymmetric constants $f_{abc}$, so
that this term is a potential counterterm. At the same time,
the term $k^d\int d^nx\ C^*_d$ belongs to the BRST cohomology
in ghost number $-2$. If we take the action $S_k=S+k^d\int
d^nx\ C^*_d$, we have $1/2(S_k,S_k)=O(k^2)$. This implies
according to the quantum action principle for the regularized theory
that $1/2(\G_k,\G_k)=O(k^2)$ and then, at order $1$ in $\hbar$ for the
divergent part, that $$(S_k,{\G^{(1)}_k}_{div})=O(k^2).$$ The $k$ independent
part of this equation gives the usual condition that the divergent
part of the $k$ independent effective action at first order must be
BRST closed, $(S,{\G^{(1)}}_{div})=0$, and contains in particular the
candidate $K$ above. The $k$ linear part of this equation requires 
$$(\left.\frac{\partial {\G^{(1)}_k}_{div}}{\partial
k^d}\right|_{k=0},S)+(\int d^nx\ C^*_d,{\G^{(1)}}_{div})=0.$$
This condition eliminates the candidate $K$ because $$(\int d^nx\
C^*_d,K)=2f_{abd}\int d^nx\ A^{*a\mu}A^b_\mu+C^{*a}C^b$$ 
is not BRST exact but represents a non trivial BRST
cohomology class in ghost number $-1$. Hence,  
there exists a purely cohomological reason why $K$ cannot
appear as a counterterm. Note that as soon as the abelian fields are
coupled to matter fields, the functionals
$\int d^nx\ C^*_d$ but also $K$ do not belong to the BRST
cohomology any more and the problem with this particular type 
of counterterms
does not arise to begin with.

\subsubsection*{Anomalies}
A related problem, which is relevant in 
\cite{Lavrov:1985hr,Anselmi:1994ry,Anselmi:1995zx,Gomis:1996jp}, is to
provide a sensible definition of the $\Delta$ operator of 
the quantum Batalin-Vilkovisky master equation. 
Indeed, its expression as a second order functional
differential operator with respect to fields and antifields, obtained
from formal path integral considerations, 
does not make sense when applied to local
functionals. In \cite{Troost:1990cu,Howe:1990pz,DeJonghe:1996xm,%
Paris:1995dx,Paris:1996jg}, 
the antifield formalism has
been discussed in the context of explicit regularization and 
renormalization schemes and the related question of anomalies
(assumed to be absent in 
\cite{Lavrov:1985hr,Anselmi:1994ry,Anselmi:1995zx,Gomis:1996jp})
has  been addressed. In particular, non trivial anomalies are shown to
be constrained by the local BRST cohomology in ghost number
$1$, which is the expression of the Wess-Zumino consistency condition 
\cite{Wess:1971cm} in this context, and well defined expressions for the
regularized $\Delta$ operator are proposed at one loop level in
\cite{Troost:1990cu} in the context of Pauli-Villars regularization
and at higher orders in \cite{Paris:1995dx,Paris:1996jg} for non-local
regularization.

In the context of dimensional renormalization, it has been shown in 
\cite{'tHooft:1972fi,Costa:1977pd,Breitenlohner:1977hr,Breitenlohner:1977hg,%
Breitenlohner:1977te,Aoyama:1981yw,%
Bonneau:1980zp,Bonneau:1980jx,Bonneau:1981ya,Bonneau:1983uj,Bonneau:1986ea,%
Tonin:1992wf} 
that anomalies can be dealt with
consistently, if evanescent terms are taken into account properly. In
particular \cite{Tonin:1992wf}, the evanescent breaking terms of the 
master equation are responsible for a non trivial $\Delta$ operator,
even though  $''\delta(0)''=0$. Furthermore, 
it is shown in \cite{Aoyama:1981yw,Tonin:1992wf} that it is 
useful to couple the 
non trivial anomalies from the beginning with coupling constants in ghost
number $-1$ in order to control the absorption of divergences 
in the presence of anomalies.

\subsubsection*{The completely extended antifield formalism}

As a combination of the previous three ideas, 
i.e., (i) that global symmetries are controlled by
coupling BRST cohomological classes in negative ghost numbers, (ii) that
renormalizability, or more precisely stability, is related to the
question if the BRST cohomological classes in ghost number $0$ are all
coupled and (iii) that renormalization in the presence of 
non trivial anomalies requires the coupling of the local BRST
cohomology classes in positive ghost numbers, one is naturally 
led to consider the solution of the classical master equation to
which all local BRST cohomology classes have been coupled. To first
order in the couplings, this extension will satisfy the standard
master equation. To all orders in the couplings, the equation
satisfied by a suitable completion of the extension will satisfy an
extended master equation which has the same structure than the
extended master equation of the formalism where only the local BRST
cohomological classes in negative ghost numbers have been
coupled. This will
be shown by similar techniques than the ones used in the extended
antifield formalism of \cite{Brandt:1998cz}. 

A remarkable property of
this extended antifield formalism is that the cohomology in
all ghost numbers of the BRST differential associated to the extended master
equation is contained completely in the solution to this master
equation. As a consequence, using this formalism, 
renormalizability in the modern sense will be shown to hold by
construction for all
gauge theories, which answers the questions raised in
\cite{Gomis:1996jp,Weinberg:1996kw}, even in the presence of anomalies. 
In other words, we will complete the general analysis
of the absorption of divergences in the presence of possibly anomalous
local or global symmetries independently of the precise form of the BRST
cohomology of the theory. 

At the same time, it will be shown that in the extended formalism, there
exist well defined differentials, both on the classical and on the
quantum level, that play the role of the quantum Batalin-Vilkovisky
$\Delta$ operator. 

From the point of view of the algebraic approach to renormalization,
the extended antifield formalism is
stable by construction for all theories. It allows to 
extend the algebraic approach to
the case of anomalous gauge theories. Two well defined isomorphic
versions of quantum BRST cohomology will be constructed in this
context: the first one on the level
of local insertions and the second one on the level of derivations in the
couplings of the theory. 

\subsubsection*{Organization of the review}

The definition of a differential graded Lie
algebra $(L,d,[\cdot,\cdot])$ is recalled in section \ref{s2}
and the existence of 
Lie-Massey brackets in
cohomology \cite{Ret} is briefly discussed. Associated to a Hodge
decomposition 
$L=K\oplus R\oplus dR$, with $K\simeq H(d,L)$
and $\{e_a\}$ a basis
of $K$, two additional algebras are
considered. If $sK^*$ is the suspension of the dual $K^*$ of $K$ with
basis the ``couplings'' $\{\xi^a\}$, the first one, $L(\xi)$, is the
tensor product of $L$ with the polynomials in the couplings, $\wedge
(sK^*)$. The second one is 
the commutator algebra of graded right
derivations in the couplings, $RDer[\wedge (sK^*)]$. 
On $L(\xi)$, an ``extended
master equation'' involving a nilpotent graded right derivation
$\Delta$ on $\wedge (sK^*)$ is constructed in such a way that the
cohomology defined by the solution of the extended master equation in
$L(\xi)$ is isomorphic to the commutator cohomology of
$[\Delta,\cdot]$ in $RDer[\wedge (sK^*)]$. The perturbative 
techniques used for the
construction are similar to those of \cite{Brandt:1998cz}. 
It is then shown how to compute these cohomologies by using a spectral
sequence associated to the polynomial degree in the couplings. At the
end of section \ref{s2}, we discuss some examples of 
differential graded Lie
algebras, where the general construction can be applied. 

In section \ref{s3}, we review relevant features of the antifield
formalism, such as: the spectrum of fields and ghosts, the
antisymplectic structure between fields and antifields, the master
equation, local BRST cohomology and the gauge fixing procedure. 

Section \ref{s4} is devoted to show that anomalies and divergences
must satisfy cohomological restrictions. In order to do so, we use the
framework of dimensional regularization as discussed in
\cite{Tonin:1992wf}. 
The
standard restrictions are derived first, while higher order
restrictions are obtained by coupling arbitrary BRST cocycles to the
action.   
As an application in the physically relevant case of the standard
model, it is shown explicitly how antifield dependent
counterterms can be eliminated through higher order cohomological
restrictions. 

The application of the general construction of section \ref{s2} to the
case of the (classical) antifield formalism is discussed in detail in section
\ref{s5}. Essential couplings are defined as the couplings corresponding
to independent local BRST cohomological classes. 
The stability of the extended antifield formalism
constructed in this way is then proved. 

In section \ref{s6}, the absorption of the divergences in the context
of the extended antifield formalism is considered, first in
the context of ``dimensional'' regularization. Using suitable BRST
breaking counterterms, the renormalized effective action is shown to
satisfy a deformed extended master equation, and this to all orders in
$\hbar$ independently of any assumptions on the anomalies of the
theory. Alternatively, because stability of the extended antifield
formalism has been proved in the previous section on the classical
level, the machinery of algebraic renormalization on the final
renormalized level can be applied instead of using first a
regularization. The deformed extended master equation for the
effective action is rederived in this context. Then the
renormalization of classical BRST cohomological classes is
discussed. At the end of section \ref{s6}, two well defined quantum BRST
cohomologies associated to the deformed extended master equation 
are introduced, one for local insertions and one for right
derivations, and their cohomologies are shown to be isomorphic. 

The dependence of the effective action on the parameters introduced
through the gauge fixing is analyzed in section \ref{s7}. 
It is shown
that, by suitably redefining the essential couplings by gauge
parameter dependent terms of higher order in $\hbar$, the variation of
the effective action can be made trivial in the sense that it is given by a
quantum BRST coboundary, while 
the anomaly operator can be shown to be independent
of the gauge parameters. 

In section \ref{s8}, the general form of the renormalization group
equation and the relation between the renormalization group $\beta$
functions and the anomaly coefficients is derived. If the theory is 
expressed in terms of the running couplings, we show that the variation 
of the 
effective action with respect to the renormalization scale is a quantum 
BRST coboundary and that the anomaly operator is independent of the 
renormalization scale. 
Then, the Callan-Symanzik equation is derived and the vector field
built out of the associated $\beta$ functions is shown to define a non
trivial quantum BRST cocycle. 

In section \ref{s9}, the anomaly appearing 
in the renormalization of a local BRST cohomological class with a
descent of length $d$ and a lift of length $l$ is shown to be 
characterized by a descent which is shorter or equal to $d$ and a
lift which is longer or equal to $l$. In a first part, 
the characterization of local BRST cohomological groups
according to the lengths of their descents and their lifts is
explained. Then, the main result on the lengths of the
descents and the lifts of the anomalies is proved. An alternative 
derivation of
the main results clarifying the underlying mechanism is presented in
section \ref{alt}, where differentials controlling
the one loop anomalies arising in the renormalization of 
BRST cohomological groups are
introduced.

Anomalous effective Yang-Mills theories and a new approach to the
Adler-Bardeen theorem, independently of the gauge fixing, of power
counting restrictions and without relying on the Callan-Symanzik
equation, are discussed in section \ref{s10} as an application of the
results of section \ref{s9}.

\newpage

\mysection{Cohomology of differential graded Lie algebra and
  associated sh Lie algebra}\label{s2}

\subsection{Differential graded Lie algebras}\label{dgla}

\paragraph{Definition}
Let $L=\bigoplus_{g\in \mathbb Z} L^g$ be a graded differential Lie algebra 
over a field $k$ (typically $\mathbb R$ or $\mathbb C$)
with an even or an odd bracket:
\bea
d: L^g\longrightarrow L^{g+1},\ d^2=0,\cr
[\cdot,\cdot ] : L^{g_1}\otimes L^{g_2}\longrightarrow
L^{g_1+g_2+\epsilon},
\eea 
where $\epsilon=-1,0,1$
such that 
\bea
[x,y]=-(-)^{(|x|+\epsilon)(|y|+\epsilon)}[y,x],\\
\ [x,[y,z]]=[[x,y],z]
+(-)^{(|x|+\epsilon)(|y|+\epsilon)}[y,[x,z]],
\label{jacoid}\\
d[x,y]=[dx,y]+(-)^{|x|+\epsilon}[x,dy],\label{propder}
\eea
for homogeneous elements $x,y,z \in L$.

\vspace{.25cm}

{\footnotesize

\noindent {\bf Remarks :}

(i) In principle, it is sufficient to consider the case $\epsilon =0$,
because this case can be obtained from the case $\epsilon \neq 0$ by
defining a new grading which is equal to the old one plus
$\epsilon$. However, because in various applications it is more
natural to have a bracket of degree $\epsilon$, we will keep
$\epsilon$ in the expressions below

(ii) A particular case is that of an inner differential
$d=[D,\cdot]$, where $D\in L^{1-\epsilon}$ satisfies the equation
$\frac{1}{2}[D,D]=0$.} 

\vspace{.25cm}

\paragraph{Lie-Massey brackets}
Because the differential is a derivation of the bracket
(\ref{propder}), the bracket of $2$ cocycles is again a
cocycle, and changing one of the cocycles by a coboundary modifies the
bracket only by a coboundary. 
It follows that the bracket $[\cdot,\cdot]$ induces a well
defined bracket $[\cdot,\cdot]_M$ in cohomology, 
\bea
[\cdot,\cdot]_M: H^{g_1}(d)\otimes H^{g_2}(d)\longrightarrow
H^{g_1+g_2+\epsilon}(d)\cr
[[x_1],[x_2]]_M=[[x_1,x_2]].
\eea 

There exist also higher order maps induced in cohomology, 
the Lie-Massey brackets \cite{Ret}. (In this context, the map
$[\cdot,\cdot]_M$ is called the two place Lie-Massey bracket.) 
For instance, consider cocycles $x_i,x_j,x_k$, 
such that $[[x_i],[x_j]]_M=0$, or,
equivalently $[x_i,x_j]=d x_{ij}$.
Note that by this equation, the $x_{ij}$'s are defined only up to
cocycles.
A three place bracket of such cocycles is defined by 
\bea
[x_i,x_j,x_k]=[x_{ij},x_k]
-(-)^{(g_j+\epsilon)(g_k+\epsilon)}[x_{ik},x_j]
-(-)^{g_i+\epsilon}[x_i,x_{jk}].
\eea
Because of the derivation
property (\ref{propder}) and the Jacobi identity (\ref{jacoid}), 
$d[x_i,x_j,x_k]=0$.
The corresponding cohomology class is defined to be the value of the
three place Lie Massey bracket, $[[x_i],[x_j],[x_k]]_M=
[[x_i,x_j,x_k]]$. The set of elements $\{x_i,x_{ij}\}$ is called the
defining system of $[[x_i],[x_j],[x_k]]_M$. Because 
the $x_{ij}$ are only defined up to cocycles, $[[x_i],[x_j],[x_k]]_M$
is not uniquely defined, but it is in fact a set of cohomology classes
depending on the defining system. 
A uniquely defined three place Lie Massey bracket
$[[x_i],[x_j],[x_k]]_M$ is obtained for elements $[x_i]$ that belong
to the kernel of the two place Lie-Massey bracket, $[x_i]\in {\rm
  ker_2}^g \subset H^{g}(d)$, iff  
$[[x_i],[y]]_M=0$, for all $[y]\in H(d)$. Indeed, in this case, it is
straightforward to verify that  $[[x_i,x_j,x_k]]\in 
H^{g_i+g_j+g_k+2\epsilon-1}(d)$ does not depend on the
choice of the defining system.  Hence, there is a well-defined
map
\bea
[\cdot,\cdot,\cdot]_M:{\rm ker_2}^{g_1}\otimes {\rm ker_2}^{g_2}
\otimes {\rm ker_2}^{g_3}\longrightarrow
H^{g_1+g_2+g_3+2\epsilon-1}(d)\nonumber\cr
[[x_i],[x_j],[x_k]]_M=[[x_i,x_j,x_k]].
\eea

Higher order brackets can be systematically defined along similar
lines \cite{Ret}. We
will not discuss them here, because we will follow a different route
and construct related maps in a systematic way through homological
perturbation theory below. 

\paragraph{Hodge decomposition and associated graded Lie algebras}
Suppose one has the Hodge decomposition $L = K \oplus R
\oplus dR$, with $K\simeq H(d,L)$ and let $\{e_a\}$ denote the elements
of a basis of $K$. This means that 
\bea
dx=0\Longrightarrow x=\lambda^a e_a+d y,\cr
\mu^a e_a=
dz\Longrightarrow \mu^a=0=dz,
\eea 
with $\lambda^a,\mu^a\in k$. 

Let us now consider the space $\wedge(sK^*)$, i.e., the
exterior algebra over the suspension of the dual $K^*$ of $K$. In
other words, we associate to 
each $e_a$ a variable $\xi^a$ of grading
$deg(\xi^a)=-|e_a|+1 -\epsilon$ and take linear combinations
$\lambda(\xi)=\sum_k \lambda_{a_1\dots a_k} \xi^{a_1}\dots\xi^{a_k}$ with
$\lambda_{a_1\dots a_k}\in k$. We also consider the space
$L(\xi)\equiv \wedge(sK^*)\otimes L$, with grading given by the sum
of the gradings,
\bea
tot(\xi^{a_1}\dots\xi^{a_k}x)=
\sum^k_{i=1}deg(\xi^{a_i})+|x|.
\eea 
By abuse of notation, the differential $1\otimes d$ on $L(\xi)$ is
still denoted by $d$, 
\beq
d (\xi^{a_1}\dots\xi^{a_k}x)=
(-)^{\sum^k_{i=1}deg(\xi^{a_i})}\xi^{a_1}\dots\xi^{a_k}d x,
\eeq 
and the 
bracket is extended according
to 
\bea
[\xi^{a_1}\dots\xi^{a_k}x,
\xi^{b_1}\dots\xi^{b_m}y]=
(-)^{(\sum_{i=1}^mdeg(\xi^{b_i}))(|x|+\epsilon)}
\xi^{a_1}\dots\xi^{a_k}
\xi^{b_1}\dots\xi^{b_m}[x,y].
\eea
It follows that $L(\xi)$ is a differential graded Lie algebra with
respect to the total degree $tot$.

Let us also introduce the even graded Lie algebra $RDer[\wedge(sK^*)]$
of graded right derivations on $\wedge(sK^*)$, 
\bea
RDer[\wedge(sK^*)]=\{\frac{\partial^R\cdot }{\partial
\xi^a}\lambda^a(\xi), \lambda^a(\xi)\in \wedge(sK^*)\}.
\eea 
If $\frac{\partial^R\cdot}{\partial
  \xi^a}\lambda^a(\xi): 
\wedge(sK^*)^g\longrightarrow \wedge(sK^*)^{g+\lambda}$, the
grading of $\frac{\partial^R\cdot}{\partial
\xi^a}\lambda^a(\xi)$ is defined to be $\lambda$;  
the graded commutator is defined
by 
\bea
[\frac{\partial^R\cdot}{\partial
\xi^a}\lambda^a(\xi),\frac{\partial^R\cdot}{\partial
\xi^b}\mu^b(\xi)]_{C}=\frac{\partial^R\cdot}{\partial
\xi^a}(\frac{\partial^R\lambda^a}{\partial
\xi^b}\mu^b-(-)^{\lambda\mu}\frac{\partial^R\mu^a}{\partial
\xi^b}\lambda^b).
\eea

\subsection{Main theorem}\label{main}

For every $n\geq 2$, one can construct, on the one hand, 
constants $f^b_{a_1\dots a_n}\in k$, 
with associated right derivation 
\bea
\Delta=
\frac{\partial^R\cdot}{\partial \xi^b}f^b(\xi)\in RDer[\wedge(sK^*)]^1,
\eea
where 
$f^b(\xi)=\sum_{n\geq 2}f^b_{a_1\dots
a_n}\xi^{a_1}\dots\xi^{a_n}$,
and, on the other hand, vectors $e_{a_1\dots a_n}\in L$, 
with associated element 
\bea
e(\xi)=\sum_{n\geq 1}e_{a_1\dots
a_n}\xi^{a_1}\dots\xi^{a_n}\in L(\xi)^{1-\epsilon},
\eea
such that: 

\begin{itemize}

\item the derivation $\Delta$ is a differential,
\beq
\Delta^2=0;
\label{mt1}
\eeq
[with the associated left derivation 
$\Delta^L=(-)^{b+1-\epsilon}
f^b(\xi)\frac{\partial^L}{\partial \xi^b}\in LDer[\wedge(sK^*)]^1$
also being a differential, $(\Delta^L)^2=0$]
\item the element $e(\xi)$ satisfies the Maurer-Cartan type equation
\beq
(d+\Delta^L)e(\xi)+\frac{1}{2}[e(\xi),e(\xi)]=0,\label{mt2}
\eeq
so that 
\beq
\bar d= d+[e(\xi),\cdot]+\Delta^L: L(\xi)^g\longrightarrow
L(\xi)^{g+1},\ \ \ \bar d^2=0;
\eeq
\begin{itemize}
\item {\footnotesize for an inner differential $d=[D,\cdot]$, where
$D\in L^{1-\epsilon}$ satisfies the ``classical master equation''
\beq
\frac{1}{2}[D,D]=0,
\eeq
$D(\xi)=D+e(\xi)\in L(\xi)^{1-\epsilon}$ satisfies the
``quantum master equation''
\beq
\frac{1}{2}[D(\xi),D(\xi)]+\Delta D(\xi)=0,\label{mt3}
\eeq
and 
$\bar d=[D(\xi),\cdot]+\Delta^L;$}
\end{itemize}
\item if 
$d_\Delta=[\Delta,\cdot]_C$, then
\beq
H^*(\bar d, L(\xi))\simeq H^*(d_\Delta,RDer[\wedge(sK^*)]),\label{iso}
\eeq
the isomorphism being given by 
\beq
H^*(d_\Delta,RDer[\wedge(sK^*)])\ni 
[\frac{\partial^R\cdot}{\partial
\xi^a}\lambda^a(\xi)]\longleftrightarrow [ \frac{\partial^R e(\xi)}{\partial
\xi^a}\lambda^a(\xi)]\in H^*(\bar d, L(\xi)).\label{mt4}
\eeq
\end{itemize}

{\bf Remarks:}

(i) {\it Relation of the construction to Lie-Massey brackets:} 
For all $r\geq 2$, the constants $f^b_{a_1\dots a_r}$ define higher
order maps $l_r$ in cohomology, 
\bea
l_r:\otimes^r H(d)\longrightarrow H(d),\nonumber\\
l_r([e_{a_1}],\dots,[e_{a_r}])=f^b_{a_1\dots a_r} [e_b].
\eea
For a given $r\geq 2$, let us suppose that the
$[e_{a_1}],\dots,[e_{a_r}]$ are such that the structure constants with
strictly less than $r$ indices vanish, for all choices of $a_i$'s. 
From the explicit form of the identity (\ref{mt2}),
it then follows that 
\bea
\frac{1}{2}\sum_{k=1}^{r-1}
(e_{(a_1\dots
a_k},e_{a_{k+1}\dots
a_r)})(-)^{(k(1+\epsilon)+|e_{a_1}|+\dots+|e_{a_k}|)((r-k+1)(1+\epsilon) 
+\epsilon+|e_{a_{k+1}}|+\dots +|e_{a_{k+1}}|)}\nonumber\\
+d e_{a_1\dots a_r}+ e_b f^b_{a_1\dots a_r}=0,
\eea
the round bracket for the indices in the first term denoting 
symmetrization with respect to the grading of the $\xi^{a_i}$'s. 
We thus see that, under the above assumption, 
\bea
-l_r([e_{a_1}],\dots,[e_{a_r}])=\nonumber\\
\frac{1}{2}\sum_{k=1}^{r-1}[(e_{(a_1\dots
a_k},e_{a_{k+1}\dots
a_r)})(-)^{(k(1+\epsilon)+|e_{a_1}|+\dots+|e_{a_k}|)((r-k+1)(1+\epsilon) 
+\epsilon+|e_{a_{k+1}}|+\dots +|e_{a_{k+1}}|)}].
\eea
By comparing with the definitions in \cite{Ret}, 
we identify the maps
$l_r$, under the above assumption,  
as the value of the $r$-place Lie-Massey bracket
$[[e_{a_1}],\dots,[e_{a_r}]]_M$ for the defining system 
$\{e_{a_{i_1}\dots a_{i_k}},\ k=1,\dots,r-1,\   
1\leq i_1<\dots<i_k\leq r\}$, up to a sign.

(ii) The differential $\Delta$ encodes a strongly homotopy Lie
algebra on $K$ \cite{LaSta}.

(iii) {\it Iteration of the construction:}
Because $(RDer[\wedge(sK^*)],[\cdot,\cdot]_C,d_\Delta)$ is a again a 
graded differential Lie algebra, the main theorem can be applied to this graded
differential Lie algebra, yielding a new graded differential Lie
algebra, to which the main theorem can be applied, $\dots$.  

\subsection{Proof of the main theorem}\label{proof}

\subsubsection{Auxiliary acyclic extended graded differential Lie algebra}

Let us now consider the graded differential Lie algebra 
$RDer[\wedge(s K^*)]^\epsilon$, which is obtained from the even
graded Lie algebra $RDer[\wedge(s K^*)]$ by taking as new
grading of all the elements the old one minus $\epsilon$. Explicitly,
we introduce additional variables $\xi^*_a$ of degree
$-tot(\xi^a)-\epsilon=|e_a|-1$ replacing the 
$\frac{\partial^R\cdot}{\partial \xi^a}$, and the space 
$RDer[\wedge(s K^*)]^\epsilon$ is composed of elements of the form 
$\xi^*_b\lambda^b(\xi)$. The even commutator bracket $[\cdot,\cdot]_C$
is replaced by 
\bea
[\cdot,\cdot]_{\xi}: 
(RDer[\wedge(s K^*)]^\epsilon)^{g_1}\otimes 
(RDer[\wedge(s K^*)]^\epsilon)^{g_2}
\longrightarrow (RDer[\wedge(s K^*)]^\epsilon)^{g_1+g_2+\epsilon},\cr
[\lambda,\mu]_{\xi}=\frac{\partial^R\lambda}{\partial\xi^a}
\frac{\partial^L\mu}{\partial\xi^*_a}-(-)^{a(a+\epsilon)}
\frac{\partial^R\lambda}{\partial\xi^*_a}
\frac{\partial^L\mu}{\partial\xi^a},\ \ \ \ \ \ \ \ \ \ \ \ \label{for}
\eea 
where the shorthand notation $a$ has been used for the grading of
$\xi^a$, and $\lambda,\mu\in RDer[\wedge(s K^*)]^\epsilon$. It follows
that $RDer[\wedge(s K^*)]^\epsilon$ is a graded Lie algebra with the 
same $\epsilon$ as the original one.

The auxiliary extended space is then defined to be
\bea
L(\xi,\xi^*)\equiv L\otimes \wedge(s K^*)\oplus RDer[\wedge(s K^*)]^\epsilon.
\eea
The differential $d$ and the bracket $[\cdot,\cdot]$ on $L(\xi)=L\otimes
\wedge(s K^*)$ are extended trivially to $L(\xi,\xi^*)$, while the
bracket $[\cdot,\cdot]_{\xi}$ is extended by the formula (\ref{for}),
with $\lambda,\mu$ replaced by elements $A,B\in L(\xi,\xi^*)$. This
means in particular that $RDer[\wedge(s K^*)]^\epsilon$ acts on $L(\xi)$ 
through the bracket $[\cdot,\cdot]_\xi$. The bracket 
\beq
[\cdot,\cdot\tilde]=
[\cdot,\cdot]+[\cdot,\cdot]_{\xi} 
\eeq
is then a graded Lie bracket of
degree $\epsilon$ and the auxiliary differential graded Lie algebra is
$( L(\xi,\xi^*),d,[\cdot,\cdot\tilde])$. 

In the general case where the differential $d$ is
not inner, a further extension is needed: define an element $\bar D$ 
of degree $1-\epsilon$ and consider the direct sum of the one
dimensional vector space generated by $\bar D$ with $L(\xi,\xi^*)$, 
$\bar L(\xi,\xi^*)=k\bar D\oplus L(\xi,\xi^*)$. The space $\bar
L(\xi,\xi^*)$ is turned into a graded Lie algebra by extending the
bracket $[\cdot,\cdot\tilde]$ according to 
\bea
\frac{1}{2}[\bar D,\bar D\tilde]=0,\ \ \
\ \ \ \ \  \ \ \ \ \ \ \ \ \ \ \ \ \ \ \ \ \cr
[\bar D,
\xi^{a_1}\dots\xi^{a_m}x\tilde]=
s(
\xi^{a_1}\dots\xi^{a_m}x)
=-(-)^{tot(\xi^{a_1}\dots\xi^{a_m}x)+\epsilon}
[\xi^{a_1}\dots\xi^{a_m} x,\bar D\tilde ],\cr
[\bar D,\xi^*_b\lambda^b(\xi)\tilde]=0=[\xi^*_b\lambda^b(\xi),
\bar D\tilde]=0.\ \ \ \ \  \ \ \ \ \ \ \ \ \ \ \ \ \ \ \ 
\eea 
It follows that $(\bar
L(\xi,\xi^*),d=[\bar D,\cdot\tilde],[\cdot,\cdot\tilde])$ is a graded
differential Lie algebra with an inner differential. 

\vspace{.25cm}

{\footnotesize In the case where
the differential $d$ on $L$ is inner, one does not need to further 
extend the space, 
$\bar L(\xi,\xi^*)=L(\xi,\xi^*)$ and $\bar D=D\in
L(\xi,\xi^*)$.}

\vspace{.25cm}

Define now 
\beq
\bar D^1 =\bar D +e_a\xi^a\in \bar L^{1-\epsilon}(\xi,\xi^*)
\eeq 
and 
\beq
\tilde\delta=d+
e_a\frac{\partial^L}{\partial
\xi^*_a}=[\bar D,\cdot\tilde]+[e_a\xi^a,\cdot]_\xi.
\eeq
It follows directly that
$\tilde\delta^2=0$ in $L(\xi,\xi^*)$, but also that the cohomology of
$\tilde\delta$ is trivial,
\beq
H(\tilde\delta,L(\xi,\xi^*))=0.
\eeq 
Indeed, the condition $\tilde\delta[
\xi^{a_1}\dots\xi^{a_m} x +\xi^*_a\lambda^a(\xi)]=0$ implies
separately $dx=0\Leftrightarrow x=e_b\mu^b+dy$ and $\lambda^a(\xi)=0$,
so that every cocycle is a coboundary, $\xi^{a_1}\dots\xi^{a_m} x
+\xi^*_a\lambda^a(\xi)=\xi^{a_1}\dots\xi^{a_m} x=
\tilde\delta[(-)^{tot(\xi^{a_1}\dots\xi^{a_m})}
\xi^{a_1}\dots\xi^{a_m}(y+\xi^*_b\mu^b)]$. 

\subsubsection{Master equation in $\bar L(\xi,\xi^*)$ 
with acyclic differential through HPT}

Take as resolution degree in $L(\xi,\xi^*)$ the number of $\xi$'s and
define $\tilde d^1=[\bar D^1,\cdot\tilde]=\tilde\delta +
[e_a\xi^a,\cdot]$. The bracket
$\frac{1}{2}[\bar D^1,\bar D^1\tilde]
=\frac{1}{2}[e_a\xi^a,e_b\xi^b]$ is of
resolution degree $2$, so that $(\tilde
d^1)^2=\frac{1}{2}[[e_a\xi^a,e_b\xi^b],\cdot]$ is of resolution degree
$1$. 

Suppose that we have constructed $\bar D^k=\bar D+\bar D_1+\dots
\bar D_k$, with $\bar D_i\in \bar
L^{1-\epsilon}(\xi,\xi^*)$ for $i=1,\dots, k$ of resolution degree $i$ 
such that 
$\frac{1}{2}[\bar D^k,\bar D^k\tilde]=R_{k+1}+R_{r>k+1}$, where
$R_{k+1},R_{r>k+1}\in L(\xi,\xi^*)^{2-\epsilon}$ have respectively
resolution degree $k+1$ and $r>k+1$. From
$0=[\bar D^k,\frac{1}{2}[\bar D^k,\bar 
D^k\tilde]\tilde]=\tilde\delta R_{k+1} +
R^\prime_{r>k+1}$, we find at resolution degree $k+1$ that 
$\tilde \delta R_{k+1}=0$ and hence that $R_{k+1}=-\tilde\delta \bar
D_{k+1}$,
where $\bar D_{k+1}\in  L(\xi,\xi^*)^{1-\epsilon}$ is of resolution degree
$k+1$. It follows that if $\bar D^{k+1}=\bar D^k+\bar D_{k+1}$,
$\frac{1}{2}[\bar D^{k+1},\bar D^{k+1}\tilde]=
R^{\prime\prime}_{r>k+1}$. Hence,
the construction can be continued recursively to get a
$\bar D(\xi,\xi^*)=\bar D+\sum_{k\geq 1}\bar 
D_k\in \bar L(\xi,\xi^*)^{1-\epsilon}$ of the form 
\beq
\bar D(\xi,\xi^*)=\bar D+ e(\xi)+
\xi^*_bf^b(\xi),
\eeq
such that
\bea
\frac{1}{2}[\bar D(\xi,\xi^*),\bar D(\xi,\xi^*)\tilde]=0.\label{me}
\eea 
The corresponding
differential in $L(\xi,\xi^*)$ is
\bea
\tilde d= [\bar 
D(\xi,\xi^*),\cdot\tilde].
\eea
The cohomology of $\tilde d$ is trivial, 
\beq
H^*(\tilde d,
L(\xi,\xi^*))=0.
\eeq 
Indeed, developing
the cocycle $l$ in $\tilde d l=0$ according to the resolution degree,
$l=\sum_{r\geq M} l_r$ we get at lowest order $\tilde\delta
l_M=0\Longrightarrow l_M=\tilde\delta k_M$, so that $l-\tilde d k_M$
starts at resolution degree $M+1$ and one can continue recursively to
show that $l=\tilde d k$ for some $k\in L(\xi,\xi^*)$. 

\subsubsection{Decomposition}

If $\bar V(\xi)=k\bar D\oplus V(\xi)$,
the solution $\bar D(\xi,\xi^*)$ splits into 
\bea
D(\xi)=\bar D+ e(\xi)\in \bar V(\xi),\cr
\Delta^*=\xi^*_bf^b(\xi)
\in RDer[\wedge(s K^*)]^\epsilon,
\eea
The master equation (\ref{me}) in $RDer[\wedge(s K^*)]^\epsilon$ gives
\bea
\frac{1}{2}[\Delta^*,\Delta^*]_\xi=0.
\eea
This equation is equivalent to 
$\Delta^2=0$ and proves (\ref{mt1}).
Explicitly, for each $r\geq 3$, it gives the set of
higher order Jacobi identities 
\bea
\sum_{m=2}^{r-1}mf^c_{(a_1\dots a_{m-1}|b|}f^b_{a_m\dots a_r)}=0,
\eea
where the round parenthesis denote graded symmetrization with respect
to the grading of the $\xi^a$.

The associated differential in $RDer[\wedge(s K^*)]^\epsilon$ is
$d_{\Delta^*}=[\Delta^*,\cdot]_\xi$. 
The graded differential Lie algebra $(RDer[\wedge(s K^*)]^\epsilon,
d_{\Delta^*},[\cdot,\cdot]_\xi)$ can be
transformed to the even graded differential Lie algebra 
$(RDer[\wedge(s K^*)],d_\Delta,
[\cdot,\cdot]_C)$ through the substitution $\xi^*_b\leftrightarrow
\frac{\partial^R}{\partial \xi^b}$. 
The master equation (\ref{me}) in $V(\xi)$ reduces to equation (\ref{mt2}), 
while in the case of an inner differential $d=[D,\cdot]$, it reduces to 
the quantum master equation (\ref{mt3}). 

The isomorphism between $H^*(\bar d, V(\xi))$ and $H^*(d_\Delta,
RDer[\wedge(s K^*)])$ follows directly from the triviality of 
the cohomology of $\tilde 
d$. Indeed, for $l=a(\xi)+\xi^*_a\nu^a(\xi)$ and
$k=b(\xi)+\xi^*_a\lambda^a(\xi)$, the relation $\tilde d
l=0\Longleftrightarrow l=\tilde d k$ gives, in the case where 
$\nu^a(\xi)=0$, 
\bea
\bar d a(\xi)=0\Longleftrightarrow\left\{\begin{array}{c}
a(\xi)=\bar d
b(\xi) +\frac{\partial^R e(\xi)}{\partial \xi^a}\lambda^a(\xi),\cr
[\Delta^*,\xi^*_b\lambda^b(\xi)]_\xi=0.
\end{array}\right.
\eea
In
order to find the cohomology, we need to know when this decomposition
is direct, i.e., we need to solve $\bar d
b(\xi) +\frac{\partial^R e(\xi)}{\partial \xi^a}\lambda^a(\xi)=0$,
under the condition $[\Delta^*,\xi^*_b\lambda^b(\xi)]_\xi=0$. This is
equivalent to $\tilde d (b(\xi) +\xi^*_a\lambda^a(\xi))=0$, and, 
using again the triviality of
the cohomology of $\tilde d$, this implies that 
$b(\xi) +\xi^*_a\lambda^a(\xi)=\tilde d (c(\xi)+\xi^*_c\mu^c(\xi))$,
or explicitly, 
\bea
\left\{\begin{array}{c}
\bar d
b(\xi) +\frac{\partial^R e(\xi)}{\partial \xi^a}\lambda^a(\xi)=0,\cr
[\Delta^*,\xi^*_b\lambda^b(\xi)]_\xi=0,
\end{array}\right.\Longleftrightarrow
\left\{\begin{array}{c}
b(\xi)=\bar d c(\xi)+\frac{\partial^R e(\xi)}{\partial
\xi^a}\mu^a(\xi),\cr
\xi^*_a\lambda^a(\xi)=[\Delta^*,\xi^*_c\mu^c(\xi)]_\xi,
\end{array}\right.
\eea
This implies that the decomposition
is direct iff $\xi^*_a\lambda^a(\xi)$ is not a coboundary and proves
(\ref{mt4}). 

\subsection{Spectral sequence for the computation of 
the cohomologies of $d_\Delta$ and $\bar d$}

The purpose of this subsection is to discuss briefly the general
computation of the cohomology $H(d_\Delta,RDer[\wedge(s K^*)])$ (and
thus also of $H(\bar d,L(\xi))$ because of the isomorphism
(\ref{iso}), (\ref{mt4})). 
In order to do so, we give some details on exact couples and spectral
sequences and apply these concepts to the present problem. 

Let $\lambda=\frac{\partial^R\cdot}{\partial\xi^A}\lambda^A$ be a
right derivation. We assume that the $\lambda^A$ are formal power
series in $\xi^A$. In the following, we provide this space with an 
obvious filtration. It will however not have finite length in general,
and for 
particular theories, better filtrations have to be found in order 
to do a complete computation. Since the techniques will be similar, 
it is nevertheless useful to show how they work for this filtration. 

\subsubsection{Grading and filtration on the space of right derivations}

Let $N_\xi=\frac{\partial^R}{\partial \xi^a}\xi^a$ be 
the operator counting the
number of $\xi$'s. A general right derivation admits the following
decomposition according to the eigenvalues of $N_\xi$:
$\lambda=\lambda_{-1}+\lambda_0+\lambda_{1}+\dots$, where
$[\lambda_p,N_\xi]= p \lambda_p$. 
Hence, $RDer[\wedge(s K^*)]$ is a graded space, 
$RDer[\wedge(s K^*)]=\oplus_{p=-1} RDer[\wedge(s K^*)]^p$. (It is actually a 
bigraded space, the other grading, for which $d_{\Delta}$ is
homogeneous of degree $1$ being the grading $tot$.)

The graded right commutator 
satisfies $[[\lambda_m,\mu_n],N_\xi]=(m+n)
[\lambda_m,\mu_n]$. The decomposition of $\Delta$
starts at eigenvalue $1$: $\D=\D_{1}+\D_{2}+\dots$~; the
corresponding decomposition of $d_{\Delta}$ being $d_{\Delta}=
[\D_{1},\cdot]+[\D_{1},\cdot]=\dots\equiv d_1+d_2+\dots$. 
 It follows that the cocycle condition
$d_{\Delta}\lambda=0$ decomposes as 
\bea
d_1\lambda_{-1}&=0,\cr
d_1\lambda_{0}+d_2\lambda_{-1}&=0,\cr
d_1\lambda_{1}+d_2\lambda_{0}+d_3\lambda_{-1}&=0,\cr
&\vdots,
\eea
while the coboundary condition $\lambda=d_{\Delta}\mu$ decomposes
as
\bea
\lambda_{-1}=0,\cr
\lambda_{0}=d_1\mu_{-1},\cr
\lambda_{1}=d_1\mu_{0}+d_2\mu_{-1},\cr
\lambda_{2}=d_1\mu_{1}+d_2\mu_{0}+d_3\mu_{-1},\cr
\hspace*{-1cm}\vdots,
\eea
In order to construct the spectral sequence associated to this
problem, we follow \cite{BoTu}. 

Consider the spaces $K_{p}$ of derivations having $N_{\xi}$
degree greater than $p$, i.e., $\lambda\in K_p$ if
$\lambda=\lambda_{p}+\lambda_{p+1}+\dots$. The space of all right
derivations is $RDer[\wedge(s K^*)]=K_{-1}$, $K_{p+1}\subset K_{p}$
and $d_{\Delta}
K_p\subset K_p$. The sequence of spaces $K_p$ is a decreasing
filtration of $RDer[\wedge(s K^*)]$, with $K_p/K_{p+1}\simeq 
RDer[\wedge(s K^*)]^p$.  

We have the short exact sequence\footnote{A diagram is said to be exact if the
  image of a map is equal to the kernel of the next map.}: 
\beq
0\longrightarrow \oplus_{p=-1} K_{p+1}\stackrel{i}{\longrightarrow}
  \oplus_{p=-1}K_p\stackrel{j}{\longrightarrow} \oplus_{p=-1} K_p/K_{p+1}
\longrightarrow 0,
\eeq
where $\oplus_{p=-1} K_p/K_{p+1}\simeq \oplus_{p=-1}RDer[\wedge(s K^*)]^p$. 
The following diagram is exact at each corner: 
\begin{eqnarray}
\begin{array}{ccc}
H(d_{\Delta},\oplus_{p=-1}K_{p+1})\stackrel{i_0}{\longrightarrow}
H(d_{\Delta},\oplus_{p=-1}K_p)\\
k_0\nwarrow\ \swarrow j_0\\
E_0,
\end{array}\label{dec}
\end{eqnarray}
where $E_0=\oplus_{p=-1} K_p/K_{p+1}\simeq \oplus_{p=-1} 
RDer[\wedge(s K^*)]^p$. 
In this diagram, $H(d_{\Delta},K_{p})$ is defined by the cocycle condition 
$d_{\Delta}(\lambda_p+\lambda_{p+1}+\dots)=0$, and the coboundary
condition $\lambda_p+\lambda_{p+1}+\dots=d_{\Delta}(\mu_p+\mu_{p+1}
+ ...)$. The maps $i_0$ and $j_0$ are induced by $i$ and $j$,
$i_0[\lambda_{p+1}+\lambda_{p+2}+\dots]=[\lambda_{p+1}+\lambda_{p+2}+\dots]$ 
and $j_0[\lambda_p+\lambda_{p+1}+\dots]=[j(\lambda_p+\lambda_{p+1}+\dots)]
=[\lambda_p]$. 
They are well defined, because $i_0$ maps cocycles to cocycles and
coboundaries to coboundaries, while
$j(d_{\Delta}(\mu_p+\mu_{p+1}\dots))
\in K_{p+1}$. 
The map $k_0$ is defined by $k_0
[\lambda_p]=[d_{\Delta}\lambda_p]$. It does not depend on the choice 
of representative for $[\lambda_p]\in K_p/K_{p+1}$, because
$[d_{\Delta}(\lambda_{p+1}+\dots)]=0\in H(d_{\Delta},K_{p+1})$. 

Let us check explicitly that this diagram is exact:
\begin{itemize}
\item ${\rm ker}\ j_0$ is given by elements $[\lambda_p+\lambda_{p+1}+...]\in 
H(d_{\Delta},K_{p})$
such that $d_{\Delta}(\lambda_p+\lambda_{p+1}+...)=0$ and
$\lambda_p=0$. This is the same than $i_{0}H(s,K_{p+1})$, which is
given by $[\lambda_{p+1}+\lambda_{p+2}+\dots]$, with
$d_{\Delta}(\lambda_{p+1}+\lambda_{p+2}+...)=0$, the equivalence
relation being the equivalence relation in $H(s,K_{p})$ by definition
of $i_0$. 
\item ${\rm ker}\ k_0$ is given by elements $[\lambda_p]$ such that 
$[d_{\Delta}\lambda_p]=0\in H(d_{\Delta},K_{p+1})$, i.e. such that 
$d_{\Delta}\lambda_p=d_{\Delta}(\mu_{p+1}+\mu_{p+2}+....)$. By
the identification
$\lambda_{p+1}=-\mu_{p+1},\lambda_{p+2}=-\mu_{p+2},\dots$, this is
indeed the same than $j_0 H(d_{\Delta},K_p)$ given by $[\lambda_p]$
with $d_{\Delta}(\lambda_p+\lambda_{p+1}+\dots)=0$. 
\item ${\rm ker}\  i_0$ is given by elements 
$[\lambda_{p+1}+\lambda_{p+2}+...]$
  such that $d_{\Delta}(\lambda_{p+1}+\lambda_{p+2}+...)=0$ 
and $\lambda_{p+1}+\lambda_{p+2}+...=d_{\Delta}
(\mu_{p}+\mu_{p+1}+...)$, while $k_0[\mu_p]$ is given by 
$[\lambda_{p+1}+\lambda_{p+2}+...]$ of the form 
$[d_{\Delta}\mu_{p}]$ so that $\lambda_{p+1}+\lambda_{p+2}+...=
d_{\Delta}\mu_{p}+d_{\Delta}(\mu_{p+1}+\dots)$, which is indeed
the same.
\end{itemize}

\subsubsection{Exact couples and associated spectral sequence}
To every exact couple $(A_0,B_0)$, i.e., exact diagram of the form 
\begin{eqnarray}
\begin{array}{ccc}
A_0\stackrel{i_0}{\longrightarrow}
A_0\\
k_0\nwarrow\ \swarrow j_0\\
B_0,
\end{array}\label{dec1}
\end{eqnarray}
one can associated a derived exact couple 
\begin{eqnarray}
\begin{array}{ccc}
A_1\stackrel{i_1}{\longrightarrow}
A_1\\
k_1\nwarrow\ \swarrow j_1\\
B_1.
\end{array}\label{dec2}
\end{eqnarray}
In this diagram, the spaces and maps are defined as follows: 
$A_1=i_0A_0$; $B_1=H(d_0,B_0)$, where $d_0=j_0\circ k_0$ ( $d_0^2=0$ because 
$k_0\circ j_0=0$); for $a_1=i_0a_0$, $i_1 a_1=i_1 (i_0a_0)
=i_0^2a_0$; $j_1 a_1=[j_0a_0]$ (this map is well defined: $j_0a_0$ is a
cocycle, because $k_0\circ j_0=0$, furthermore the map does not depend on
the representative chooser for $a_0$, 
because if $i_0a_0=0$, $a_0=k_0 b_0$ for some $b_0$ and
$j_1 a_1 =[j_0\circ k_0 b_0]=0$); $k_1 [b_0]=k_0b_0$ ($k_0b_0 =i_0a_0$ for
some $b_0$ because $d_0b_0=j_0(k_0b_0)=0$ implies $k_0b_0=i_0a_0$, 
furthermore $k_0 d_0c_0=0$
because $k_0\circ j_0=0$).  

Let us also check explicitly exactness of this diagram:
\begin{itemize}
\item ${\rm ker}\ j_1$ is given by elements $a_1=i_0a_0$ such that
  $[j_0a_0]=0$, i.e., $j_0a_0=j_0k_0 b_0$ and then $a_0-k_0b_0
  =i_0c_0$, 
implying that
  $a_1 =i^2_0c_0$. $i_1
  A_1$ is given by elements $a_1=i_1 c_1 =i^2_0c_0$. 
  It follows that ${\rm ker}\ j_1\subset i_1
  A_1$, while the inverse inclusion follows from $j_0\circ i_0=0$.  
\item ${\rm ker}\ k_1$ is given by elements $[b_0]$ such that 
$k_0b_0=i_0a_0=0$, i.e., such that $b_0=j_0c_0$, for some $c_0$, 
while ${\rm im}\ j_1$ is given by elements $[b_0]$ such that
$[b_0]=[j_0e_0]$, i.e $b_0=j_0(e_0 +k_0f_0)$. It follows that ${\rm ker}\
k_1={\rm im}\ j_1$.
\item ${\rm ker}\  i_1$ is given by elements $a_1=i_0a_0$ such
  that $i_0(i_0a_0)=0$, i.e., $i_0a_0=k_0b_0$ 
(which implies in particular $d_0b_0=0$).  
${\rm im}\ k_1$ is given by elements $a_1 =i_0a_0=k_0b_0$ for some
$b_0$ with $d_0b_0=0$, so both spaces are indeed the same. 
\end{itemize}

Clearly, this construction can be iterated by taking as the starting
exact couple the derived couple. We thus get 
a sequence of exact couples 
\begin{eqnarray}
\begin{array}{ccc}
A_r\stackrel{i_r}{\longrightarrow}
A_r\\
k_r\nwarrow\ \swarrow j_r\\
B_r.
\end{array}\label{decr}
\end{eqnarray}
and the associate spectral sequence
$(B_r,d_r)$, for $r=0,1,\dots$, i.e., spaces $B_r$ and differentials
$d_r$ satisfying $B_{r+1}=H(d_r,B_r)$.

\subsubsection{Spectral sequence associated to $d_\Delta$}
Let us now apply the general theory to the case of the exact couple
(\ref{dec}) and give explicitly the differentials $d_r$ and the spaces 
$B_r$ (called $E_r$) in this case for $r=0,1,2,3$.  

We have $E_0=\oplus_{p=-1} K_p/K_{p+1}\simeq \oplus_{p=-1} 
RDer[\wedge(s K^*)]^p$. The
differential $d_0$ is defined by $d_0[\lambda_p]_0=j_0[d_{\Delta}
\lambda_p]$, where 
$[d_{\Delta}\lambda_p]\in H(d_{\Delta},K_{p+1})$. It follows that 
$d_0[\lambda_p]_0=[d_1\lambda_p]$.
This means that $E^p_1$ is defined by elements $[[\lambda_p]_0]_1$
with the cocycle condition 
\beq
d_1 \lambda_p=0
\eeq 
and the coboundary condition
\beq
\lambda_p=d_1\mu_{p-1}. 
\eeq
Because 
$d_1=\frac{\partial^R}{\partial \xi^a}f^{a}_{bc}\xi^b\xi^c$, and  
$d_1^2=0$ implies that the $f^{a}_{bc}$ are the structure constants of a 
graded Lie algebra, this
group is just a graded version of standard Lie algebra 
(Chevalley-Eilenberg) 
cohomology with representation space the adjoint representation. 

Take now $[[\lambda_p]_0]_1\in E^p_1$. The differential 
$d_1[[\lambda_p]_0]_1=j_1k_1[[\lambda_p]_0]_1=j_1k_0[\lambda_p]_0=j_1
[d_{\Delta}\lambda_p]=[j_0i_0^{-1}[d_{\Delta}\lambda_p]]_1$. This
means that $[d_{\Delta}\lambda_p]$ has to be considered as an
element of $H(d_{\Delta},K_{p+2})$ so that 
$d_1[[\lambda_p]_0]_1=[[d_2\lambda_p]_0]_1$.
Hence $E^p_2$ is defined by elements $[[[\lambda_p]_0]_1]_2$ with 
the cocycle condition
\bea
d_2\lambda_p+d_1\lambda_{p+1}=0,\\
d_1\lambda_p=0,
\eea
and the coboundary condition
\bea
\lambda_p=d_2\mu_{p-2}+d_1\mu_{p-1},\\
0=d_1\mu_{p-2}.
\eea
We thus find that $E_2^p=H^p(d_2,H(d_1))$. 

The differential $d_2$ in $E^p_2$ is defined by 
$d_2[[[\lambda_p]_0]_1]_2=j_2k_2[[[\lambda_p]_0]_1]_2=
j_2k_1[[\lambda_p]_0]_1=j_2k_0[\lambda_p]_0
=[j_1i_1^{-1}k_0[\lambda_p]_0]_2
=[[j_0i_0^{-1}i_1^{-1}k_0[\lambda_p]_0]_1]_2$. 
In order to make sure that $k_0[\lambda_p]_0$ belongs to
$i_1i_0H(d_{\Delta},K_{p+1})$ we use 
$\lambda_p+\lambda_{p+1}$ as a representative for $[\lambda_p]_0$. It
follows that $d_2[[[\lambda_p]_0]_1]_2=[[[d_3\lambda_p
+d_2\lambda_{p+1}]_0]_1]_2$. The cocycle condition for an element 
$[[[[\lambda_p]_0]_1]_2]_3\in E^p_3$ is then given by 
\bea
d_3\lambda_p+d_2\lambda_{p+1}=d_2\mu_{p+1}+d_1\mu_{p+2},\\
d_2\lambda_p+d_1\lambda_{p+1}=0,\\
d_1\lambda_{p}=0,
\eea
with $d_1\mu_{p+2}=0$. The redefinition
$\lambda_{p+1}\rightarrow\lambda_{p+1}-\mu_{p+1}$ and
$\lambda_{p+2}=-\mu_{p+2}$, then gives as cocycle condition
\bea
d_3\lambda_p+d_2\lambda_{p+1}+d_1\lambda_{p+2}=0,\\
d_2\lambda_p+d_1\lambda_{p+1}=0,\\
d_1\lambda_{p}=0.
\eea
The coboundary condition is
$[[[\lambda_p]_0]_1]_2=d_3[[[\mu_{p-3}+\mu_{p-2}]_0]_1]_2$, where
$d_1\mu_{p-3}=0,d_2\mu_{p-3}+d_2\mu_{p-2}=0$, hence
$[[\lambda_p]_0]_1=[[d_3\mu_{p-3}+d_2\mu_{p-2}]_0]_1
+d_2[[\sigma_{p-2}]_0]_1$, 
with $d_1\sigma_{p-2}=0$ which gives 
\bea
\lambda_p=d_3\mu_{p-3}+d_2\mu_{p-2}+d_2\sigma_{p-2}+d_1\rho_{p-1},\\
0=d_2\mu_{p-3}+d_1\mu_{p-2},\\
0=d_1 \mu_{p-3},\ 0=d_1\sigma_{p-2}.
\eea
The redefinition $\mu_{p-2}\rightarrow\mu_{p-2}+\sigma_{p-2}$ and 
$\rho_{p-1}=\mu_{p-1}$, then gives the coboundary condition
\bea
\lambda_p=d_3\mu_{p-3}+d_2\mu_{p-2}+d_1\mu_{p-1},\\
0=d_2\mu_{p-3}+d_1\mu_{p-2},\\
0=d_1 \mu_{p-3}.
\eea
This construction can be continued in the same way for higher $r$'s.

The original problem was the computation of 
$H(d_{\Delta},RDer[\wedge(s K^*)])=H(d_{\Delta},K_{-1})$.
From exactness of the couples (\ref{decr}), it follows that 
\bea
H(d_{\Delta},K_{-1})\simeq j_0 E^{-1}_0
\oplus {\rm ker}\ j_0\nonumber\\
\simeq
{\rm ker}\ k_0(\subset E^{-1}_0)\oplus i_0 
H(d_{\Delta},K_{0})\nonumber\\
\simeq
{\rm ker}\
k_0(\subset E^{-1}_0)\oplus {\rm ker}\
k_1(\subset E^{0}_1)\oplus i_1 i_0H(d_{\Delta},K_{1})\\
\vdots \\
\simeq \oplus_{r=0}^R{\rm ker}\
k_r(\subset E^{r-1}_r)\oplus i_R\dots i_0 H(d_{\Delta},K_{R}).
\eea
Furthermore, $E_0\simeq E_1\oplus F_0\oplus d_0 F_0$ 
and $E^{-1}_0\simeq
E^{-1}_1 \oplus F_0^{-1}$. $F_0$ does not belong to ${\rm ker}\ k_0$
because $d_0 F_0\neq 0$. Thus
${\rm ker}\ k_0(\subset E^{-1}_0))\simeq {\rm ker}\ k_1(\subset
E^{-1}_1)$. Similarly, $E_1\simeq E_2\oplus F_1\oplus d_1 F_1$ and 
$(d_1 F_1)^{-1}=(d_1 F_1)^{0}=0$. Again, $d_1[F_1]_1\neq 0$ implies
that $F_1$ does not belong to  ${\rm ker}\ k_1$. This means that 
${\rm ker}\ k_1(\subset
E^{-1}_1)\simeq {\rm ker}\ k_2(\subset
E^{-1}_2)$ and ${\rm ker}\ k_1(\subset
E^{0}_1)\simeq {\rm ker}\ k_2(\subset
E^{0}_2)$.
Going on in the same way, we conclude that 
${\rm ker}\ k_r(\subset E^{r-1}_r))\simeq {\rm ker}\ k_R(\subset
E^{r-1}_R))$.
We thus get 
\bea
H(d_{\Delta},K_{-1})\simeq \oplus_{r=0}^R {\rm ker}\
k_R(\subset E^{r-1}_R)\oplus i_R\dots i_0 H(d_{\Delta},K_{R}).
\eea
 
This construction is most useful 
if it would stop at some point. Indeed, suppose that $K_{R}=0$. 
Because $k_R[\dots[\lambda_p]_0\dots]_R$
belongs to $i_{R}\dots i_0 H(d_{\Delta},K_{p+R+1})=0$, it follows
that 
\beq
H(d_{\Delta},K_{-1})\simeq \oplus_{r=0}^{R-1} E^{r-1}_R.
\eeq

\subsection{Examples}

\subsubsection{Lie algebra cochains and chains}

As a first example, we consider the tensor product of chains and
cochains on a semi-simple Lie algebra ${\cal G}$ with generators 
$T_I$, This space can be identified with the space 
\beq
L=\wedge({\cal G})\otimes\wedge({\cal G}^*)=\wedge ({\cal P}_I,C^I),
\eeq
where ${\cal P}_I$ and $C^I$ are
Grassmann odd variables. The grading is obtained by giving degree $-1$
to ${\cal P}_I$ and degree $1$ to $C^I$. 
The graded Lie bracket on $L$ (with $\epsilon =0$) is taken to be 
\beq
\{\cdot,\cdot\}=\frac{\partial^R\cdot}{\partial {\cal P}_I}
\frac{\partial^L\cdot}{\partial C^I}
+\frac{\partial^R\cdot}{\partial C^I}
\frac{\partial^L\cdot}{\partial {\cal P}_I}.
\eeq
The differential $d$ on $L$ is inner, 
\bea
\Omega=\frac{1}{2}{f_{IJ}}^K{\cal P}_K C^JC^I,\ d=\{\Omega,\cdot\},\\
d=C^J {f_{JI}}^K{\cal P}_K\frac{\partial^L}{\partial {\cal
    P}_I}+\frac{1}{2} C^JC^I{f_{IJ}}^K\frac{\partial^L}{\partial
  C^K}. 
\eea

It is the standard Chevalley-Eilenberg differential with
representation space the space of chains transforming under the
extension of the adjoint representation. The interesting feature of this
example is that the cohomology $H(d,L)$ can be computed exactly: it is
generated by the primitive invariant chains and the primitive
invariant cochains, which
are completely known for semi-simple Lie algebras (see e.g. \cite{GHV}).

Note that this example has more structure than discussed in the
general case. Indeed, $(L,d,\{\cdot,\cdot\})$ is a graded differential
Poisson algebra for the exterior product of chains and cochains.

In the particular case of $su(2)$ a basis $\{e_a\}$ of $K\simeq H(d,L)$ is
given by 
\bea
e_1=1,e_2=\frac{1}{3}{f_{IJK}}C^IC^JC^K,e_3=\frac{1}{3}
f^{IJK}{\cal P}_I{\cal P}_J{\cal P}_K, e_4=e_2e_3,
\eea
where the indices are lowered and raised with the Killing metric
$g_{IJ}={f_{IL}}^K{f_{JK}}^L$ and its inverse. Direct computation
shows that only one bracket of the elements of the basis is non
vanishing and that it 
is $d$ exact, $\{\xi^2e_2,\xi^3e_3\}+\{\Omega,\xi^2\xi^3e_{23}\}=0$, where
$e_{23}= -\frac{2}{3}{f^{IJ}}_K{\cal P}_I{\cal P}_JC^K$. Furthermore,
$\{e_{23},e_a\}=0=\{e_{23},e_{23}\}$. It follows that $\Delta=0$ and that
$\Omega(\xi)=\Omega+\xi^ae_a+\xi^2\xi^3e_{23}$ satisfies 
\bea
\frac{1}{2}\{\Omega(\xi),\Omega(\xi)\}=0.
\eea

In the case of a general semi-simple Lie algebra ${\cal G}$, it is not
difficult to show that the first order structure functions $f^a_{bc}$ 
all vanish. The computation of the higher order structure functions is
more involved. The complete analysis of the sh Lie structure of the
cohomology $H(d,L)$ and of the associated ``quantum master equation''
will be discussed elsewhere \cite{prep}.    

\subsubsection{Schouten-Nijenhuis bracket in $S({\cal G})\otimes
  \wedge {\cal G}^*$}

The second example is discussed in section 4 of \cite{Koszul*}. The
space $L$ is taken to be the symmetric tensors on ${\cal G}$ tensor
product with the cochains, 
\beq
L=S({\cal G})\otimes\wedge({\cal G}^*)=\wedge(x_I,C^I), 
\eeq
where the $x_I$ are generators of ${\cal G}$
considered as even coordinates on ${\cal G}^*$, with degree $0$, while
the Grassmann odd generators $C^I$ have degree $1$. The
Schouten-Nijenhuis bracket of grading $\epsilon =-1$ 
is defined by 
\beq
(\cdot,\cdot)=\frac{\partial^R\cdot}{\partial x_I}
\frac{\partial^L\cdot}{\partial C^I}-\frac{\partial^R\cdot}{\partial C^I}
\frac{\partial^L\cdot}{\partial x_I},
\eeq
Again, the differential $d$ on $L$ is inner, 
\bea
\Omega=\frac{1}{2}{f_{IJ}}^Kx_K C^JC^I,\ d=(\Omega,\cdot),\\
d=C^J {f_{JI}}^Kx_K\frac{\partial^L}{\partial x_I}
+\frac{1}{2} C^JC^I{f_{IJ}}^K\frac{\partial^L}{\partial
  C^K}. 
\eea
It is the standard Chevalley-Eilenberg differential with
representation space in this case the space of symmetric tensor on 
${\cal G}$ transforming under the
extension of the adjoint representation.  
The cohomology $H(d,L)$ can also be computed exactly in this case: it
is generated by primitive invariant symmetric tensors and 
primitive invariant cochains, which
are again completely classified for semi-simple Lie algebras (see
e.g. \cite{GHV}). 

In this example, $L$ is a Gerstenhaber algebra: 
the Schouten-Nijenhuis bracket is a graded derivation of the exterior product.
Furthermore, the bracket $(\cdot,\cdot)$ measures the failure
of the second order differential operator 
${\cal D}=\frac{\partial^R}{\partial
  x_K}\frac{\partial^R}{\partial C^K}$ to be a derivation. 

\subsubsection{Hamiltonian BFV formalism}

Another, physically relevant example of a differential graded Lie
algebra is the formalism of Batalin, Fradkin and Vilkovisky for
constrained Hamiltonian systems 
\cite{Fradkin:1975cq,Fradkin:1977wv,Fradkin:1977hw,Batalin:1977pb,%
Fradkin:1978xi,Batalin:1983pz} 
(for reviews, see e.g. 
\cite{Henneaux:1985kr,Gitman:1990qh,Govaerts:1991gd,Stasheff:1991eb,%
Henneaux:1992ig}). The
even graded Lie bracket $\{\cdot,\cdot\}$ 
is the extension of the standard Poisson
bracket to the ghost and ghost momenta, while the inner differential
is generated by the BRST charge $\Omega$, built out of the constraints
and the structure functions arising in their Poisson bracket algebra.  

\subsubsection{Lagrangian BV formalism}

The general formalism to control gauge symmetries through a
differential during quantization in the Lagrangian framework,
developed by Becchi, Rouet, Stora 
\cite{Becchi:1974xu,Becchi:1974md,Becchi:1975nq,BecchiTira}, 
Tyutin \cite{Tyutin:1975qk}, Zinn-Justin
\cite{Zinn-Justin:1974mc,Zinn-Justin:1989mi}, Batalin
and Vilkovisky 
\cite{Batalin:1981jr,Batalin:1983wj,Batalin:1983jr,Batalin:1984ss,%
Batalin:1985qj} is refereed to as the ``antifield formalism''
below. It is another physically important 
example of a differential graded Lie algebra. Since
the main objective of this review is to analyze what can be gained by the
general construction of section \ref{main} in this context, the
different ingredients of the antifield formalism will be briefly reviewed in
the next section. More details can be found in the
original papers, in the reviews 
\cite{Henneaux:1992ig,Troost:1993mr,Troost:1994xw,Gomis:1995he,Barnich:2000zw}
and in the references cited therein.

\newpage

\mysection{Antifield formalism and local BRST cohomology}\label{s3}

\subsection{Antibracket and master equation}

The starting point for local gauge field theories is an action 
\beq
S_0[\phi^i]=\int d^nx\ L_0,
\eeq
where the Lagrangian $L_0$ depends on the fields $\phi^i$ and a
finite number of their derivatives. The left hand side of the 
equations of motion are determined by the Euler-Lagrange derivatives
of $L_0$, 
\beq
\frac{\delta L_0}{\delta\phi^i}=0. 
\eeq
In gauge theories, the left hand side of the equations of motion are
not all independent, but there exist on-shell non vanishing relations among
them. These are the non trivial Noether identities in one to one
correspondence with the non trivial gauge symmetries.   

The original set of fields
$\{\phi^i\}$ is extended to the set $\{\phi^a\}$ by 
introducing in addition (i)
ghost fields  
for the non trivial Noether identities, (ii) ghosts for ghosts
associated to the non trivial reducibility identities 
of Noether identities, (iii) ghosts for ghosts for
ghosts for the second stage non trivial reducibility identities$\dots$,
and (iv) antifields $\phi^*_a$ associated to all of the above fields.
In the following, we will denote the fields and antifields
collectively by $z^\alpha$.

In the classical theory, the relevant space is the space of local
functionals ${\cal F}$ in the fields and antifields. 
Under appropriate vanishing
conditions on the fields, antifields and their derivatives at infinity,  
this space is isomorphic to the space of functions in the fields, 
the antifields and a finite number of their derivatives, 
up to total divergences. Introducing the horizontal one forms $dx^\mu$
and the horizontal differential $d=dx^\mu\partial_\mu$, with
$\partial_\mu=\frac{\partial}{\partial
  x^\mu}+\sum_{k=0}z^\alpha_{,(\mu\nu_1\dots\nu_k)}
\frac{\partial}{\partial z^\alpha_{,(\nu_1\dots\nu_k)}}$, the space of
  local functionals is isomorphic to the cohomology of the horizontal
  differential $d$ in form degree $n$ in the space
  $\Omega$ of form valued local functions, ${\cal F}\simeq H^n(d,\Omega)$.
The grading is the ghost number obtained by
assigning degree $0$ to the original fields, degree $1$ to the ghosts, $2$ to
the ghosts for ghosts. The ghost number of an antifield is defined to
be minus the ghost number of the corresponding field minus $1$. 
The odd graded Lie bracket with $\epsilon=1$ of local functionals
$A_i=\int d^nx\ a_i$ is defined by 
\bea
(A_1,A_2)=\int d^nx\ \frac{\delta^R a_1}{\delta\phi^a}
\frac{\delta^L a_2}{\delta\phi^*_a}-\frac{\delta^R a_1}{\delta\phi^*_a}
\frac{\delta^L a_2}{\delta\phi^a}.
\eea
It is called ``antibracket'' in this context. 
In terms of generating functionals for Green's functions introduced
below, the antibracket is defined by a similar expression where the
Euler-Lagrange derivatives of the integrands, $\frac{\delta
  a}{\delta z^\alpha}$ is replaced by the functional derivatives 
$\frac{\delta A}{\delta z^\alpha(x)}$ for the corresponding
functionals. This generalization is consistent because for local
functionals $\frac{\delta A}{\delta z^\alpha(x)}=\frac{\delta
  a}{\delta z^\alpha}(x)$.

The central object of the formalism is the construction of a solution
$S$ to the classical master equation 
\bea
\frac{1}{2}(S,S)=0.\label{g1}
\eea
This solution is obtained by first constructing a (possibly reducible)
generating set of non trivial gauge symmetries, as well as
generating sets of non trivial reducibility identities for the gauge
symmetries, of non trivial reducibility identities of the second stage
for the previous reducibility identities$\dots$. 

If the solution $S$ of the master equation (\ref{g1}) is required
to be in ghost number $0$, Grassmann even, minimal and proper,
i.e., to contain in addition to the starting point action $S_0$ the
gauge transformations related to the generating set of non trivial
gauge symmetries as well as the various reducibility identities in a
canonical way, one can show existence and locality of this solution, with
uniqueness holding up to canonical field-antifield redefinitions.
The BRST differential is then $s=(S,\cdot)$, so that $({\cal
F},(\cdot,\cdot),s)$ is a graded differential Lie algebra with an inner 
differential.

\subsection{Local BRST cohomology}

The cohomology of $s$ in the space of local functionals, $H^*(s,{\cal
  F})$ is called local BRST cohomology. With the algebraic
characterization of local functionals discussed above, it 
is isomorphic to the cohomology of $s$ modulo $d$ in form degree $n$, 
$H^{*,n}(s|d,\Omega)$, in the space of form valued local functions
$\Omega$. This cohomological group is in turn related to the
cohomology of $s$ in $\Omega$, $H(s,\Omega)$ through descent
equations. 

In negative ghost numbers $g$, the groups $H^{g,n}(s|d,\Omega)$
describe the generalized non trivial global symmetries of the
theory. In ghost number zero, they describe the observables, i.e., the
equivalence classes of local functionals that are gauge invariant when
the gauge covariant equations of motion $\delta L_0/\delta\phi^i=0$
hold, where two such functionals, that coincide when these equations
hold, have to be identified. The BRST cohomology in ghost number $0$
also describes the infinitesimal deformations of the master equation,
the obstructions to deformations being described by local BRST
cohomological classes in ghost number $1$. As will be discussed below,
the local BRST cohomology in ghost number $0$ also constrains the
divergences and the counterterms of the quantum theory, while the classes in
ghost number $1$ describe the anomaly candidates. 

\subsection{Gauge fixing}

In order to get well defined propagators, needed as a starting point
for perturbation theory, the gauge has to be fixed. 
The gauge fixing can be done in two
steps: first one adds a cohomological trivial non minimal
sector. This amounts to extending the
minimal solution of the master equation to $S^\prime=S+\int d^nx\
B^a\bar C^*_a$.
The canonical BRST differential
extended to the antifields and the non minimal sector is 
$s=(S^\prime,\cdot)_{\phi,\phi^*}$.
The second step is to perform an anticanonical transformation generated
by a gauge fixing fermion $\Psi[\phi^a]$: the gauge fixed action to
be used for quantization is $S_{\rm gf}[\phi^a,\tilde\phi^*_a]=
S^\prime[\phi^a,\tilde\phi^*_a+\frac{\delta^L\Psi}{\delta
  \phi^a}]$, with $\Psi$ chosen in such a way that the propagators
of the theory are well defined. 
For instance, in Yang-Mills type theories, standard linear gauges are 
obtained from 
\bea
\Psi=\int d^nx\ \bar
C_a(\partial^\mu A^a_\mu +\frac{1}{2}\alpha B^a).\label{psi}
\eea
The  cohomology of the associated
BRST differential $s=(S_{\rm gf},\cdot)_{\phi,\tilde\phi^*}$ in the space of
local functions or in the space of local functionals is
isomorphic to the cohomology of the canonical BRST differential in the
respective spaces and can be
obtained from it through the shift of antifields
$\phi^*=\tilde\phi^*+\delta^L\Psi/\delta\phi$.
The dependence of the gauge fixed action on the fields and 
antifields of the non
minimal sector is explicitly given by 
\bea
\begin{array}{cc}
\frac{\delta^R
S_{\rm gf}}{\delta\bar C^a}=-(S_{\rm
gf},\frac{\delta^R\Psi}{\delta\bar C^a})_{\phi,\tilde\phi^*},\ 
&\frac{\delta^R
S_{\rm gf}}{\delta B^a}=-(S_{\rm
gf},\frac{\delta^R\Psi}{\delta B^a})_{\phi,\tilde\phi^*}+\tilde{\bar C^*_a},
\nonumber\\
\frac{\delta^R
S_{\rm gf}}{\delta\tilde{\bar C^*_a}}=B^a,\
&\frac{\delta^R
S_{\rm gf}}{\delta\tilde{B^*_a}}=0.\label{5bis}
\end{array}
\eea 

The dependence of the gauge fixed solution of the master equation 
$S_{\rm gf}$ on any parameter $\alpha^i$ appearing in the gauge
fixing fermion $\Psi$ alone, and not in the minimal solution $S$ of
the master equation, is given by 
\beq
\frac{\partial S_{\rm gf}}{\partial^R \alpha^i}=
-(S_{\rm gf},\frac{\partial^R\Psi}{\partial\alpha^i}).\label{gaugedep}
\eeq

In the following, it will always be understood that the gauge fixed
action with tilded antifields is used in manipulations involving the
Green's functions, even if we do not always indicate this explicitly,
when we are interested in statements concerning the local BRST
cohomology.  

\newpage

\mysection{Higher order cohomological restrictions and
  renormalization}\label{s4} 

\subsection{Regularization}\label{sreg}

We will assume that there is a regularization with the properties of
dimensional regularization as explained in reference
\cite{Tonin:1992wf}, i.e.,

\begin{itemize}
\item the regularized action $S_\tau=\Sigma_{n=0}\tau^n S_n$ 
is a polynomial or a power series in
$\tau$, the $\tau$ independent part corresponding to the starting
point action $S_0=S$, so that the algebraic relations that hold for
the classical action $S$ hold in the regularized theory for $S_0$.  

\item if the renormalization has been carried out up to $n-1$ loops, 
the divergences of the effective action at $n$ loops 
are poles in $\tau$ up to the order $n$ with residues that are local
functionals, 
and 

\item the regularized quantum action principle holds 
\cite{Breitenlohner:1977hr
}.
\end{itemize}

Let $\tilde
S=S_\tau+\rho^*\theta_\tau$, with
$\theta_\tau=\frac{1}{2\tau}(S_\tau,S_\tau)$, so that
$\theta_0=(S,S_1)$, and
$\rho^*$ a global source in ghost number $-1$. On the classical level,
we have, using $(\rho^*)^2=0$,
\bea
\frac{1}{2}(\tilde S,\tilde S)=\tau\frac{\partial \tilde S}
{\partial \rho^*},\label{cl}\\
(\tilde S,\frac{\partial \tilde S}{\partial \rho^*})=0,\label{cl1}
\eea
which translates, according to the quantum action principle, 
into the corresponding equations for the regularized generating functional
for 1PI vertex functions:
\bea
\frac{1}{2}(\tilde \Gamma,\tilde \Gamma)=\tau\frac{\partial \tilde
  \Gamma}
{\partial \rho^*},\label{actp}\\
(\tilde\Gamma,\frac{\partial \tilde
  \Gamma}
{\partial \rho^*})=0.
\eea

Using $(\rho^*)^2=0$, these equations reduce to 
\bea
\frac{1}{2}(\Gamma,\Gamma)=\tau\frac{\partial \tilde
  \Gamma}
{\partial \rho^*},\label{3}\\
(\Gamma,\frac{\partial \tilde
  \Gamma}
{\partial \rho^*})=0.
\eea
At one loop, we get 
\bea
(S_\tau,\Gamma^{(1)})=\tau\theta^{(1)},\label{g}\\
(S_\tau,\theta^{(1)})+(\Gamma^{(1)},\theta_\tau)=0,\label{l}
\eea
where $\Gamma^{(1)}$ and $\theta^{(1)}$ are respectively the one loop
contributions of $\Gamma$ and $\partial \tilde
  \Gamma/{\partial \rho^*}$. By assumption, we have
\bea
\Gamma^{(1)}=\sum_{n=-1}\tau^n\Gamma^{(1)n},\\ 
\theta^{(1)}=\sum_{n=-1}\tau^n\theta^{(1)n},
\eea
where $\Gamma^{(1)-1}$
and $\theta^{(1)-1}$ are local functionals.

\subsection{Lowest order cohomological restrictions}

At ${1}/{\tau}$, equations (\ref{g}) and (\ref{l}) give
\bea
(S,\Gamma^{(1)-1})=0\label{3.4},\\
(S,\theta^{(1)-1})+(\Gamma^{(1)-1},\theta_0)=0\label{3.3}.
\eea
Using 
$\theta_0=(S,S_1)$ and equation (\ref{3.4}), equation (\ref{3.3}) 
reduces to 
\bea
(S,\theta^{(1)-1}-(S_1,\Gamma^{(1)-1}))=0.
\eea
In addition, we get, from the term independent of $\tau$ in 
equation (\ref{g}),
\bea
(S,\Gamma^{(1)0})=\theta^{(1)-1}-(S_1,\Gamma^{(1)-1})
\label{3.6}.
\eea
The term linear in $\tau$ gives
\bea
(S,\Gamma^{(1)1})=\theta^{(1)0}-(S_1,\Gamma^{(1)0})
-(S_2,\Gamma^{(1)-1}).
\eea
The one loop renormalized effective action is 
$\Gamma^1_R=
S+\hbar \Gamma^{(1)0}+O(\hbar^2,\tau)$,
where the notation $O(\hbar^2,\tau)$ means that those terms which are
not of order at least two in $\hbar$ are of order at least one in
$\tau$ and vanish when the regularization is removed
($\tau\longrightarrow 0$), so that 
\bea
{1\over 2}(\Gamma^1_R,\Gamma^1_R)=\hbar A_1+O(\hbar^2,\tau),
\eea
with, using equation (\ref{3.6}),
\bea
A_1=\theta^{(1)-1}-(S_1,\Gamma^{(1)-1}),\label{a}
\eea 
and the consistency conditions for local functionals
\bea
(S,\Gamma^{(1)-1})=0\Longrightarrow \Gamma^{(1)-1}
=c_1^iC_i+(S,\Xi_1),\\
(S,A_1)=0\Longrightarrow A_1=a_1^iA_i+(S,\Sigma_1),
\eea
where $[C_i]$ and $[A_i]$ are respectively a basis of representatives for
$H^{0}(s)$ and $H^{1}(s)$. These are the standard
cohomological restrictions on the divergences and the anomalies of
the quantum theory.  

\subsection{First order cohomological restrictions}\label{s43}

The idea is now to get additional cohomological restrictions on
anomalies and counterterms, using the bracket induced in cohomology
discussed in section \ref{dgla}. In order to do so, we couple to the
starting point action an arbitrary local BRST cohomological class. 

Let $D$ be a BRST cocycle in any ghost number $g$ and consider 
$S^j=S+jD$, where the source $j$ is of ghost number $-g$. The
regularized action is $S^j_\tau=S_\tau+jD_\tau$, with $D$ a polynomial
in $\tau$ starting with $D$. If $\tilde S^j=S^j_\tau+\rho^*\theta^j_\tau$,
where $\theta^j_\tau={1\over2\tau}(S_\tau,S_\tau)+{1\over\tau}
(jD_\tau,S_\tau)$, we have
\bea
{1\over 2}(\tilde S^j,\tilde S^j)=\tau 
{\partial \tilde S^j\over\partial\rho^*}+O(j^2),
\eea
and the corresponding equation for the regularized generating
functional $\tilde \Gamma_j$
\bea
{1\over 2}(\tilde \Gamma_j,\tilde \Gamma_j)=\tau 
{\partial \tilde \Gamma_j\over\partial\rho^*}+O(j^2).
\eea
At one loop, we get for the term independent of $\rho^*$,
\bea
(\Gamma^{(1)}_j,S^j_\tau)=\tau\theta^{(1)}_j+O(j^2).
\eea
The term linear in $j$ of order ${1\over\tau}$ gives
\bea
(D^{(1)-1},S)+(D,\Gamma^{(1)-1})=0,\label{heur}
\eea
with $D^{(1)-1}=(\partial\Gamma^{(1)-1}_j/\partial j)|_{j=0}$.
This gives our first theorem.
\begin{theorem}\label{th41}
The antibracket of the divergent one loop part $\Gamma^{(1)-1}$,
which is BRST closed and local, 
with any local BRST cocycle is BRST exact in the
space of local functionals.
\end{theorem}

The theorem
can be reformulated by saying that 
the antibracket map induced in the local BRST cohomology groups 
\bea
([\Gamma^{(1)-1}],[D])_M=[0]\label{e45}
\eea
for all $[D]\in H^g(s)$.
This equation represents a cohomological
restriction on the coefficients $c^i_1$ that can appear~; it can be
calculated classically from the knowledge of $H^0(s)$ and the
antibracket map  from $H^0(s)\otimes H^g(s)$ to
$H^{g+1}(s)$. 
According to the previous section, the theorem holds in particular
when $D=\Gamma^{(1)-1}$ or $D=A_1$. 

In the same way, the consistency condition is
\bea
(\Gamma_j,{\partial \tilde \Gamma_j\over\partial\rho^*})+O(j^2)=0,
\eea
and gives at one loop,
\bea
(\Gamma^{(1)}_j,\theta^j_\tau)+(S^j_\tau,\theta^{(1)}_j)+O(j^2)=0.
\eea
The term linear in $j$ of order ${1\over\tau}$ gives
\bea
(D^{(1)-1},\theta_0)-\left(\left.{\partial {\theta^j}_0\over\partial j}
\right|_{j=0},
\Gamma^{(1)-1}\right)\nonumber\\+(D,\theta^{(1)-1})
-\left(\left.{\partial \theta_j^{(1)-1}\over\partial j}\right|_{j=0},S
\right)=0.
\eea
Using $\theta_0=(S,S_1)$, ${\partial \theta^j_0/\partial
  j}|_{j=0}=(D_1,S)+(D,S_1)$,  equations (\ref{3.4}), (\ref{a})
and (\ref{heur}),
we get
\bea
(D,A_1)-\left(\left.{\partial \theta_j^{(1)-1}\over\partial
    j}\right|_{j=0}-(D_1,\Gamma^{(1)-1})-(D^{(1)-1},S_1),S\right)=0.
\label{3.9}
\eea
This gives our second result.
\begin{theorem}\label{th42}
The antibracket of the BRST closed first order anomaly $A_1$
with any local BRST cocycle is BRST exact in the
space of local functionals.
\end{theorem}

The theorem
can again be reformulated by saying that 
the antibracket map 
\bea
([A_1],[D])_M=[0]\label{e410}
\eea
for all $[D]\in H^g(s)$~; it represents a classical cohomological
restriction on the coefficients $a^i_1$ that can appear. 

\subsection{Higher orders}

Let $B^0=S$ and $B^1=\Gamma^{(1)-1}$. We have the following theorem.
\begin{theorem}
The first order counterterms can be completed into a local deformation
of $S$, i.e., there exist local functionals $B^n$ such that 
\bea 
\frac{1}{2}(S^{j^\infty},S^{j^\infty})=0,\\
S^{j^\infty}=S+\sum_{n=1}j^nB^n.
\eea
\end{theorem}
\proof{The theorem is true for $j^0,j^1$ and $j^2$, if we take
$D=\Gamma^{(1)-1}=B^1$ in (\ref{heur}) and $B^2=1/2
(\partial\Gamma^{(1)-1}_j/\partial j)|_{j=0}$. 
Suppose the theorem true at order $j^k$ i.e., we have  
\bea
\frac{1}{2}(S^{j^k},S^{j^k})=O(j^{k+1}),
\\S^{j^k}=S+\sum^k_{n=1}j^nB^n.
\eea
and 
\bea
B^n={1\over n}(\partial^{n-1}\Gamma^{(1)-1}_{j^{n-1}}/\partial j^{n-1})|_{j=0}.
\eea
At the regularized level, consider the action 
\bea
S_\tau^{j^k}=S_\tau+\sum^k_{n=1}j^nB_\tau^n
\eea
and $\tilde
S^{j^k}=S_\tau^{j^k}+\rho^*\theta_{\tau}^{j^k}$, with
$\theta_{\tau}^{j^k}=\frac{1}{2\tau}
(S_\tau^{j^k},S_\tau^{j^k})+O(j^{k+1})$, so that
\bea
\frac{1}{2}(\tilde S^{j^k},\tilde S^{j^k})=
\tau\frac{\partial\tilde S^{j^k}}{\partial
\rho^*}+O(j^{k+1}). 
\eea
The corresponding equation for 
$\tilde \Gamma_{j^k}$ based on the action $\tilde S^{j^k}$ is 
\bea
\frac{1}{2}(\tilde \Gamma_{j^k},\tilde \Gamma_{j^k})=
\tau\frac{\partial\tilde \Gamma_{j^k}}{\partial
\rho^*}+O(j^{k+1}). 
\eea
At one loop, we get, for the part independent of
$\rho^*$,  
\bea
(S_\tau^{j^k},\Gamma^{(1)}_{j^k})=
\tau\theta^{(1)}_{j^k}+O(j^{k+1}).
\eea
At order $j^k$, this equation gives 
\bea
\left (S_\tau,\left.{\partial^k
\Gamma^{(1)}_{j^k}\over\partial j^k}\right|_{j=0}\right )
+\left (B_\tau^1,\left.{\partial^{k-1}
\Gamma^{(1)}_{j^k}\over\partial j^{k-1}}\right|_{j=0}\right )
+\dots\nonumber\\+\left
(B_\tau^k,\left.\Gamma^{(1)}_{j^k}\right|_{j=0}\right )=
\tau\left.{\partial^k
\theta^{(1)}_{j^k}\over\partial j^k}\right|_{j=0}.
\eea
At order $1/\tau$, we get, using 
\bea
\left.{\partial^{n-1}
\Gamma^{(1)-1}_{j^k}\over\partial j^{n-1}}\right|_{j=0}=
\left.{\partial^{n-1}
\Gamma^{(1)-1}_{j^{n-1}}\over\partial j^{n-1}}\right|_{j=0}=nB^n,
\eea
for $n=1,\dots,k-1$ and defining $\left.{\partial^k
\Gamma^{(1)-1}_{j^k}\over\partial j^k}\right|_{j=0}=(k+1)B^{k+1}$, 
the relation
\bea
(S,(k+1)B^{k+1})+(B^1,kB^k)+\dots+(B^{k},B^1)=0,
\eea
or equivalently
\bea
0=\sum_{m=0}^k(B^m,(k+1-m)B^{k+1-m})=
{(k+1)\over 2}\sum_{m=0}^{k+1}(B^m,B^{k+1-m}),\label{id}
\eea
which proves the theorem.
} 

\vspace{1cm}

Let $E^0=A_1=\theta^{(1)-1}-(B^1,S_1)$.
\begin{theorem}
The lowest order contribution to the anomaly $E^0$ can be extended to
a local cocycle of the deformed solution of the master equation
$S^{j^\infty}$, i.e., there exist local functionals $E^m$ such that 
\bea
(S^{j^\infty},E^{j^\infty})=0,\\
E^{j^\infty}=\sum_{m=0}j^mE^m.
\eea
\end{theorem}

\proof{The theorem holds for $j^0$ and $j^1$ by taking in (\ref{3.9})
$D=B^1$, and defining 
\bea
E^1=\left.{\partial \theta_j^{(1)-1}\over\partial
    j}\right|_{j=0}-(D_1,\Gamma^{(1)-1})-(D^{(1)-1},S_1)\\=
\left.{\partial \theta_j^{(1)-1}\over\partial
    j}\right|_{j=0}-(B_1,B^1_1)-(2B^2,S_1).
\eea
Let us define
\bea
E^m=\left.{\partial^{m} \theta_{j^{m}}^{(1)-1}\over\partial
   j^m}\right|_{j=0}-\sum_{n=0}^{m}((n+1)B^{n+1},B^{m-n}_1).\label{def}
\eea
The consistency condition is 
\bea
(\Gamma_{j^k},{\partial \tilde\Gamma_{j^k}\over\partial \rho^*})=O(j^{k+1}).
\eea
At one loop, we have,
\bea
(\Gamma^{(1)}_{j^k},\theta_\tau^{j^k})+
(S_\tau^{j^k},\theta_{j^k}^{(1)})=O(j^{k+1}).
\eea
The term of order $j^k$ of this equation gives 
\bea
\sum_{m=0}^{k}
\left[\left(\left.{\partial^m \Gamma^{(1)}_{j^k}\over\partial j^m}
\right|_{j=0},\left.{\partial^{k-m} \theta_\tau^{j^k}\over\partial j^{k-m}}
\right|_{j=0}\right)
+\left(B^m_\tau,\left.{\partial^{k-m} \theta_{j^k}^{(1)}\over\partial j^{k-m}}
\right|_{j=0}\right)\right]=0.
\eea
At order $1/\tau$, we get 
\bea
\sum_{m=0}^{k}\Bigg[\left((m+1)B^{m+1},
\sum_{l=0}^{k-m}(B^l,B^{k-m-l}_1)\right)
\nonumber\\
+\left (B^m,\left.{\partial^{k-m} \theta_{j^k}^{(1)-1}\over\partial j^{k-m}}
\right|_{j=0}\right )\Bigg]=0.\label{bas}
\eea
Using the Jacobi identity, the first term is given by 
\bea
\sum_{m=0}^{k}\sum_{l=0}^{k-m}\left[\left(\left((m+1)B^{m+1},
B^{k-m-l}\right),B^l_1\right)
-\left(B^l,\left((m+1)B^{m+1},B^{k-m-l}_1\right)\right)\right].
\eea
Changing the sum $\sum_{m=0}^{k}\sum_{l=0}^{k-m}$ to the equivalent 
sum $\sum_{l=0}^{k}\sum_{m=0}^{k-l}$, the first term of this equation
vanishes on account of (\ref{id}), while the second term, using the
definition (\ref{def}), combines with the second term of (\ref{bas}) to
give
\bea
\sum_{m=0}^{k}(B^m,E^{k-m})=0,
\eea
which proves the theorem.
}

The investigation in this section is a 
first step in order to analyze
the cohomological restrictions on anomalies and counterterms at higher
orders in $\hbar$. To see this, we note that if we put
$j=(-\hbar/\tau)$, the action $S^{(-\hbar/\tau)^\infty}$ satisfies the
(deformed) master equation
$1/2(S^{(-\hbar/\tau)^\infty},S^{(-\hbar/\tau)^\infty})=0$, while the
corresponding effective action is finite at order $\hbar$. Its
divergences at order $\hbar^2$ are poles up to order $2$ in $\tau$
with residues that are local functionals. A systematic analysis of the
subtraction procedure at higher orders in $\hbar$ will be presented
in the context of the extended antifield formalism below.

\subsection{Cohomological restrictions through mixed antibracket map}

In the antifield formalism, there is an additional map relating two
different types of cohomologies, the BRST cohomology for local
functionals $H^{*,n}(s|d,\Omega)$ and the BRST cohomology for local
functions, $H^{*}(s,\Omega^0)$. It is defined in terms of the bracket 
$(\cdot,\cdot)_{alt}$ from local functionals tensor product with local
functions to local functions defined by 
\bea
(\cdot,\cdot)_{alt}: H^{n}(d,\Omega)\otimes\Omega^0\longrightarrow
\Omega^0,\ \ \ \ \ \ \ \ \ \ \ \ \ \ \\
(A,b)_{alt}=\sum_{k=0}\partial_{\mu_1}\dots\partial_{\mu_k}
\frac{\delta^R a}{\delta\phi^a}
\frac{\partial^L b}{\partial\phi^*_{a,(\mu_1\dots\mu_k)}}
-\sum_{k=0}\partial_{\mu_1}\dots\partial_{\mu_k}
\frac{\delta^R a}{\delta\phi^*_a}
\frac{\partial^L b}{\partial\phi^a_{,(\mu_1\dots\mu_k)}},
\eea
where $A=\int d^n x \ a$. It is straightforward to verify that this
bracket induces a well defined mixed map $(\cdot,\cdot)_{m}$ 
in cohomology, i.e., that it maps
cocycles to cocycles and that the resulting cohomology class does not
depend on the choice of the representatives:
\bea
(\cdot,\cdot)_{m}: H^{g_1,n}(s|d,\Omega)\otimes H^{g_2}(s,\Omega^0)
\longrightarrow H^{g_1+g_2+1}(s,\Omega^0),\\
([A],[b])_{m}=[(A,b)_{alt}].\ \ \ \  \ \ \ \ \ \ \ \ 
\eea

By using a local source $j(x)$ instead of a coupling constant $j$ to
couple the representative $d$ of a class $[d]\in H^{g}(s,\Omega^0)$,
theorems \ref{th41} and \ref{th42} of section \ref{s43} become 
\begin{theorem}
The mixed antibracket map of the first order divergences or of
the first order anomalies with any
non integrated BRST cohomology class $[d]\in H^{g}(s,\Omega^0)$ vanishes:
\bea
([\G^{(1)-1}],[d])_m=0,\\
([A_1],[d])_m=0.
\eea
\end{theorem}

\subsection{Application 1: Elimination of antifield dependent counterterms 
in Yang-Mills theories with $U(1)$ factors}

In this section, we will discuss the elimination by higher order
cohomological restrictions, imposed by the mixed antibracket map of the 
previous subsection,
of a type of antifield dependent counterterms arising in non
semi-simple Yang-Mills theories. These counterterms 
have been discussed for the first time in 
\cite{Bandelloni:1978ke,Bandelloni:1978kf}, 
were analyzed from a cohomological point of view 
in \cite{Barnich:1994ve,Barnich:1995mt}
and reconsidered in the concrete context of the standard
model in \cite{Grassi:1999mc}. These counterterms 
(also called instabilities because they are not present in the
starting point action) have the
following general structure \cite{Barnich:1994ve,Barnich:1995mt}: 
$$
K^\prime=f^\Delta_\alpha\int d^nx\ j^\mu_\Delta A_\mu^\alpha
+(A^{*\mu}_a
X^a_{\mu\Delta}+y^*_iX^i_\Delta)C^\alpha,
$$ 
where $f^\Delta_\alpha$ are constants, $A_\mu^\alpha$ abelian gauge
fields, $j^\mu_\Delta$ non trivial conserved currents and
$\delta_\Delta A^a_\mu=X^a_{\mu\Delta},\delta_\Delta y^i=X^i_\Delta$ 
the generators of the corresponding
symmetries on all the gauge fields $A_\mu^a$ and the matter fields
$y^i$. 
In order to eliminate these
instabilities by cohomological means, we will show that:

{\it It is
sufficient that there exists a set
of local, non integrated, off-shell gauge invariant polynomials $O_\Gamma(x)$
constructed out of the $A^a_\mu,y^i$ and their derivatives,
that break the global symmetries $\delta_\Delta$  
in the following sense: the variation of $O_\Gamma(x)$ under the gauged
global symmetries $\delta_\Delta$ with gauge parameter given by
$f^\Delta_\alpha \epsilon^\alpha$ should not be equal on shell 
to an ordinary gauge transformation (involving the abelian gauge
parameters $\epsilon^\alpha$ alone) of some local polynomials 
$P_\Gamma(x)$ constructed 
out of the $A^a_\mu,y^i$ and their derivatives.}

Indeed,
using the
extended action
\footnote{The author thanks P.A. Grassi for suggesting the use of
external sources instead of external couplings in this example.} 
$S_{k(x)}=S+\int d^nx\ k^\Gamma(x) O_\Gamma(x)$, which
satisfies $1/2(S_{k(x)},S_{k(x)})=0$ and the corresponding regularized
action principle, it follows from the equation independent of the sources
$k(x)$ that the divergences ${\G^{(1)}}_{div}$ of the theory without
$k(x)$ are, as usual, required to be BRST invariant. 
The terms linear in $k(x)$ then imply
\bea
(\left.\frac{\delta {\G^{(1)}_{k(x)}}_{div}}{\delta
k^\Gamma(x)}\right|_{k(x)=0},S)+(O_\Gamma(x),{\G^{(1)}}_{div})=0,
\label{sudu}
\eea
or, equivalently,
\bea
([{\G^{(1)}}_{div}],[O_\Gamma])_m=0.\label{supdup}
\eea
The second term of this equation gives
for the antifield dependent counterterms 
${\G^{(1)}}_{div}=K^\prime$
above $(O_\Gamma(x),K^\prime)=(C^\alpha f^\Delta_\alpha
\delta_{\Delta}) O_\Gamma(x)$, 
because we have $O_\Gamma(x)$ can be chosen to be
independent of the antifields. From (\ref{sudu}), it then follows that 
$(C^\alpha f^\Delta_\alpha \delta_{\Delta}) O_\Gamma(x)$ 
must be given on-shell by a 
gauge transformation,
involving the abelian ghosts alone, of polynomials $P_\Gamma(x)$. 
This follows
by using the explicit form of the BRST differential, and after
evaluation, putting to zero the antifields, and the non abelian
ghosts. Hence, the counterterms $K^\prime$ are excluded a priori 
whenever it is possible
to construct $O_\Gamma(x)$'s for which the corresponding $P_\Gamma(x)$
do not exist so that (\ref{sudu}) cannot be satisfied.

{\bf Remark}: In this example, we use external sources and the mixed
antibracket map instead of coupling constants and the standard
antibracket map because the restrictions we get are
stronger and the discussion is simplified: we need not worry 
about possible integrations by parts (in
momentum space, this means that the restrictions we get are valid for
all values of the external momentum and not only for zero external
momentum). 

The condition (\ref{supdup}) 
means that besides the arguments of 
\cite{Bandelloni:1978ke,Bandelloni:1978kf,Gomis:1996jp,Grassi:1999mc}, there
exists an elegant cohomological mechanism to eliminate this
type of antifield dependent counterterms.
 
In the concrete case of the standard model, the global symmetries
$\delta_\Delta$ correspond to lepton and baryon number
conservation. There is only one abelian ghost $C^\alpha$, the abelian gauge
transformation of the matter fields being $\delta_{abelian} y=i{\cal
Y}y$, where ${\cal Y}={\cal Y}^i_jy^j{\partial\over\partial y^i}$ 
is the hypercharge. As an example of $O_\Gamma$'s we can take any three
linearly independent operators out of the lepton number non
conserving gauge invariant operators of dimension $5$ in the
matter fields given in eq.(20) of \cite{Weinberg:1979sa}
(they can also be found in eq. (21.3.54) of \cite{Weinberg:1996kr}) and one
baryon number non conserving operator out of the six dimension $6$
gauge invariant operators given in eqs. (1)-(6) in 
\cite{Weinberg:1979sa,Wilczek:1979hc}.
Because these operators are build out of the undifferentiated matter
fields alone, a sufficient
condition for (\ref{sudu}) to hold is the existence of
$P^\prime_{\Gamma}(x)$'s build out of the undifferentiated $y^i$ such
that 
\bea
f^\Gamma n_\Gamma O_\Gamma={\cal Y}P^\prime_{\Gamma},\label{ult}
\eea
(with no summation over $\Gamma$), 
where $n_\Gamma$ is the lepton number of the $O_\Gamma$'s for 
$\Gamma=1,2,3$ and the
baryon number for $O_4$.
This follows by identifying the term in the abelian ghost and putting, in 
addition to the non abelian ghosts and the antifields, 
the derivatives of the abelian ghost, the derivatives
of the matter fields and all the gauge fields to zero and using the fact
that the equations of motion necessarily involve derivatives. 
Because the $O_\Gamma$'s we have chosen
are all of homogeneity $4$ in the $y^i$ and ${\cal Y}$ is of
homogeneity $0$, we can assume that the homogeneity of the 
$P^\prime_{\Gamma}$'s is also $4$. By decomposing the space $M_4$ of
monomials of homogeneity $4$ in the $y^i$ into eigenspaces of the
hermitian operator ${\cal Y}$ with definite eigenvalues
$M_4=M_4^0+\oplus_{n \neq 0} M_4^n$, it follows that (\ref{ult}) has
no non trivial solutions. Indeed, decomposing 
$P^\prime_{\Gamma}=P^{0\prime}_{\Gamma}+
\sum_{n\neq 0} P^{n\prime}_{\Gamma}$, (\ref{ult})
reads $f^\Gamma n_\Gamma O_\Gamma=\sum_{n\neq
0}n P^{n\prime}_{\Gamma}$. Applying ${\cal Y}$ $k$ times and using the
fact that gauge invariance of $O_\Gamma$ implies ${\cal
Y}O_{\Gamma}=0$, we get $\sum_{n\neq
0}n^k P^{n\prime}_{\Gamma}=0$. We then can conclude that
$P^{n\prime}_{\Gamma}=0$ for $n\neq 0$, which implies $f^\Gamma=0$. 

As usual, this one loop reasoning can be extended recursively to
higher orders, or alternatively, it can be 
discussed independently of the assumption that there exists 
an invariant regularization scheme in the context of
algebraic renormalization. 

\newpage

\mysection{Extended antifield formalism. Classical theory}\label{s5}

\subsection{Coupling constants}

The solution $S$ of the classical master equation usually depends on
some coupling constants. Differentiating 
(\ref{g1}) with respect to such a coupling
constant $g$, implies that $(S,\frac{\partial^R S}{\partial
g})=0$, so that $[\frac{\partial^R S}{\partial
g}]\in H^0(s)$. Note that the presence of coupling constants implies
that the cohomology of $s$ has to be computed in the space of local
functionals depending on the coupling constants. We will not specify
more precisely the functional dependence on these couplings, although 
in the applications below, we have in mind mostly a polynomial or a
formal power series dependence. 

Let us adapt the considerations in \cite{Weinberg} (see also 
\cite{Weinberg:1995mt} chapter 7.7) to the present context.
\begin{definition} 
A set of coupling constants $g^i$ is essential iff 
the relation $\frac{\partial^R S}{\partial g^i}\lambda^i=(S,\Xi)$
implies $\lambda^i=0$, where $\lambda^i$ may depend on all the
couplings of the theory.
\end{definition}
In other words, essential couplings correspond to independent elements 
$[\frac{\partial^R S}{\partial g^i}]$
of $H^0(s)$ computed in the space of local functionals over the ring
of functions in the couplings. (In this context, one does not want to
consider as independent cohomology classes local functionals that
differ only by a factor depending on the couplings alone.)
It follows that essential couplings stay
essential after anticanonical field-antifield redefinitions, because
these redefinitions do not affect the cohomology.  

In the following, we suppose that $S$ depends only on essential
couplings. 
Note that because of equation (\ref{gaugedep}),
the couplings introduced through the gauge fixing alone are all
redundant.

\subsection{Application of the main theorem}

Let us now apply and recall the results of sections \ref{main} 
and \ref{proof} in the present case. 

\subsubsection{Anti constant ghosts and acyclic differentials}\label{s521}

Let $\{[S_A]\}$ be a basis of $H^*(s,{\cal
F})$ over the ring of functions in the essential coupling constants of the 
theory, so that the equation $(S,A)=0$ implies
$A=S_A \lambda^A+(S,B)$, where $\lambda^A$ 
is independent of the fields and anti-fields, but can 
depend on the coupling constants of the theory,
with $S_A \lambda^A+(S,B)=0$ iff 
$\lambda^A=0$.
For each $S_A$ of the above basis,
we introduce a constant ``ghost'' $\xi^A$ and a constant ``antifield''
$ \xi^*_A$ such that $gh\  \xi^A=-gh\ S_A$, $gh\
\xi^*_A=-gh\  \xi^A-1$. We consider the space ${\cal E}$ of functionals 
${\cal A}$ of the form 
\bea
{\cal A}=A[\phi,\phi^*,\xi]+ \xi^*_A
\lambda^A( \xi),
\eea
i.e., ${\cal A}$ contains a local functional $A$ which admits in 
addition to the dependence on the coupling constants, a dependence
on the constant ghosts $ \xi^A$, and a non integrated piece linear in
the constant antifields $ \xi^*_A$ depending only on the constant
ghosts (and the coupling constants). 

The differential $\tilde \delta$ is defined by $\tilde \delta A=(S,A)$, 
$ \tilde \delta  \xi^*_A=S_A$, and $\tilde \delta \xi^A=0$.
\begin{corollary}
The cohomology of $\tilde\delta$ is trivial, $H(\tilde\delta,{\cal
E})=0$.
\end{corollary}
We define the resolution degree to be the degree in the ghosts
$\xi^A$, which implies that $\tilde\delta$ is of degree $0$.

The extended antibracket is defined by 
\bea
(\cdot,\cdot\tilde)=(\cdot,\cdot)+(\cdot,\cdot)_\xi\nonumber\\
=(\cdot,\cdot)
+\frac{\partial^R}{\partial  \xi^A}\frac{\partial^L}{\partial
 \xi^*_A}
-\frac{\partial^R}{\partial  \xi^*_A}\frac{\partial^L}{\partial
 \xi^A}
\eea
and satisfies the same graded antisymmetry and graded Jacobi identity
as the usual antibracket.
The extended
antibracket has two pieces, the old piece $(\cdot,\cdot)$, 
which is of degree $0$, and
the new piece $(\cdot,\cdot)_\xi$, which is of degree $-1$. 
\begin{corollary}\label{t1}
There exists a solution $\tilde {\cal S}\in {\cal E}$ of ghost number $0$ 
to the master equation 
\bea
\frac{1}{2}(\tilde {\cal S},\tilde {\cal S}\tilde)=0.\label{m1}
\eea
with initial condition $\tilde {\cal S}=S+S_A \xi^A+ \dots$, where the
dots denote terms of resolution degree higher or equal to $2$. The
cohomology of the differential $\tilde s=(\tilde {\cal S},\cdot
\tilde)$ in ${\cal E}$ is trivial. 
\end{corollary}

The solution $\tilde {\cal S}$ is of the form 
\bea
\tilde {\cal S}=S+\sum_{k=1} S_{A_1\dots
A_k}\xi^{A_1}\dots
\xi^{A_k}+\sum_{m=2}\xi^*_B f^B_{A_1\dots A_m}
\xi^{A_1}\dots \xi^{A_m},
\eea
which implies the graded symmetry of the generalized structure
constants $f^B_{A_1\dots A_m}$ and the functionals $S_{A_1\dots
A_k}$. 
The $\xi^*_A$ independent part of the master equation (\ref{m1})
gives, at resolution degree $r\geq 1$, the relations
\bea
(S,S_{A_1\dots A_r})+\sum_{k=1}^{r-1}\frac{1}{2}(S_{(A_1\dots
A_k},S_{A_{k+1}\dots A_r)})(-)^{(A_1+\dots+A_k)(A_{k+1}\dots A_r +1)}
\nonumber\\
+\sum_{k=1}^{r-1}kS_{(A_1\dots A_{k-1}|B|}f^B_{A_{k}\dots
A_r)}=0,\label{m2} 
\eea
where $(\ )$ denotes graded symmetrization.
The first relations read explicitly
\bea
(S,S_{A_1})=0,\\
(S,S_{A_1A_2})+\frac{1}{2}(S_{A_1},S_{A_2})(-)^{A_1(A_2+1)}
+S_Bf^B_{A_1A_2}=0,\\
(S,S_{A_1A_2A_3})+(S_{(A_1},S_{A_2A_3)})(-)^{A_1(A_2+A_3+1)}\nonumber\\
+S_Bf^B_{A_1A_2A_3}+2S_{(A_1|B|}f^B_{A_2A_3)}=0,\\
\vdots.\nonumber
\eea
The $\xi^*_A$ dependent part of the master equation (\ref{m1}) gives,
for $r\geq 3$, 
the generalized Jacobi identities 
\bea
\sum_{m=2}^{r-1}mf^C_{(A_1\dots A_{m-1}|B|}f^B_{A_m\dots A_r)}=0,
\label{j2}
\eea
the first identities being
\bea
2f^C_{(A_1|B|}f^B_{A_2A_3)}=0,\\
2f^C_{(A_1|B|}f^B_{A_2A_3A_4)}+3f^C_{(A_1A_2|B|}f^B_{A_3A_4)}=0,\\
\vdots\nonumber.
\eea

\subsubsection{Ambiguity of the construction}

The above solution $\tilde {\cal S}$ is not unique. For a given
initial condition, there is at each
stage of the construction of $\tilde S$, for $k\geq 2$, the 
liberty to add the exact term $\tilde \delta K_k$ to $\tilde {\cal
S}_k$. While this liberty will not affect the structure constants of
order $k$, since a $\tilde\delta$ exact term does not involve a
$\xi^*$ dependent term, it will in general affect the structure
constants of order strictly higher than $k$. 
Furthermore, 
there is a freedom in the choice of the initial condition:
instead of $S_1=S_A\xi^A$, one could have chosen 
$S^\prime_1=\sigma_A^BS_B\xi^A+(S,K_A)\xi^A$ with an invertible matrix
$\sigma_A^B$. If we consider the following anticanonical 
redefinitions:
\bea
z^\prime=\exp (\cdot,K_A\xi^A)z,\\
{\xi^\prime}^B=\sigma_A^B\xi^A,\
{\xi^\prime}^*_B={\sigma^{-1}}_B^A\xi^*_A, 
\eea
we have that
$S+S^\prime_1=S(z^\prime)+S_B(z^\prime){\xi^\prime}^B+O(\xi^2)$. 
We can then consider the solution $\tilde {\cal S}^\prime$ in terms of
the new variables. This is equivalent to taking as initial condition 
$S(z^\prime)+S_B(z^\prime){\xi^\prime}^B$ and making the same choices
for the terms of degree higher than $2$ in the new variables
than we did before in the old variables. It is thus always 
possible to make
the choices in the construction of $\tilde {\cal S}$ for $k\geq 2$ in
such a way that the structure constants
$f^B_{A_1\dots A_m}$ do not depend on the choice
of representatives for the cohomology classes and 
transform tensorially with respect to a change of basis in
$H^*(s,{\cal F})$. 
Hence, we have shown 
\begin{corollary}
Associated to a solution $\tilde {\cal S}$ of the
master equation (\ref{m1}),  
there exist multi-linear, graded symmetric
maps in cohomology, defined through the structure constants
$f^B_{A_{1}\dots A_r}$:
\bea
l_r:\wedge^r H^*(s)\longrightarrow H^*(s)\\
l_r([S_{A_1}],\dots,[S_{A_r}])=[S_{B}]f^B_{A_{1}\dots A_r}\label{gl2}
\eea
\end{corollary}

\subsubsection{Essential couplings and constant ghosts}\label{s523}

In the construction so far, there has been a kind of redundancy
because we have coupled to the solution of the master equation with
new independent couplings all local BRST cohomological classes,
although the classes $[\frac{\partial^R S}{\partial g^i}]$ in ghost
number $0$ are already coupled through the essentials couplings
$g^i$. 

Since the $[\frac{\partial^R S}{\partial g^i}]$ are linearly
independent, one can construct a basis $[S_A]$ of $H^*(s)$ such that
the $[\frac{\partial^R S}{\partial g^i}]$ are the first elements. Let
us denote the remaining elements by $[S_\alpha]$, so that 
$\{[S_A]\}=\{[\frac{\partial^R S}{\partial g^i}],[S_\alpha]\}$. 
The construction of the generating functional $\tilde {\cal S}$ then
starts with $S(g^i)+\frac{\partial^R S}{\partial
g^i}\xi^i+S_\alpha\xi^\alpha$. 

Consider the action $\bar S=S(g^i+\xi^i)$. A basis of the cohomology
of $\bar S$ is given by $\{[\frac{\partial^R \bar S}{\partial
g^i}],[\bar S_\alpha]\}$, with associated differential
$\bar{\tilde\delta}=(\bar S,\cdot)+\frac{\partial^R \bar S}{\partial
\xi^i}\frac{\partial^L}{\partial \xi^*_i}+\bar
S_\alpha\frac{\partial^L}{\partial  \xi^*_\alpha}$, which is acyclic
in the space where the only dependence on $\xi^i$ is through the
combination $g^i+\xi^i$.
If we take as starting point the action $\bar S+\bar S_\alpha\xi^\alpha$ and
start the perturbative construction of the solution of the master
equation, with resolution degree the degree in the ghost $\xi^\alpha$
alone, the ghosts $\xi^i$ only appear through the combination $g^i+\xi^i$, 
because of the properties of $\bar{\tilde\delta}$.
The solution $\bar{\tilde {\cal S}}$ will then be of
the form 
\bea
\bar{\tilde {\cal S}}=\bar S+\sum_{k=1}\bar
S_{\alpha_1\dots\alpha_k}\xi^{\alpha_1}\dots\xi^{\alpha_k}
+\sum_{m=2}(\xi^*_\beta \bar f^\beta_{\alpha_1\dots\alpha_m}+\xi^*_i
\bar f^i_{\alpha_1\dots\alpha_m})\xi^{\alpha_1}\dots\xi^{\alpha_m},
\label{sol1}
\eea
where the $\bar S_{\alpha_1\dots\alpha_k}$, 
$\bar f^i_{\alpha_1\dots\alpha_m}$, $\bar f^i_{\alpha_1\dots\alpha_m}$
depend on the combination $g^i+\xi^i$. 

Now, the solution $\bar{\tilde {\cal S}}$ satisfies the initial
condition $\bar{\tilde {\cal S}}^1=S(g)+\frac{\partial^R S}{\partial
g^i}\xi^i+S_\alpha\xi^\alpha$ in the old resolution degree 
and the master equation
(\ref{m1}). 
We can then  derive the
higher order maps $l_r$ from the solution (\ref{sol1}) and get 
\bea
l_r([\frac{\partial^R S}
{\partial g^{i_1}}],\dots,[\frac{\partial^R S}
{\partial g^{i_n}}],[S_{\alpha_{n+1}}],\dots,[S_{\alpha_r}])
\nonumber\\=\frac{1}{n!}[S_\beta]
\frac{{\partial^R}^n \bar f^\beta_{\alpha_{n+1}\dots\alpha_r}}{\partial
g^{i_n}\dots\partial g^{i_1}} (g)+\frac{1}{n!}[\frac{\partial^R S}
{\partial g^{j}}]
\frac{{\partial^R}^n \bar f^j_{\alpha_{n+1}\dots\alpha_r}}{\partial
g^{i_n}\dots\partial g^{i_1}} (g).\label{exp1}
\eea

In the following, we will make the redefinition
$g^i+\xi^i\longrightarrow \xi^i$, and identify the essential
couplings with some of the constant ghosts. 
Alternatively, the remaining constant ghosts can be considered as 
generalized essential coupling constants since they couple the
remaining BRST cohomology classes, which play the role of generalized
observables in this formalism.

\subsubsection{Decomposition of $\tilde s$}

The space ${\cal E}$ admits the direct sum decomposition ${\cal
E}=F\oplus G$, where $F={\cal E}|_{\xi^*=0}$ is the space 
of functionals in the field and
antifields with $\xi$ dependence, but no $\xi^*$ dependence, while $G$
is the space of power series in $\xi$ with a linear $\xi^*$ dependence.

The differential $\tilde s$ in ${\cal E}$ induces two
well-defined differentials, $\bar s$ in $F$ and $s_Q$ in $G$ given
explicitly by 
\bea
\bar s=(S(\xi),\cdot)+(-)^{D}f^D\frac{\partial^L
}{\partial\xi^D}
\eea
and 
\bea
s_{\Delta^*}=(\Delta^*,\cdot)_\xi, 
\ \Delta^*=\xi^*_Cf^C(\xi).
\eea
Indeed, for ${\cal
A}=A(\xi)+\xi^*_D\lambda^D(\xi)$, the master equation (\ref{m1})
implies $(\tilde{\cal S},(\tilde{\cal S},{\cal A}\tilde)\tilde)=0$ and
hence $(\tilde{\cal S},\bar s A(\xi)+s_Q
\xi^*_D\lambda^D(\xi)\tilde)=0$ and then $(\bar s)^2A(\xi)+(s_{\Delta^*})^2
\xi^*_D\lambda^D(\xi)=0$, which splits into two equations because the
decomposition of ${\cal E}$ is direct. 
If $\tilde S=S(\xi)+\xi^*_Cf^C(\xi)$, ${\cal
A}=A(\xi)+\xi^*_C\lambda^C(\xi)$ and ${\cal
B}=B(\xi)+\xi^*_C\mu^C(\xi)$. The extended master equation (\ref{m1})
can be written compactly as 
\begin{eqnarray}
\frac{1}{2}(S(\xi),S(\xi))+\frac{\partial^R
S(\xi)}{\partial\xi^C}f^C=0,
\label{j0}\\
\frac{1}{2}(\xi^*_Cf^C(\xi),\xi^*_Df^D(\xi))_\xi=0,\label{j1}
\end{eqnarray}
so that (\ref{j0}) summarizes (\ref{m2}) and (\ref{j1}), which is 
equal to $\frac{1}{2}(\Delta^*,\Delta^*)_\xi=0$, or explicitly 
$\frac{\partial^R f^D(\xi)}{\partial \xi^C}f^C(\xi)=0$, 
summarizes the generalized Jacobi identities (\ref{j2}).

\begin{corollary}\label{t3}
The cohomology groups $H(\bar s,F)$ and $H(s_{\Delta^*},G)$ are isomorphic. 
\end{corollary}
More precisely,
\bea
\bar s A(\xi)=0 \Longleftrightarrow 
\left\{\begin{array}{c}
A(\xi)=\bar s B(\xi)+\frac{\partial^R
S(\xi)}{\partial\xi^C}\mu^C(\xi),
\cr
(\xi^*_Cf^C(\xi),\xi^*_D\mu^D(\xi))_\xi=0,
\end{array}\right.
\label{26}
\eea
and
\bea
\left\{\begin{array}{c}\bar s B(\xi)+\frac{\partial^R
S(\xi)}{\partial\xi^C}\mu^C(\xi)=0,\cr
(\Delta^*,\xi^*_B\mu^B(\xi))_\xi=0,
\end{array}\right.\Longleftrightarrow
\left\{\begin{array}{c}
B(\xi)=\bar s C(\xi)+\frac{\partial^R
S(\xi)}{\partial\xi^C}\nu^C(\xi),\cr
\xi^*_D\mu^D(\xi)=(\Delta^*,\xi^*_E\nu^E(\xi))_\xi,
\end{array}\right.\label{27}
\eea
so that 
\bea
m: H(s_{\Delta^*},G)\longrightarrow H(\bar s,F),\nonumber\\
m([\xi^*_D\mu^D(\xi)])=[\frac{\partial^R S(\xi)}{\partial\xi^C}\mu^C(\xi)]
\eea
is one-to-one and onto. 

\subsubsection{Discussion}

(i) In order to compare the starting point cohomology $H^*(s,{\cal F})$
with the cohomology $H^*(\bar s,F)$, we can put the additional
couplings $\xi^\alpha$ to zero in (\ref{26}). The cocycle condition
then reduces to the standard cocycle condition of the non extended
formalism, $s A_{\xi^\alpha=0}=0$. The same operation in the general
solution gives $A_{\xi^\alpha=0}=sB_{\xi^\alpha=0}
+\frac{\partial^R S}{\partial \xi^i}\mu^i_{\xi^\alpha=0} 
+S_\alpha\mu^\alpha_{\xi^\alpha=0}$. Contrary to the ordinary $s$
cohomology, the coefficients
$\mu^A_{\xi^\alpha=0}$ are not
free however, but they come from $\mu^A$'s which are constrained to
satisfy the cocycle condition in (\ref{26}). In particular, at 
order $1$ in the new couplings $\xi^\alpha$, (\ref{26}) implies that 
$\mu^\alpha_{\xi^\alpha=0}$ is in the kernel of the map $l_2$, 
$f^A_{\beta\alpha}\mu^\alpha_{\xi^\alpha=0}=0$. 
We thus see that the cohomology has become ``smaller'' through
the introduction of the additional couplings because the extended
differential encodes higher order cohomological restrictions.
 
(ii) At first sight, it might seem a little strange to introduce new
couplings in order to get information on the renormalization of the
theory without these couplings: that it is convenient and extremely
useful to do so was already realized in the first papers
\cite{Becchi:1974xu,Becchi:1974md,Becchi:1975nq,BecchiTira} on the
subject: the additional (space-time dependent) couplings 
in these papers are just the sources of the BRS transformations, and
can of course be set to zero after renormalization, if one is only
interested in the renormalization of the effective action itself.

(iii) The result (\ref{26}) implies also that the $\bar s$ 
cohomology is
contained completely in the solution $S(\xi)$ and can be obtained from
it by applying $\frac{\partial^R\cdot}{\partial \xi^A}\lambda^A(\xi)$,
where the coefficients $\lambda^A(\xi)$ are constrained to be
$s_{\Delta^*}$ cocycles. 

\subsubsection{``Quantum'' Batalin-Vilkovisky formalism on the classical
level}

If we define 
\bea
{\Delta^L_c}=(-)^{D}f^D(\xi)\frac{\partial^L}{\partial \xi^D},\ \Delta_c=
\frac{\partial^R}{\partial \xi^D}f^D(\xi)
\eea
on $F$, the following properties of the quantum
Batalin-Vilkovisky formalism hold in $F$: 
the operator ${\Delta^L_c}$ is nilpotent, 
\bea
{\Delta^L_c}^2=0,\label{nil}
\eea
(as a consequence of (\ref{j1}) or (\ref{j2}).) Furthermore, 
\bea
{\Delta^L_c}(A(\xi),B(\xi))=
({\Delta^L_c} A(\xi),B(\xi))+(-)^{|A|+1}(A(\xi),{\Delta^L_c} B(\xi)). 
\label{der}
\eea
Similar properties also hold for the right derivation $\Delta_c$. 

To the standard solution of the master equation $S$ in ${\cal F}$
corresponds in $F$ the solution $S(\xi)$ of the extended
master equation 
\bea
\frac{1}{2}(S(\xi),S(\xi))+{\Delta_c} S(\xi)=0,\label{qme}
\eea
(which is just rewriting (\ref{j0}) using the 
definition of ${\Delta_c}$). 
Because 
\bea
\bar s= (S(\xi),\cdot)+{\Delta^L_c},
\eea
the $\bar s$ cohomology corresponds to 
the quantum BRST cohomology $\sigma$ discussed for instance in  
\cite{Henneaux:1989jq,Henneaux:1992ig}. 
Corollary
\ref{t3} shows how to compute the ``quantum'' BRST cohomology out of 
the standard BRST cohomology and the higher order maps encoded in
$s_{\Delta^*}$. 

In this analogy, putting $\xi=0$ corresponds to the classical
limit $\hbar\longrightarrow 0$ of the quantum Batalin-Vilkovisky
formalism.

Note however that (i) the space $F$ is not directly an algebra,
because the product of two local functionals is not well defined,
contrary to the formal discussion of the quantum Batalin-Vilkovisky
formalism, where one assumes the space to be an algebra, (ii) the
above ``quantum'' Batalin-Vilkovisky formalism is purely classical and
depends only on the BRST cohomology and the higher order maps of the
theory. 

\subsection{Deformations and stability}

We consider now one parameter deformations of the extended
master equation (\ref{qme}), i.e., in the space $F[t]$ of power series in
$t$ with coefficients that belong to $F$, we want to construct
$S_t(\xi)=S(\xi)+tS_1(\xi)+t^2S_2(\xi)+\dots$ 
such that 
\bea
\frac{1}{2}(S_t(\xi),S_t(\xi))+{\Delta_c} S_t(\xi)=0.
\eea

A deformation $S_t(\xi)=S(\xi)+tS_1(\xi)$ to first order in $t$, i.e.,
such that
$\frac{1}{2}(S_t(\xi),S_t(\xi))+{\Delta_c} S_t(\xi)=O(t^2)$ is called an
infinitesimal deformation. 
The term 
linear in $t$ of an  infinitesimal deformation, $S_1(\xi)$,
is a cocycle of the extended BRST differential $\bar s$. 
If $S_1(\xi)$ is
a $\bar s$ coboundary, we call the infinitesimal deformation trivial,
while the parts of $S_1(\xi)$  corresponding to the $\bar s$
cohomology are non trivial. 

\begin{theorem}\label{t4}
Every infinitesimal deformation of the solution $S$ to 
the extended master equation
can be extended to a complete deformation $S_t$.
This extension is obtained by (i) performing a 
$t$ dependent anticanonical field-antifield redefinition
$z\rightarrow z^\prime$, by (ii) performing a $t$ 
dependent coupling constant redefinition
$\xi\rightarrow \xi^\prime$, which does not affect $\Delta_c$, 
and (iii) by adding to $S(z^\prime,\xi^\prime)$ a suitable extension
determined by both coupling constant and the field-antifield
redefinition and vanishing whenever the latter does.

Furthermore, the deformed solution 
considered as a function of the new variables 
$S_t(z(z^\prime,\xi^\prime),\xi(\xi^\prime))$ 
satisfies the extended master equation in terms of the new variables 
and the cohomology $H(\bar s^\prime,F^\prime)$ of 
the differential $\bar
s^\prime=(S_t,\cdot)_{z^\prime}+{\Delta^L_c}^\prime$ 
in the space $F^\prime$ of functionals depending on
$z^\prime,\xi^\prime$ is isomorphic to the cohomology $H(\bar s,F)$.
\end{theorem}
\proof{Equation (\ref{26}) implies that $S_1(\xi)=\bar s
B+ \frac{\partial^R S(\xi)}{\partial \xi^C}\mu^C(\xi)$ with
$(\xi^*_Df^D(\xi),\xi^*_C\mu^C(\xi))_\xi=0$. In other words, 
$S_1(\xi)=(\tilde S,B(\xi)+\xi^*_C\mu^C(\xi)\tilde)$. In the extended
space ${\cal E}$, with $z^\alpha=(\phi^a,\phi^*_a)$, 
consider the anticanonical transformation 
\bea
{z^\prime}^\alpha=\exp t(\cdot, B(\xi)+\xi^*_C\mu^C(\xi)\tilde)\ z^\alpha
\nonumber\\=z^\alpha+t(z^\alpha,B(\xi))+O(t^2),
\eea
\bea
{\xi^\prime}^A=\exp t(\cdot, B(\xi)+\xi^*_C\mu^C(\xi)\tilde)\ \xi^A=
\exp t(\cdot,\xi^*_C\mu^C(\xi) )_\xi\ \xi^A\nonumber\\=
\xi^A+t\mu^A(\xi)+O(t^2),
\eea
\bea
{\xi^\prime_A}^*=\exp t(\cdot, B(\xi)+\xi^*_C\mu^C(\xi)\tilde)\ {\xi^*_A}
\nonumber\\={\xi^*_A}-t\frac{\partial^L
}{\partial \xi^A}(B(\xi)+\xi^*_C\mu^C(\xi))\label{anti}
+O(t^2).
\eea
Note that $z^\prime=z^\prime(z,\xi)$, $\xi^\prime=\xi^\prime(\xi)$ and
${\xi^*_A}^\prime={\xi^*_A}^\prime(z,\xi,\xi^*)=g_A(z,\xi)
+\xi^*_Bg^B_A(\xi)$,
for a function $g_A(z,\xi)=-t\frac{\partial^L B}{\partial
\xi^A}+O(t^2)$ determined by (\ref{anti}) through both $B$ and $\mu$
and a function 
$g^B_A(\xi)=\delta^B_A-t(-)^{A(B+1)}\frac{\partial^L \mu^B}{\partial
\xi^A}+O(t^2) $ determined by (\ref{anti}) through $\mu$ alone. 

The master equation (\ref{m1}) holds in any variables, 
and thus also in terms of the primed variables. If we denote functions in
terms of the new variables by a prime, we get $\frac{1}{2}(\tilde
S^\prime,\tilde S^\prime\tilde)_{z^\prime,\xi^\prime}=0$. Because the
transformation is anticanonical, we also have 
\bea
\frac{1}{2}(\tilde
S^\prime,\tilde S^\prime\tilde)_{z,\xi}=0. \label{nspl}
\eea
Since
\bea
\tilde S^\prime=S^\prime+g_A{f^\prime}^A+\xi^*_Bg^B_A{f^\prime}^A,
\eea
equation (\ref{nspl}) splits into 
\bea
\frac{1}{2}(S^\prime+g_A{f^\prime}^A,S^\prime+g_A{f^\prime}^A)_z
+\frac{\partial^R}{\partial\xi^D}(S^\prime+g_A{f^\prime}^A)
g^D_E{f^\prime}^E=0,\label{45}\\
\frac{1}{2}(\xi^*_Bg^B_A{f^\prime}^A,\xi^*_Dg^D_C{f^\prime}^C)_\xi=0.
\label{nil1}
\eea
We have 
\bea
\frac{d
(S^\prime+g_A{f^\prime}^A)}{dt}|_{t=0}
=(S(\xi),B(\xi))+{\Delta^L_c} B+\frac{\partial^R
S(\xi)}{\partial \xi^D}\mu^D=S_1(\xi),
\eea
and, because $\xi^*_E\mu^E(\xi)$ is a $s_{\Delta^*}$ cocycle, the relation 
\bea
g^D_C{f^\prime}^C=f^D.\label{us}
\eea
Indeed, if we consider the above canonical transformation with $B=0$,
i.e., $\exp
t(\cdot,\xi^*\mu)_\xi)$ alone,
$\xi^*_Cg^C_Df^D(\xi^\prime)={\xi^*}^\prime_Df^D(\xi^\prime)=
\exp t(\cdot,\xi^*\mu)_\xi\xi^*_Ef^E=\xi^*_Ef^E$, because 
$(\xi^*_Ef^E,\xi^*_G\mu^G)_\xi=0$.
This shows the first part of the theorem, with
$S_t=S^\prime+g_A{f^\prime}^A$. 

In order to prove the second part, we first note that 
\bea
{\Delta^L_c}^\prime=(-)^Df^D(\xi^\prime)\frac{\partial^L}{\partial
{\xi^\prime}^D}=(-)^Df^D(\xi^\prime)\frac{\partial^L\xi^C}{\partial
{\xi^\prime}^D}\frac{\partial^L}{\partial {\xi^C}}=\Delta^L_c
\eea
because 
\bea
g^C_D=\frac{\partial^L\xi^C}{\partial
{\xi^\prime}^D}.\label{tt}
\eea
Indeed, 
we have
$\delta^A_B=({\xi^\prime}^A,{\xi^*_B}^\prime\tilde)_{z^\prime,\xi^\prime}
=({\xi^\prime}^A,{\xi^*_B}^\prime)_{\xi}
=\frac{\partial^L{\xi^\prime}^A}{\partial\xi^C}g^C_B$. 
Together with (\ref{45}), this implies
\bea
\frac{1}{2}(S^\prime+g_A{f^\prime}^A,S^\prime
+g_A{f^\prime}^A)_{z^\prime}+\Delta^\prime_c(S^\prime+g_A{f^\prime}^A)=0.
\eea
We then start from the relations 
\bea
(\tilde S^\prime,{\cal
A}^\prime\tilde)_{z^\prime,\xi^\prime}=0\Longleftrightarrow 
{\cal
A}^\prime=(\tilde S^\prime,{\cal B}^\prime\tilde)_{z^\prime,\xi^\prime},
\eea
where ${\cal A}^\prime=A^\prime+{\xi^*}^\prime_A{\lambda^\prime}^A$
and ${\cal B}^\prime=B^\prime+{\xi^*}^\prime_A{\rho^\prime}^A$.
These relations hold with the bracket taken in the old variables, 
because the transformation is
anticanonical. Writing the resulting relations explicitly, 
using (\ref{us}), 
we get that the set of relations 
\bea\left\{\begin{array}{l}
(S^\prime+g_A{f^\prime}^A,A^\prime+g_B{\lambda^\prime}^B)+
\frac{\partial^R }{\partial
\xi^C}(S^\prime+g_A{f^\prime}^A)g^C_D{\lambda^\prime}^D \\
+(-)^Af^A\frac{\partial^L}{\partial
\xi^A}(A^\prime+g_B{\lambda^\prime}^B)=0,\\
\xi^*_A\frac{\partial^R g^A_B{f^\prime}^B}{\partial \xi^C}g^C_D
{\lambda^\prime}^D+(-)^Dg^D_C{f^\prime}^C\frac{\partial^L}{\partial
\xi^D}(\xi^*_Ag^A_B{\lambda^\prime}^B)=0,
\end{array}\right.\label{ch1}
\eea
is equivalent to the set 
\bea\left\{\begin{array}{l}
A^\prime+g_B{\lambda^\prime}^B=
(S^\prime+g_A{f^\prime}^A,B^\prime+g_C{\rho^\prime}^C)+
\frac{\partial^R }{\partial
\xi^C}(S^\prime+g_A{f^\prime}^A)
g^C_D{\rho^\prime}^C \\+(-)^Af^A\frac{\partial^L}{\partial
\xi^A}(B^\prime+g_C{\rho^\prime}^C),\\
\xi^*_Ag^A_B{\lambda^\prime}^B=
\xi^*_A\frac{\partial^R g^A_B{f^\prime}^B}{\partial \xi^C}g^C_D
{\rho^\prime}^D+(-)^Dg^D_C{f^\prime}^C\frac{\partial^L}{\partial
\xi^D}(\xi^*_Ag^A_B{\rho^\prime}^B).
\end{array}\right.\label{ch2}
\eea
Using (\ref{tt}), the last equations in (\ref{ch1}) and (\ref{ch2})
are just the $s_{\Delta^*}$ cocycle and coboundary conditions, expressed in
terms of the ${\xi^*}^\prime,\xi^\prime$ variables. 
Following the same reasoning as in the proof of theorem \ref{t3}, we
get, 
\bea
(S^\prime+g_A{f^\prime}^A,A^\prime)+\Delta^L_cA^\prime=0\nonumber\\
\Longleftrightarrow
A^\prime=(S^\prime+g_A{f^\prime}^A,B^\prime+g_C{\rho^\prime}^C)
\nonumber\\+\Delta^L_c(B^\prime+g_C{\rho^\prime}^C)+
\frac{\partial^R }{\partial
{\xi^\prime}^C}(S^\prime+g_A{f^\prime}^A){\rho^\prime}^C,\\
({\xi^*}^\prime_A{f^\prime}^A,
{\xi^*}^\prime_B{\rho^\prime}^B)_{\xi^\prime}=0,
\eea
where $\frac{\partial^R }{\partial
{\xi^\prime}^C}(S^\prime+g_A{f^\prime}^A){\rho^\prime}^C$ is 
$(S^\prime+g_A{f^\prime}^A,\cdot)+\Delta^L_c$ exact iff 

\noindent ${\xi^*}^\prime_B{\rho^\prime}^B=({\xi^*}^\prime_A{f^\prime}^A,
{\xi^*}^\prime_C{\nu^\prime}^C)_{\xi^\prime}$. Since
$(S^\prime+g_A{f^\prime}^A,\cdot)+\Delta^L_c=
(S^\prime+g_A{f^\prime}^A,\cdot)_{z^\prime}
+{\Delta^L_c}^\prime=\bar s^\prime$, 
we get that
$H(\bar s^\prime,F^\prime)$ is determined by 
$\frac{\partial^R  }{\partial
{\xi^\prime}^C}(S^\prime+g_A{f^\prime}^A){\rho^\prime}^C$ 
corresponding to the class 
$\frac{\partial^R S}{\partial
\xi^C}{\rho}^C$ of $H(\bar s,F)$.}

{\bf Remark:} Note that one can prove in the same way that the relations
$\frac{1}{2}(A,A)+\Delta_c A=C$ and  
$(A,D)+\Delta^L_c D= E$ become, after the change of variables,
$\frac{1}{2}(A^\prime + g_A 
{f^\prime}^A,A^\prime + g_A 
{f^\prime}^A)+\Delta_c (A^\prime + g_A 
{f^\prime}^A)=C^\prime$, respectively
$(A^\prime + g_A {f^\prime}^A,D^\prime) +\Delta^L_c D^\prime=E^\prime$.

\newpage

\mysection{Extended antifield formalism. Quantum theory}\label{s6}

We show how the renormalization can be performed while respecting the
symmetry, encoded in the extended master equation for the 
starting point action : the corresponding renormalized
effective action satisfies a suitably deformed master equation. In
other words, the Zinn-Justin equation can be written to all orders as
a functional differential equation without breakings.

In a first part, we consider a regularization and discuss the
absorption of divergences by field-antifield and coupling constant
redefinitions. In a second part, we derive the same
results by applying the methods of algebraic renormalization, relying
on the use of the renormalized quantum action principles, in the
context of the extended antifield formalism. 

\subsection{Regularization and absorption of divergences}

We make the same assumptions on the regularization as in 
\ref{sreg} and apply it to the extended master equation (\ref{qme})
and its solution $S(\xi)$. In the following, we will always understand
the $\xi$ dependence without explicitly indicating it. Local functionals are
understood to belong to $F$. 

Let $\theta_\tau=\frac{1}{2\tau}(S_\tau,S_\tau)
+\frac{1}{\tau}{\Delta_c}
S_\tau$. Note that $\theta_\tau$ is of order $\tau^0$ because $S_0$ 
satisfies the extended master
equation. $\theta_\tau$ characterizes the breaking of the extended
master equation due to the regularization. In order to control this
breaking during renormalization, it is useful to couple it with a
global source $\rho^*$ in ghost number $-1$ and consider 
$S_{\rho^*}=S_\tau+\theta_\tau\rho^*$. On the classical, regularized
level, we have, using $(\rho^*)^2=0$, and the properties (\ref{nil})
and (\ref{der}) of ${\Delta_c}$,
\bea
\frac{1}{2}(S_{\rho^*},S_{\rho^*})+{\Delta_c}
S_{\rho^*}=\tau\frac{\partial^R S_{\rho^*}}{\partial\rho^*},
\eea
Applying the quantum action principle, we get, for the regularized
generating functional for 1PI irreducible vertex functions 
$\Gamma_{\rho^*}$ associated to $S_{\rho^*}$,
\bea
\frac{1}{2}(\G_{\rho^*},\G_{\rho^*})+{\Delta_c}
\G_{\rho^*}=\tau\frac{\partial^R \G_{\rho^*}}{\partial\rho^*},
\eea
which splits, using $(\rho^*)^2=0$, into 
\bea
\frac{1}{2}(\G,\G)+{\Delta_c}
\G=\tau\frac{\partial^R \G_{\rho^*}}{\partial\rho^*},\label{endl1}\\
(\G,\frac{\partial^R\G_{\rho^*}}{\partial\rho^*})
+{\Delta^L_c}\frac{\partial^R\G_{\rho^*}}{\partial\rho^*}=0\label{endl2}.
\eea

\subsubsection{Invariant regularization}

Before proceeding with the general analysis, let us briefly discuss
the case when the regularization respect the symmetries under
consideration. 
In this case, $\theta_\tau$ vanishes and we have 
\bea
\frac{1}{2}(S_\tau,S_\tau)+\Delta_c S_\tau=0. 
\eea
For the regularized generating functional, we get 
\bea
\frac{1}{2}(\G,\G)+\Delta_c \G=0,\label{ein}
\eea
where by assumption,
$\G=S_\tau+\hbar\sum_{n=-1}\tau^n\G^{(1)n}+O(\hbar^2)$. 
To order $\hbar/\tau$, (\ref{ein}) gives 
\bea
\bar s \G^{(1)-1}=0\Longleftrightarrow \G^{(1)-1}=\bar
s{\Xi_1}+\frac{\partial^R S_0}{\partial \xi^A}\mu^A_1,
\eea
where $\bar s=(S_0,\cdot)+\Delta^L_c$ and $s_{\Delta^*}\ \xi^*_A\mu^A_1=0$.

We then make the following 
change of fields, antifields and
coupling constants:
\bea
z^1=\exp -\frac{\hbar}{\tau}(\cdot ,{\Xi_1}+\xi^*\mu_1\tilde)z ,\\
{\xi^1}=\exp -\frac{\hbar}{\tau}(\cdot ,\xi^*\mu_1)_\xi
\xi .
\eea
If we denote by a superscript $1$ functions depending on these new
variables, we have, according to the remark after theorem \ref{t4},
that the action $S_{R1}=S^1_\tau+{g_1}_A{f^1}^A$, with ${g_1}_A$ 
is determined through the generators 
${\Xi_1}$ and $\mu^C_1$ of the first redefinition, 
satisfies the extended master
equation (\ref{qme}),
\bea
\frac{1}{2}(S_{R_1},S_{R_1})+\Delta_c S_{R_1}=0.
\eea
It allows to absorb the one loop divergences, since
$S_{R1}=S_\tau-\hbar/\tau\G^{(1)-1}+\hbar O(\tau^0)+O(\hbar^2)$.
We thus have for the corresponding regularized generating functional 
$\G_{R_1}=S_\tau+\hbar\sum_{n=0}\tau^n\G_{R_1}^{(1)n}
+\hbar^2\sum_{n=-2}\tau^n
\G_{R_1}^{(2)n}+O(\hbar^2)$,
\bea
\frac{1}{2}(\G_{R_1},\G_{R_1})+\Delta_c \G_{R_1}=0.
\eea
At order $\hbar^2/\tau^2$ , we get 
\bea
\bar s \G_{R_1}^{(2)-2}=0\Longleftrightarrow \G^{(2)-2}=\bar
s\Xi_{2,-2}+\frac{\partial^R S_0}{\partial \xi^A}\mu^A_{2,-2},
\eea
with $s_{\Delta^*}\ \xi^*_A\mu^A_{2,-2}=0$.
The appropriate change of variables is
\bea
z^{2,-2}=\exp -\frac{\hbar^2}{\tau^2}(\cdot
,{\Xi_{2,-2}}+{\xi^*}\mu_{2,-2}\tilde) z ,\\
{\xi^{2,-2}}=\exp -\frac{\hbar^2}{\tau^2}(\cdot ,
{\xi^*}\mu_{2,-2})_\xi\xi.
\eea
The regularized action 
$S_{R_{2,-2}}=S_{R_1}^{2,-2}+{g_{2,-2}}_A{f^{2,-2}}^A$ satisfies the
extended master equation and allows to absorb the poles
of order $\hbar^2/\tau^2$:
\bea
\frac{1}{2}(S_{R_{2,-2}},S_{R_{2,-2}})+\Delta_c S_{R_{2,-2}}=0,
\eea
and 
$\G_{R_{2,-2}}=S_\tau+\hbar\sum_{n=0}\G_{R_{2,-2}}^{(1)n}
+\hbar^2\sum_{n=-1}\tau^n
\G_{R_{2,-2}}^{(2)n}+O(\hbar^3)$.

In the same way, one can then proceed to 
absorb the poles of order $\hbar^2/\tau$ to get a regularized action 
$S_{R_{2,-1}}$ and an associated two loop finite effective action 
$\G_{R_{2,-1}}$, with both actions satisfying the extended master
equation.

Going on recursively to higher orders in $\hbar$, we can achieve, 
through a
succession of redefinitions, the absorptions of the infinities to
arbitrary high order in the loop expansion, while preserving the
extended master equation for the redefined action and the
corresponding generating functional,
\bea
\frac{1}{2}(S_{R_\infty},S_{R_\infty})+\Delta_c S_{R_\infty}=0,
\eea
with $\G_{R_\infty}$ finite and satisfying
\bea
\frac{1}{2}(\G_{R_\infty},\G_{R_\infty})+\Delta_c \G_{R_\infty}=0.
\eea
We have thus shown:
\begin{theorem}
In theories admitting an invariant regularization scheme, the
divergences can be absorbed by successive redefinitions in such a way
that both the subtracted and the effective action satisfy the extended
master equation. 
\end{theorem}

\subsubsection{Structural constraints and cohomology of $\bar s$}

Structural constraints have been introduced in \cite{Gomis:1996jp} to give in
particular cases a sufficient, but not a necessary condition for
renormalizability in the modern sense. An example of a structural
constraint is the requirement that in every BRST cohomological class in
ghost number $0$, there exists a representative that is independent of
the antifields. In the cases of semi-simple
Yang-Mills theories or gravity for instance, this constraint is
fulfilled. It guarantees that one
can couple these representatives to the action in such a way that the
extended action satisfies the same, unmodified master equation. 
In non anomalous theories, the infinities can then be absorbed
by successive coupling constant and field-antifield redefinitions in
such a way that the standard master equation holds, both for the
subtracted action and the effective action. 

What has been shown in the previous section is that structural
constraints are not necessary conditions for
renormalizability in the modern sense. If one uses the extended
antifield formalism, renormalizability
in the modern sense can be proved independently of any structural
constraint. This is because the extended antifield formalism is stable
by construction, due to the fact that the cohomology of the operator
$\bar s$ incorporates higher order cohomological restrictions. 

In the case where the regularization respects the extended master
equation, the extended master equation for the
effective action implies the following stability of the quantum theory:
while the expression of the generalized observables of the theory are
affected by quantum corrections, their antibracket algebra stays the
same than in the classical theory. In particular, the usual
algebra of the generators of the global symmetries (whether linear or
not) is the same in the classical and the quantum theory\footnote{The
author is grateful to F. Brandt for pointing this out.}. This is
because the antibracket algebra of the BRST cohomology classes in
negative ghost numbers just reflects the ordinary algebra of the
symmetries they represent.  

\subsubsection{One loop divergences and anomalies}

Let us now go back to the general case where the 
regularization scheme is not invariant and $\theta_\tau$ does not
vanish. 

At one loop, we get from (\ref{endl1}) and (\ref{endl2}) 
\bea
(S_\tau,\G^{(1)})+{\Delta^L_c} \G^{(1)}=\tau\theta^{(1)},\label{1lc}\\
(S_\tau,\theta^{(1)})+(\G^{(1)},\theta_\tau)+{\Delta^L_c}\theta^{(1)}=0,
\label{1la} 
\eea
where $\G^{(1)}$ and $\theta^{(1)}$ are respectively the one loop
contributions of $\G$ and
$\frac{\partial^R\G_{\rho^*}}{\partial\rho^*}$. By assumption, we have
both 
$\G^{(1)}=\sum_{n=-1}\tau^n\G^{(1)n}$ and $\theta^{(1)}
=\sum_{n=-1}\tau^n\theta^{(1)n}$, where $\G^{(1)-1},\theta^{(1)-1}$
are local functionals. 

At $\frac{1}{\tau}$, equation (\ref{1lc}) gives
\bea
\bar s \G^{(1)-1}=0,
\eea
Using this equation together with $\theta_0=\bar s S_1$, equation
(\ref{1la}) implies
\bea
\bar s (\theta^{(1)-1}-(\G^{(1)-1},S_1))=0.
\eea
Equation (\ref{1lc}) also gives at order $\tau^0$
\bea
\bar s \G^{(1)0}=\theta^{(1)-1}-(\G^{(1)-1},S_1),\label{ma}
\eea
which allows us to identify the combination
$A_1=\theta^{(1)-1}-(\G^{(1)-1},S_1)$ as the one loop anomaly and
explicitly shows its locality.
We have thus shown in the case of a non invariant
regularization scheme:
\begin{theorem}
The one loop divergences $\G^{(1)-1}$ and the one loop anomalies $A_1$
are $\bar s$ cocycles in ghost number $0$ and $1$ respectively.
\end{theorem}

\subsubsection{One loop renormalization}

According to (\ref{26}), we have 
\bea
\G^{(1)-1}=\bar s
\Xi_1+\frac{\partial^R S_0}{\partial \xi^D}\mu^D_1
\eea
and 
\bea
A_1=\bar s\Sigma_1
+\frac{\partial^R S_0}{\partial \xi^E}\sigma^E_1, \label{ma1}
\eea
with $s_{\Delta^*}\ \xi^*_A \mu^A_1=0=s_{\Delta^*} \xi^*_B\sigma^B_1$.
The appropriate change of variables is now 
\bea
z^1=\exp -\frac{\hbar}{\tau}(\cdot ,{\Xi_1}+\xi^*\mu_1\tilde)z ,\\
{\xi^1}=\exp -\frac{\hbar}{\tau}(\cdot ,\xi^*\mu_1\tilde)\xi .
\eea
The renormalized one loop action is 
\bea
S_{{R_1}}=S^{1}_\tau+{g_1}_A{f^1}^A-\hbar\tilde\Sigma_1^1=
S_\tau-\frac{\hbar}{\tau}\G^{(1)-1}+\hbar O(\tau^0)
+O(\hbar^2),
\eea
where $\tilde\Sigma_1$ remains to be determined. 
Using the remark after theorem \ref{t4}, we get 
\bea
\theta_{R_1}\equiv\frac{1}{2\tau}({S_{{R_1}}},{S_{{R_1}}})
+\frac{1}{\tau}
\Delta_c{S_{{R_1}}}
=\theta_\tau^1-\frac{\hbar}{\tau}(\bar s \tilde\Sigma_1)^1
+O(\hbar^2)\nonumber\\
=\theta_\tau-\frac{\hbar}{\tau}\bar s[\tilde\Sigma_1+(S_1,\Xi_1)
+\frac{\partial^R S_1}{\partial \xi^A}\mu^A_1]
-\frac{\hbar}{\tau}(\G^{(1)-1},S_1)
+\hbar O(\tau^0)+O(\hbar^2).\label{do}
\eea
Finally, we consider
$\xi^1_{\rho^*}=\exp -\frac{\hbar}{\tau}(\cdot ,
\xi^*\sigma_1\rho^*)\xi=\xi-\frac{\hbar}{\tau}\sigma_1\rho^*$ and
substitute $\xi$ by $\xi^1_{\rho^*}$: 
\bea
S_{{R_1}}^{\rho^*}(z,\xi,\rho^*)
\equiv S_{{R_1}}(z,\xi^1_{\rho^*}(\xi,\rho^*))\nonumber\\
=S_{{R_1}}(z,\xi)
-\frac{\hbar}{\tau}\frac{\partial^R S_{{R_1}}}{\partial\xi^A}
\sigma^A_1\rho^*\label{de}\\=S_{{R_1}}(z,\xi)-\frac{\hbar}{\tau}
\frac{\partial^R S_0}{\partial\xi^A}\sigma^A_1\rho^*+\hbar O(\tau^0)
+O(\hbar^2). \label{di}
\eea
We also have that 
\bea
\theta^{\rho^*}_{R_1}(z,\xi,\rho^*)\equiv\theta_{R_1}(z,
\xi^1_{\rho^*}(\xi,\rho^*))\nonumber\\
=\theta_{R_1}(z,\xi)
-\frac{\hbar}{\tau}\frac{\partial^R \theta_{{R_1}}}{\partial\xi^A}
\sigma^A_1\rho^*\label{du}\\
=\frac{1}{2\tau}({S^{\rho^*}_{{R_1}}},
{S^{\rho^*}_{{R_1}}})
+\frac{1}{\tau}\Delta_c{S^{\rho^*}_{{R_1}}}\label{da}
.
\eea
Equations (\ref{do}) and (\ref{di}) imply that the action 
\bea
{S_{{R_1}}}_{\rho^*}={S^{\rho^*}_{{R_1}}}
+{\theta^{\rho^*}_{{R_1}}}\rho^*,
\eea
with $\tilde\Sigma_1=\Sigma_1-(S_1,\Xi_1)
-\frac{\partial^R S_1}{\partial \xi^A}\mu^A_1$, 
yields a one loop finite effective action 
both in the $\rho^*$ independent and the $\rho^*$ linear part, 
because the terms linear in $\rho^*$ of order $\hbar/\tau$ add up 
precisely  
to $-\theta^{(1)-1}$.
The one loop renormalized and regularized action 
${S_{{R_1}}}_{\rho^*}$ satisfies 
\bea
\frac{1}{2}({S_{{R_1}}}_{\rho^*},{S_{{R_1}}}_{\rho^*})+\Delta_c
{S_{{R_1}}}_{\rho^*}=\tau\theta_{{R_1}}^{\rho^*}\nonumber\\=\tau
\frac{\partial^R{S_{{R_1}}}_{\rho^*}}{\partial \rho^*}
+\frac{\partial^R{S_{{R_1}}}_{\rho^*}}
{\partial\xi^B}\hbar\sigma^B_1-\frac{1}{\tau}
\frac{\partial^R{S_{{R_1}}}_{\rho^*}}
{\partial\xi^B}\frac{\partial^R\hbar\sigma^B_1}{\partial\xi^A}\hbar
\sigma^A_1
\rho^*,
\eea
the first equality following from (\ref{da}), and the 
last equality from the expansions (\ref{de}), 
(\ref{du}), together with the identity 
\bea
(-)^{B(A+1)}\frac{\partial^R }{\partial \xi^B}
(\frac{\partial^R S_{R_1}}{\partial \xi^A})
\sigma^A_1\sigma^B_1=0. 
\eea
Let us now define
$\Delta^1=\Delta_c-\hbar\frac{\partial^R\cdot}{\partial\xi^A}\sigma_1^A$,
with $\frac{\partial^R\cdot}
{\partial\xi^B}[\Delta^1,\Delta^1]^B=O(\hbar^2)$. We
can then write
\bea
\frac{1}{2}({S_{{R_1}}}_{\rho^*},{S_{{R_1}}}_{\rho^*})+\Delta^1
{S_{{R_1}}}_{\rho^*}=\tau
\frac{\partial^R{S_{{R_1}}}_{\rho^*}}{\partial \rho^*}
-\frac{1}{\tau}
\frac{\partial^R{S_{{R_1}}}_{\rho^*}}
{\partial\xi^B}[\Delta^1,\Delta^1]^B
\rho^*,
\eea
According to the regularized quantum action principle,
\bea
\frac{1}{2}({\G_{{R_1}}}_{\rho^*},{\G_{{R_1}}}_{\rho^*})
+\Delta^1
{\G_{{R_1}}}_{\rho^*}
=\tau
\frac{\partial^R{\G_{{R_1}}}_{\rho^*}}{\partial\rho^*}
-\frac{1}{\tau}
\frac{\partial^R{\G_{{R_1}}}_{\rho^*}}
{\partial\xi^B}[\Delta^1,\Delta^1]^B
\rho^*.
\label{hh}
\eea
The $\rho^*$ independent part at one loop and 
lowest order, $\tau^0$, in
$\tau$ gives
\bea
\bar s{\G_{{R_1}}}^{(1)0}=\frac{\partial^R
S_0}{\partial\xi^B}\sigma^B_1,
\label{90}
\eea
and shows that only the non trivial part of the anomaly remains.

\subsubsection{Two loops}

\paragraph{Equations for the two loop poles}

The one loop renormalized action admits the expansion 
\bea
{\G_{R_1}}_{\rho^*}=S_{\rho^*}
+\hbar \Sigma_{n=0}\tau^n{\G_{R_1}}_{\rho^*}^{(1)n}
+\hbar^2\Sigma_{n=-2}\tau^n{\G_{R_1}}_{\rho^*}^{(2)n}
+O(\hbar^3).
\eea
At order $\hbar^2$, (\ref{hh}) gives
\bea
(S_{\rho^*},{\G_{R_1}}^{(2)}_{\rho^*})+\frac{1}{2}
({\G_{R_1}}^{(1)}_{\rho^*},{\G_{R_1}}^{(1)}_{\rho^*})+\Delta_c
{\G_{R_1}}^{(2)}_{\rho^*}\nonumber\\
=\tau\frac{\partial^R{\G_{R_1}}_{\rho^*}^{(2)}}{\partial\rho^*}+
\frac{\partial^R{\G_{R_1}}^{(1)}_{\rho^*}}{\partial\xi^B}\sigma^B_1
-\frac{1}{\tau}\frac{\partial^R S_0}{\partial \xi^B}
\frac{\partial^R\sigma^B_1}{\partial\xi^A}\sigma^A_1
\rho^*.
\label{ht}
\eea
Let ${\G_{R_1}}_{\rho^*}=\G_{R_1}
+\frac{\partial^R{\G_{R_1}}_{\rho^*}}{\partial\rho^*}\rho^*$.
At order $1/\tau^2$, we get, according to the $\rho^*$ independent and
linear parts,
\bea
\bar s {\G_{R_1}}^{(2)-2}=0,\label{con1} \\
\bar s(\frac{\partial^R{\G_{R_1}}_{\rho^*}^{(2)-2}}{\partial\rho^*}
-(S_1,{\G_{R_1}}^{(2)-2}))
=0.\label{con2}
\eea
The first of these equations implies:
\begin{lemma}
The second order pole of the 
two loop divergences is a $\bar s$ cocycle.
\end{lemma}
At order $1/\tau$, we get 
\bea
\bar s {\G_{R_1}}^{(2)-1}=
\frac{\partial^R{\G_{R_1}}_{\rho^*}^{(2)-2}}{\partial\rho^*}
-(S_1,{\G_{R_1}}^{(2)-2}),\label{ha}\\
\bar s(\frac{\partial^R{\G_{R_1}}_{\rho^*}^{(2)-1}}{\partial\rho^*}
-(S_1,{\G_{R_1}}^{(2)-1})-(S_2,{\G_{R_1}}^{(2)-2}))
=-\frac{\partial^R S_0}{\partial \xi^B}
\frac{\partial^R\sigma^B_1}{\partial\xi^A}\sigma^A_1.\label{gsp}
\eea
Finally, the $\rho^*$ independent part of (\ref{ht}), gives at order
$\tau^0$
\bea
\bar s{\G_{R_1}}^{(2)0}+\frac{1}{2}
({\G_{R_1}}^{(1)0},{\G_{R_1}}^{(1)0})=
\frac{\partial^R {\G_{R_1}}^{(1)0}}{\partial \xi^B}\sigma_1^B
\nonumber\\+
\frac{\partial^R{\G_{R_1}}_{\rho^*}^{(2)-1}}{\partial\rho^*}
-(S_1,{\G_{R_1}}^{(2)-1})-(S_2,{\G_{R_1}}^{(2)-2}),\label{gsp1}
\eea
which allows to identify the combination 
\bea
A_2=\frac{\partial^R{\G_{R_1}}_{\rho^*}^{(2)-1}}{\partial\rho^*}
-(S_1,{\G_{R_1}}^{(2)-1})-(S_2,{\G_{R_1}}^{(2)-2})
\eea
as the local
contribution to the two loop anomaly, whereas $\frac{\partial^R
{\G_{R_1}}^{(1)0}}{\partial \xi^B}\sigma_1^B$ is the one loop
renormalized dressing of the non trivial one loop anomaly.

\paragraph{Two loop anomaly consistency condition}
Before absorbing the divergences, let us consider (\ref{gsp}), which
can be written as 
\bea
\bar s A_2=-\frac{1}{2}\frac{\partial^R S_0}{\partial
\xi^B}[\sigma_1,\sigma_1]^B, \label{cons2}
\eea
where $\xi^*_B[\sigma_1,\sigma_1]^B
\equiv(\xi^*\sigma_1,\xi^*\sigma_1)_\xi$
is an $s_{\Delta^*}$ cocycle because 
of the graded Jacobi identity for the antibracket in $\xi,\xi^*$
space. 
According (\ref{27}),
this implies that 
\bea
\frac{1}{2}(\xi^*\sigma_1,\xi^*\sigma_1)_\xi
= s_{\Delta^*} \xi^*_A\sigma^A_2,\label{999}
\eea
and 
\bea
A_2=\bar s  \Sigma_2
+\frac{\partial^R S_0}{\partial \xi^B}\sigma^B_2.\label{100}
\eea

{\bf Discussion:}
We thus see that the consistency condition (\ref{cons2}) on the 
local contribution of the two loop anomaly does not require it to be
just a
cocycle of the extended BRST differential $\bar s$, because of the
non vanishing right hand side. This is in agreement with
the analysis of \cite{Paris:1995dx,Paris:1996jg}. 
Nevertheless, in the extended antifield
formalism, the general solution (\ref{100}) of (\ref{cons2}) can be
given. Up to a trivial $\bar s$ boundary, it contains the term 
$\frac{\partial^R S_0}{\partial \xi^B}\sigma^B_2$
with the following interpretation. From the point of view of
cohomology, equation (\ref{cons2}) should
be understood as a restriction on the non trivial one loop anomalies
$\xi^*_A\sigma^A_1$ that can arise. Indeed, its consequence is
(\ref{999}), which states that the non trivial one loop anomalies
should have a trivial antibracket map\footnote{The antibracket map
here is the antibracket induced in the $s_{\Delta^*}$ cohomological 
classes from the antibracket in $\xi$ space} among themselves. This is a
cohomological statement independent of the choice of representatives.
Through these cohomological considerations, the term
$\frac{\partial^R S_0}{\partial \xi^B}\sigma^B_2$ of the general
solution for the local part of the two loop anomaly is determined 
up to an arbitrary $s_{\Delta^*}$ cocycle, (or, using the liberty to
shift the $s_{\Delta^*}$ trivial part of it in the $\bar s$
coboundary, 
up to a $s_{\Delta^*}$
cohomological class). It contains a particular solution 
depending on the choice of representatives for
the non trivial one loop anomalies and needed to make the bracket 
$(\xi^*\sigma_1,\xi^*\sigma_1)_\xi$ $s_{\Delta^*}$ exact. 
This answers, at least in the present context of the extended
antifield formalism 
the question raised in \cite{Paris:1995dx,Paris:1996jg} on the
cohomological interpretation of the two loop anomaly consistency
condition.  It is confirmed by the analysis in the next subsection 
in the context of algebraic renormalization. 
One also sees on this example how the discussion of
the quantum Batalin-Vilkovisky formalism of
\cite{Paris:1995dx,Paris:1996jg} 
is shifted to $\xi,\xi^*$ space in the extended formalism. 

Note that, as in \cite{White:1992ni}, this result has been achieved by
adding a BRST breaking counterterm, not only for the one loop 
divergences produced
by the standard action itself, but also for the one loop 
divergences produced
by the insertion of the non trivial one loop anomaly. This is
because this anomaly has been coupled to the action itself from the
start, and the
BRST breaking counterterm $\Sigma_1$ also depends on the corresponding
coupling constants.

\paragraph{Two loop renormalization}

The general solution to (\ref{con1}) is 
${\G_{R_1}}^{(2)-2}=\bar s\Xi_{2,-2}+\frac{\partial^R
S_0}{\partial \xi^A}\mu^A_{2,-2}$.  
We consider the change of variables
\bea
z^{2,-2}=\exp -\frac{\hbar^2}{\tau^2}[(\cdot,\Xi^{\rho^*}_{2,-2}) 
+(\cdot,{\xi^1_{\rho^*}}^*
\mu_{2,-2}^{\rho^*})_{\xi^1_{\rho^*}}]
z,\\
{\xi^{2,-2}}
=\exp-\frac{\hbar^2}{\tau^2}(\cdot,{\xi^1_{\rho^*}}^*
\mu_{2,-2}^{\rho^*})_{\xi^1_{\rho^*}}\xi^1_{\rho^*}, 
\eea
where $\Xi^{\rho^*}_{2,-2}(z,\xi,\rho^*)=\Xi_{2,-2}(z,
\xi^1_{\rho^*}(\xi,\rho^*))$ and 
$\mu_{2,-2}^{\rho^*}(\xi,\rho^*)
= \mu_{2,-2}(\xi^1_{\rho^*}(\xi,\rho^*))$.
The fact that we consider this change of variables in terms of
$\xi^1_{\rho^*}$ instead of $\xi$ will not change the
absorption of the $\rho^*$ independent divergences, but it will be
important in order to control the dependence on $\rho^*$ below.
Equation (\ref{ha}) 
means that there is no non trivial part $\frac{\partial^R
S_0}{\partial \xi^A} \sigma^A_{2,-2}$ in the general solution to 
(\ref{con2}) and
hence no need for a renormalization of the coupling constants of order
$\hbar^2/\tau^2$ proportional to $\rho^*$. The general solution to
(\ref{con2}) is 
$\frac{\partial^R{\G_{R_1}}_{\rho^*}^{(2)-2}}{\partial\rho^*}
-(S_1,{\G_{R_1}}^{(2)-2})=\bar
s \Sigma_{2,-2}$, where $\Sigma_{2,-2}$ can be identified with a
particular solution ${\G_{R_1}}_P^{(2)-1}$ of (\ref{ha}).
We take 
\bea
{S^{\rho^*}_{R_{2,-2}}}(z,\xi,\rho^*)=
S_{R_1}(z^{2,-2}(z,\xi^1_{\rho^*}),\xi^{2,-2}(\xi^1_{\rho^*}))
+{g_{2,-2}}_A(z,\xi^1_{\rho^*})f^A(\xi^1_{\rho^*})
\nonumber\\-\frac{\hbar^2}{\tau}\tilde
\Sigma_{2,-2}(z^{2,-2}(z,\xi^1_{\rho^*}),
\xi^{2,-2}(\xi^1_{\rho^*})),\nonumber\\
={S_{{R_1}}}-\frac{\hbar^2}{\tau^2}{\G_{R_1}}^{(2)-2}
+O(\hbar^2\tau^{-1})
+O(\hbar^3),
\eea
where $\tilde\Sigma_{2,-2}(z,\xi)$ remains to be determined.
The remark after theorem \ref{t4} again implies
\bea
\theta^{\rho^*}_{{R_{2,-2}}}\equiv
\frac{1}{2\tau}({S^{\rho^*}_{R_{2,-2}}},
{S^{\rho^*}_{{R_{2,-2}}}})
+\frac{1}{\tau}\Delta_c{S^{\rho^*}_{{R_{2,-2}}}}
\nonumber\\
={\theta^{\rho^*}_{{R_1}}}-\frac{\hbar^2}{\tau^2}\bar s
[\tilde\Sigma_{2,-2}+\frac{\partial^R S_1}{\partial
\xi^A}\mu^A_{2,-2}+(S_1,\Xi_{2,-2})]\nonumber\\
-\frac{\hbar^2}{\tau^2}(S_1,{\G_{R_1}}^{(2)-2})+\hbar^2 O(\tau^{-1})
+O(\hbar^3).
\eea
The action 
\bea
{S_{{R_{2,-2}}}}_{\rho^*}={S^{\rho^*}_{{R_{2,-2}}}}
+{\theta^{\rho^*}_{{R_{2,-2}}}}\rho^*,
\eea
with $\tilde\Sigma_{2,-2}=\Sigma_{2,-2}-\frac{\partial^R S_1}{\partial
\xi^A}\mu^A_{2,-2}+(S_1,\Xi_{2,-2})$, yields an effective action 
${\G_{{R_{2,-2}}}}_{\rho^*}$
without $\hbar^2/\tau^2$ divergences and only simple poles at order
$\hbar^2$, because the terms linear in $\rho^*$ of order 
$\hbar^2/\tau^2$ add up precisely to
$-\frac{\partial^R{\G_{R_1}}_{\rho^*}^{(2)-2}}{\partial\rho^*}$.
We have again that 
\bea
\frac{1}{2}({S_{{R_{2,-2}}}}_{\rho^*},{S_{{R_{2,-2}}}}_{\rho^*})
+\Delta_c{S_{{R_{2,-2}}}}_{\rho^*}
=\tau{\theta^{\rho^*}_{{R_{2,-2}}}}\nonumber\\
=\tau\frac{\partial^R{S_{{R_{2,-2}}}}_{\rho^*}}{\partial\rho^*}
+\frac{\partial^R{S_{{R_{2,-2}}}}_{\rho^*}}{\partial\xi^B}
\hbar\sigma^B_1
-\frac{1}{\tau}
\frac{\partial^R{S_{{R_{2,-2}}}}_{\rho^*}}
{\partial\xi^B}\frac{1}{2}[\hbar\sigma_1,
\hbar\sigma_1]^B
\rho^*.
\eea
The last equation follows from the fact that the dependence of
$S^{\rho^*}_{{R_{2,-2}}}$ and ${\theta^{\rho^*}_{{R_{2,-2}}}}$ 
on $\rho^*$ is, as before, 
through the combination $\xi^1_{\rho^*}$.
The same equation holds again for the effective action:
\bea
\frac{1}{2}({\G_{{R_{2,-2}}}}_{\rho^*},{\G_{{R_{2,-2}}}}_{\rho^*})
+\Delta_c{\G_{{R_{2,-2}}}}_{\rho^*}
=\tau\frac{\partial^R{\G_{{R_{2,-2}}}}_{\rho^*}}{\partial\rho^*}
+\frac{\partial^R{\G_{{R_{2,-2}}}}_{\rho^*}}{\partial\xi^B}
\hbar\sigma^B_1
\nonumber\\-\frac{1}{\tau}
\frac{\partial^R{\G_{{R_{2,-2}}}}_{\rho^*}}
{\partial\xi^B}\frac{1}{2}[\hbar\sigma_1,
\hbar\sigma_1]^B
\rho^*.
\eea
The expansion of this effective action is 
\bea
{\G_{R_{2,-2}}}_{\rho^*}=S_{\rho^*}
+\hbar \Sigma_{n=0}\tau^n{\G_{R_{2,-2}}}_{\rho^*}^{(1)n}
+\hbar^2\Sigma_{n=-1}\tau^n{\G_{R_{2,-2}}}_{\rho^*}^{(2)n}
+O(\hbar^3).
\eea

The divergences $\G_{{R_{2,-2}}}^{(2)-1}$ and
$\frac{\partial^R{\G_{R_{2,-2}}}_{\rho^*}^{(2)-1}}{\partial\rho^*}$, 
now satisfy $\bar s
\G_{{R_{2,-2}}}^{(2)-1}=0$ and $\bar s A^\prime_2
=
\frac{\partial^R S_0}{\partial
\xi^A}\frac{1}{2}[\sigma^1,\sigma^1]^A$, with $A^\prime_2=
\frac{\partial^R{\G_{R_{2,-2}}}_{\rho^*}^{(2)-1}}{\partial\rho^*}-
(\G_{{R_{2,-2}}}^{(2)-1},S_1)$. The general solutions are 
$\G_{{R_{2,-2}}}^{(2)-1}=\bar s \Xi_{2,-1}
+\frac{\partial^R S_0}{\partial
\xi^A}\mu_{2,-1}$ and $A^\prime_2=\bar s\Sigma_{2,-1}
+\frac{\partial^R S_0}{\partial
\xi^A}\sigma_2^A$. As in the one loop case, one first subtracts a
suitably defined BRST breaking counterterm, then one makes the
field-antifield 
and coupling constant redefinition determined by 
$\Xi^{\rho^*}_{2,-1}$ and
$\mu^{\rho^*}_{2,-1}$, and finally, one substitutes 
$\xi^1_{\rho^*}$ everywhere by
$\xi^2_{\rho^*}=\xi^1_{\rho^*}-\frac{\hbar^2}{\tau}\sigma_2\rho^*$, 
giving a total $\rho^*$ dependence through the combination 
$\xi^2_{\rho^*}=\xi
-\frac{\hbar}{\tau}\sigma_1\rho^*
-\frac{\hbar^2}{\tau}\sigma_2\rho^*$. 

Using the same arguments as in the one loop case, one finally
finds\footnote{Note that the following equations in this and the next
two subsection have been corrected with respect to the ones in the
original paper \cite{Barnich:1999qz} and that theorem \ref{t9} below
has been accordingly improved.} 
that the two loop renormalized and regularized action
${S_{R_2}}_{\rho^*}$ satisfies, by defining 
$\Delta^2=\Delta_c-\hbar\frac{\partial^R\cdot}{\partial\xi^A}\sigma_1^A-
\hbar^2\frac{\partial^R\cdot}{\partial\xi^A}\sigma_2^A$, with
$\frac{\partial^R\cdot}{\partial\xi^B}
[\Delta^2,\Delta^2]^B=0(\hbar^3)$,
\bea
\frac{1}{2}({S_{{R_{2}}}}_{\rho^*},{S_{{R_{2}}}}_{\rho^*})
+\Delta^2{S_{{R_{2}}}}_{\rho^*}
=\tau
\frac{\partial^R{S_{{R_{2}}}}_{\rho^*}}{\partial\rho^*}
-\frac{1}{\tau}\frac{\partial^R{S_{{R_{2}}}}_{\rho^*}}{\partial\xi^B}
\frac{1}{2}[\Delta^2,\Delta^2]^B\rho^*,
\eea
the same equation holding for the 
two loop renormalized effective action 
${\G_{{R_{2}}}}_{\rho^*}$.

\subsubsection{Higher orders}

It is then possible to continue recursively to higher loops 
to get a completely subtracted and regularized action 
${S_{{R_{\infty}}}}$. It is obtained from 
\bea
S_\tau-\sum_{n=1}\frac{\hbar^n}{\tau^{n-1}}\sum_{k=0}^{n-1}\tau^k
\tilde\Sigma_{n,k-n},
\eea
with suitably chosen BRST breaking counterterms 
$\tilde\Sigma_{n,k-n}$,
by successive canonical field-antifield and 
coupling constants redefinitions.  
It satisfies
\bea
\frac{1}{2}({S_{{R_\infty}}}_{\rho^*},{S_{{R_\infty}}}_{\rho^*})
+\Delta^\infty{S_{{R_\infty}}}_{\rho^*}
=\tau\frac{\partial^R{S_{{R_\infty}}}_{\rho^*}}{\partial\rho^*},
\label{end1}
\eea
with 
\bea
\Delta^\infty = \Delta_c-\sum_{n=1}
\hbar\frac{\partial^R\cdot}{\partial\xi^A}\sigma_n^A,\\ 
\frac{\partial^R\cdot}{\partial\xi^B}
[\Delta^\infty,\Delta^\infty]^B\equiv (\Delta^\infty)^2=0.
\eea
The corresponding 
completely renormalized and regularized effective action 
${\G_{{R_\infty}}}_{\rho^*}$ satisfies the same equation. 
\bea
\frac{1}{2}({\G_{{R_\infty}}}_{\rho^*},{\G_{{R_\infty}}}_{\rho^*})
+\Delta^\infty {\G_{{R_\infty}}}_{\rho^*}
=\tau\frac{\partial^R{\G_{{R_\infty}}}_{\rho^*}}{\partial\rho^*}
.\label{end}
\eea
One can then take safely the limit $\tau\longrightarrow 0$,
because there are no more divergences left and put $\rho^*$ to zero.  
The renormalized 
effective action 
$\Gamma^\infty=(\lim_{\tau\rightarrow
0}{\G_{{R_\infty}}}_{\rho^*})|_{\rho^*=0}$ satisfies 
\bea
\frac{1}{2}(\G^\infty,\G^\infty)
+\Delta^\infty\G^\infty
=0.
\eea
\begin{theorem}\label{t9}
The absorption of the divergences in the extended antifield formalism 
involves, besides redefinitions of the solution of the extended master
equation,
determined by anticanonical field-antifield and coupling constant 
renormalizations, only the subtraction of suitably chosen BRST breaking
counterterms. The renormalization can be done in 
such a way that the effective action satisfies an extended master
equation with a differential $\Delta^\infty$ that is a deformation of
the differential $\Delta_c$ of the classical extended master
equation. This statement contains the cohomological information
on the anomaly consistency condition to all orders.
\end{theorem}

\subsubsection{The quantum Batalin-Vilkovisky $\Delta$ operator}

In \cite{Troost:1990cu,Paris:1995dx,Paris:1996jg}, explicit expression
for the $\Delta$
operator have been obtained in the context of Pauli-Villars and non
local regularization respectively. The aim of this section is to get
such an expression in the context of the present ``dimensional'' 
renormalization. 
The expression we will get here will be defined on all the
generalized observables of the theory, and not only on $S$ alone,
since they are contained in the solution $S(\xi)$ of the extended
master equation. 

As discussed for instance in section 4 of 
\cite{DeJonghe:1996xm} in the context of
the BPHZ renormalized antifield formalism, even though there is a well
defined expression for the anomaly, there is no room for the
formal Batalin-Vilkovisky $\Delta$ operator in the final renormalized
theory. 
Contact with the quantum Batalin-Vilkovisky formalism in the present 
set-up has thus to be done on the renormalized theory before the
regulator $\tau$ is removed. Moreover, as in the previous 
discussion of the renormalization, it turns out to 
be important not to put to zero the
fermionic variable $\rho^*$, which couples the breaking of the extended
master equation due to the regularization.
Let us introduce the notation 
$W={S_{R_\infty}}_{\rho^*}$ for
the completely renormalized and regularized action and define 
$\Delta_d=\frac{\tau}{i\hbar}\frac{\partial^R\cdot
}{\partial\rho^*}$
so that (\ref{end1}) becomes
\bea
\frac{1}{2}(W,W)
+(\Delta^\infty -i\hbar \Delta_d) W
=0.\label{bla}
\eea
The operator $\Delta_d$ is of ghost number $1$, it is nilpotent,
$\Delta_d^2=0$, it anticommutes with $\Delta^\infty$, 
$\{\Delta^\infty,\Delta_d\}=0$, so that $\Delta^T=\Delta^\infty-i\hbar
\Delta_d$ 
is a differential $(\Delta^T)^2=0$. $\Delta_d$ is also a graded
derivation of the antibracket, i.e., it satisfies equation (\ref{der})
(with $\Delta_c$ replaced by $\Delta_d$ or $\Delta^T$).

{\bf Discussion:}
Starting from the path integral expression
\bea
Z(J,\phi^*,\xi,\rho^*)=\int {\cal D}\phi
\exp\left(\frac{i}{\hbar}[W+\int d^nx\ J_a\phi^a]\right),
\eea
with associated effective action ${\G_{{R_\infty}}}_{\rho^*}$,
standard formal path integral manipulations using integrations 
by parts give
\bea
\frac{1}{2}({\G_{{R_\infty}}}_{\rho^*},{\G_{{R_\infty}}}_{\rho^*})
+\Delta^\infty{\G_{{R_\infty}}}_{\rho^*}={\cal A}^\prime\circ
{\G_{{R_\infty}}}_{\rho^*},
\eea
where 
\bea
{\cal A}^\prime=\frac{1}{2}(W,W)+\Delta^\infty
W-i\hbar''\Delta W''.
\eea
This expression involves the second order functional 
derivative operator 
$\Delta=(-)^{A+1}\frac{\delta^R}{\delta\phi^a(x)}
\frac{\delta^R}{\delta\phi^*_a(x)}$.
The quotation marks mean that the above definition of $\Delta$ cannot
be used since $\Delta$ is ill defined when acting on local functionals
and thus on $W$. Using (\ref{end}) for the left hand side, we get 
${\cal A}^\prime=i\hbar\Delta_d W$. Using furthermore (\ref{bla}), 
it follows that $-i\hbar''\Delta
W''=0$, as was to be expected in ``dimensional'' regularization, where
$''\delta(0)''=0$. 

In equation (\ref{bla}), obtained by an analysis of the
renormalization procedure, there appears the operator 
$\Delta_d$, which is unexpected from the
point of view of formal path integral manipulations, not taking the
regularization and renormalization into account. Furthermore, the 
operator $\Delta_d$ has the same algebraic properties as the formal
operator  $\Delta$, when acting on local functionals.
In ``dimensional regularization'', one has traded the operator
$\Delta$, vanishing on local functionals, for the operator
$\Delta_d$. We thus find, in the context of dimensional
regularization, that the role of the Batalin-Vilkovisky $\Delta$
operator is played by the operator $\Delta_d$, 
introduced originally in \cite{Aoyama:1981yw}. 

Furthermore, the terms of higher order in
$\hbar$ of $\Delta^\infty$ contain the information about the anomalies
of the theory, while (\ref{bla}) suggests that the operator $\Delta_c$, 
can be understood as a classical part of the 
Batalin-Vilkovisky $\Delta$
operator. We also note that both
$\Delta_c$ and $\Delta_d$ arise in a similar way from an extended 
action satisfying a standard master equation in an extended space with
an enlarged bracket: this was shown for $\Delta_c$ in section \ref{s521}.  
In \cite{Tonin:1992wf} in the context of the standard Batalin-Vilkovisky 
formalism, it was shown that $\Delta_d$ also arises
from an ``improved'' classical
master equation, if 
the space of fields and
antifields is enlarged to include the global pair of 
variables $\rho,\rho^*$, the antibracket is 
extended to this pair and the
regularized action is extended to $S_\tau+\theta_\tau\rho^*+\tau\rho$.

\subsection{Algebraic renormalization in the extended antifield formalism}

The algebraic approach to control symmetries in renormalization theory
is based on the use of the renormalized quantum action principles 
\cite{Lowenstein:1971jk,Lowenstein:1971vf,Lam:1972mb,Lam:1973qa,Clark:1976ym},
that hold independently of the particular scheme being used.

More precisely, let $S_{\rm gf}$ be the classical gauge fixed 
action of the theory. If
$\G^\infty$ denotes the
renormalized generating functional for one particle irreducible
vertices, one has
\bea
\G^\infty=S_{\rm gf}+O(\hbar).
\eea
Similarly, if $\D\circ\G^\infty$, respectively $\D(x)\circ\G^\infty$,
denotes the renormalized insertion of an (integrated) local polynomial into 
$\G$, one has 
\bea
\D\circ\G^\infty=\D+O(\hbar).
\eea

Let $g$ be a parameter of $S_{\rm gf}$ and let $\phi(x)$ be the source for the 
field operators in the generating functional for one particle
irreducible vertex functions with $\rho(x)$ an external source coupling to 
a polynomial in the fields and their derivatives. 
We will use the quantum action principle in the following forms: 
\bea
{\rm (non\ linear)\ field\ variations:}\nonumber\\ 
{\delta\G^\infty\over\delta \phi(x)}
{\delta\G^\infty\over\delta \rho(x)}
=\D^{\prime\prime}(x)\circ\G^\infty,\nonumber\\
\D^{\prime\prime}(x)\circ
\G^\infty
={\delta S_{\rm gf}\over\delta \phi(x)}{\delta S_{\rm gf}
\over\delta \rho(x)}+O(\hbar).
\eea
\bea
{\rm coupling\ constants:}\ 
{\partial\G^\infty\over\partial g}=\D\circ\G^\infty,\nonumber\\ 
\D\circ\G^\infty={\partial S_{\rm gf}\over\partial g}+O(\hbar).
\eea

In the following, we will assume the validity of these equations, even
in the context of power counting non renormalizable theories. The precise
functional dependence of $\G^\infty$ 
on the couplings will not be discussed, we just
assume it to be sufficiently regular to allow for the manipulations
below. 

In the algebraic approach to the usual version of the 
BRST-Zinn-Justin-Batalin-Vilkovisky set-up,
there are two main issues to be considered (see
e.g. \cite{Bonneau:1990xu,Piguet:1995er}): stability and anomalies.

\subsubsection{Stability}
  
The problem of stability (in the physical sector) is 
the question if to every local BRST cohomological class $H^{0,n}(s|d)$ 
in ghost number
$0$, there corresponds an independent coupling of the standard master
equation. If this is the case, every infinitesimal deformation of the
action (by invariant, in this context finite, counterterms) 
can be absorbed by a coupling constant and anticanonical
field-antifield redefinition in such a way that the master equation is
still satisfied. 

The
extended formalism solves this problem by construction, because
all standard cohomological classes have been coupled with independent
couplings. Indeed, in the extended formalism, the differential
controlling the ``instabilities'', i.e., the divergences and/or
counterterms, is the differential $\bar s$. According to theorem
\ref{t4}, every infinitesimal deformation can be absorbed in such a
way that the deformed action still satisfies the extended master
equation. 

Since no additional arguments like power counting have been used to
achieve this stability, one can say that the use of 
the extended antifield formalism guarantees
``renormalizability in the modern sense'' \cite{Gomis:1996jp} for all
gauge theories. 

Of course, it will be often convenient in practice not to couple all 
the local BRST cohomological classes but only a subset needed to
guarantee that the theory is stable, especially if one uses
additional conditions like power counting. 

\subsubsection{Anomalous Zinn-Justin equation}
\label{sec2}

In the standard set-up, the question of anomalies is mostly reduced to the
computation of the local BRST cohomological group $H^{1,n}(s|d)$ 
in ghost number $1$ and to a
discussion of the coefficients of the corresponding classes. In the
presence of anomalies, there is no differential 
on the quantum level associated to the anomalously broken Zinn-Justin
equation for the effective action. In the
extended antifield formalism however, because all the local BRST
cohomological classes in positive ghost numbers have been coupled to the
solution of the master equation, the broken Zinn-Justin equation can
be written as a functional differential equation and an 
associated differential exists, even in the presence of anomalies. To
show this, is the object of the remainder of this subsection
\footnote{We rederive section 4 of \cite{Barnich:1998ke} in a more
appropriate notation, 
insisting on the existence of the quantum BRST differential
in the anomalous case and its relation to its classical counterpart $\bar
s$. Note that the relation after (4.9) of that paper 
should read $s_Q \frac{\partial^R\cdot}{\partial
\xi^A}\sigma^A=\frac{1}{2}\frac{\partial^R\cdot}{\partial \xi^A}
[\sigma,\sigma]^A$ instead of $s_Q \frac{\partial^R\cdot}{\partial
\xi^A}\sigma^A=0$.}.

The quantum action principle
applied to (\ref{qme}) gives
\bea
\frac{1}{2}(\G,\G)+\D_c\G=\hbar{\cal A}\circ\G\label{2},
\eea
where $\G$ is the renormalized generating functional for 1PI vertices
associated to the solution $S$ of the extended master equation
and the local functional ${\cal A}$ is an element of 
$F$ in ghost number $1$. 
Applying
$(\G,\cdot)+\D^L_c$ to (\ref{2}), the l.h.s vanishes identically
because of the 
graded Jacobi identity for the antibracket and the properties of
$\D_c$, so that one gets
the consistency condition $(\G,{\cal A}\circ\G)+\D_c^L{\cal A}\circ\G
=0$. 
To lowest order
in $\hbar$, this gives $\bar s {\cal A}=0$, the general solution of
which can be written as 
\bea
{\cal A}=-{\partial^R S\over\partial \xi^A}\Delta^A_1+\bar s \Sigma_1,
\eea
with $[\Delta_c,\Delta_1]=0$, because of the relation between the
$\bar s$ and the $s_{\Delta_c}$ cohomologies discussed in the previous
section. 

If
one now defines ${S}^1=S-\hbar\Sigma_1$, the
corresponding generating functional admits the expansion 
$\G^1=\G-\hbar\Sigma_1+O(\hbar^2)$ and satisfies
$\frac{1}{2}(\G^1,\G^1)+\D^1\G^1=O(\hbar^2)$, where 
$\Delta^1=\Delta_c+\hbar 
\Delta_1$.  On the
other hand, the quantum action principle applied to
$\frac{1}{2}(S^1,S^1)+\D^1 S^1=O(\hbar)$ implies
$\frac{1}{2}(\G^1,\G^1)+\D^1\G^1=\hbar\bar{\cal A}\circ\G^1$, for a local
functional $\bar{\cal A}$. Comparing the two
expressions, we deduce that 
\bea
\frac{1}{2}(\G^1,\G^1)+\D^1\G^1=\hbar^2{\cal A}^\prime\circ\G^1,
\eea
for a local functional ${\cal A}^\prime$.
Applying now
$(\G^1,\cdot)+(\Delta^1)^L$, one gets as consistency condition
\bea
\frac{1}{2}[\D^1,\D^1]\G^1+\hbar^2((\G^1,{\cal A}^\prime\circ\G^1)
+\Delta^1{\cal A}^\prime\circ\G^1)=0,
\eea
giving to lowest order 
\bea
1/2[\D_1,\D_1]S+\bar s 
{\cal A}^\prime=0.
\eea
Since $1/2[\D_1,\D_1]$ is 
a $s_{\Delta_c}$ cocycle because of the graded Jacobi identity for the graded 
commutator, equation (\ref{27})
imply that the general solution to this equation is 
\bea
\frac{1}{2}[\D_1,\D_1]+[\Delta_c,\D_2]=0, \\
{\cal A}^\prime=\bar s \Sigma_2-{\partial^R
  S\over\partial\xi^B}\D^B_2.
\eea 
The redefinition $S^2=S^1-\hbar^2\Sigma_2$ then allows to achieve 
\bea
\frac{1}{2}(\G^2,\G^2)+\D^2\G^2=
\hbar^3{\cal A}^{\prime\prime}\circ\G^2,
\eea
for a local functional ${\cal A}^{\prime\prime}$, with 
$\D^2=\D^1+\hbar^2\D_2$.
The reasoning can be
pushed recursively to all orders with the result 
\bea
\frac{1}{2}(\G^\infty,\G^\infty)+\D^\infty\G^\infty=0,\label{qsm}
\eea
where ${\G}^\infty$ is associated to the action 
${S}^\infty=S-\Sigma_{k=1}\hbar^k\Sigma_k$ and 
$\D^\infty=\D_c+\hbar \D_1 +\hbar^2 \D_2+\dots$ satisfies 
$(\D^\infty)^2=0$.
The associated quantum BRST differential is
\bea
s^q=(\G^\infty,\cdot)+(\D^\infty)^L. 
\eea
In the limit $\hbar$ going to zero, it 
coincides with the classical differential $\bar s$. 

In the extended antifield formalism, the anomalous Zinn-Justin
equation can thus be written as a functional differential equation for the
renormalized effective action. The derivations $\D_1,\D_2,\dots$ are
guaranteed to exist due to the quantum action principles. They
satisfy a priori cohomological restrictions due to the fact that 
the differential $\D^\infty$ is a formal deformation with deformation
parameter $\hbar$ of the differential $\D_c$. In the context of chiral 
Yang-Mills theories, where $\Delta_c=0$, 
an anomalous master equation of the form
(\ref{qsm}) for the renormalized effective action has appeared for the 
first time in \cite{Costa:1977pd}. 

\subsubsection{Renormalization of local BRST cohomological
  classes}\label{s623} 

Associated quantum extension of a representative $\lambda_0S $ of a 
classical BRST cohomological classes of the extended formalism 
satisfying $\bar s\lambda_0S=0$, or equivalently
$[\Delta_c,\lambda_0]=0$,  are given by $\lambda_0
  \Gamma^\infty$. By acting with
$\lambda_0$ on (\ref{qsm}),
one gets 
\bea
s^q\lambda_0
  \Gamma^\infty=-[\Delta^\infty,\lambda_0]
\Gamma^\infty. 
\eea
The derivation $[\Delta^\infty,\lambda_0]=\sum_{n\geq L}\hbar^n \mu_n$ 
on the right hand side is of order at least $\hbar$, $L\geq 1$, because 
$[\Delta_c,\lambda_0]=0$ and corresponds to the anomaly in the invariant
renormalization of $\lambda_0
  S$. The question then is whether there
exists modified quantum extension $\lambda^\infty
  \Gamma^\infty$, with
${\lambda}^\infty=\lambda_0+\hbar\lambda_1+\dots$, such that 
\bea
s^q\lambda^\infty
  \Gamma^\infty=0\Longleftrightarrow
  [\Delta^\infty,{\lambda}^\infty]=0.\label{cocsup} 
\eea
This is for instance the case if
$\lambda_0$ corresponds to the 
trivial cohomological class, $\lambda_0=[\Delta_c,\nu_0]$. The searched
for extension can then simply be taken to be
$\lambda^\infty=[\Delta^\infty,\nu_0]$. 

Because $[\Delta^\infty,[\Delta^\infty,\lambda_0]]=0$, 
the lowest order part of the anomaly, $\mu_L$ is an $s_{\Delta_c}$
cocycle, $[\Delta_c,\mu_L]=0$. Suppose that $\mu_L$ is a trivial
solution to this equation, $\mu_L=-[\Delta_c,\lambda_L]$. The modified
quantum extension $(\lambda_0+\hbar^L\lambda_L)\Gamma^\infty$
allows to push the anomaly to order $L+1$. 

Hence, to lowest order in $\hbar$, the non trivial
part of the anomaly in the renormalization of a classical cohomological 
class
$H^g(s_{\Delta_c})$ is constrained to belong to
$H^{g+1}(s_{\Delta_c})$. All the
quantum information on this anomaly is encoded in the derivations 
$\Delta_1,\Delta_2,\dots$ and the whole discussion 
has been shifted from local functionals to derivations of
functions of the coupling constants. 

\subsubsection{Quantum BRST cohomologies}

The following lemma turns out to be extremely useful in the
sequel. It concerns the insertion of BRST exact local
functionals. 

\begin{lemma}\label{00}
The insertion of a BRST exact local functional $\bar
s\Xi=(S_{\rm gf},\Xi)+\Delta_c^L\Xi$ is equal to
$s^q=(\G^\infty,\cdot)+(\Delta^\infty)^L$ applied to a quantum
extension of this local
functional, up to a local
insertion of higher order in $\hbar$, 
\beq
[\bar s \Xi]\circ\G^\infty=s^q \Xi^Q\circ \G^\infty
+\hbar I\circ\G^\infty,\label{A1}
\eeq
where 
$\Xi^Q=\Xi+O(\hbar)$ and 
$I$ are local functionals.
\end{lemma} 
\proof{If $S^\infty$ is the sum of $S_{\rm gf}$ and the
BRST finite breaking local counterterms needed to achieve  (\ref{qsm}),
the action $S_\rho=S^\infty+\Xi\rho$ satisfies
$\frac{1}{2}(S_\rho,S_\rho)+\Delta^\infty S_\rho=\bar s
\Xi\rho+O(\hbar)+O(\rho^2)$. 
Applying the quantum
action principle, we get
$\frac{1}{2}(\G_\rho,\G_\rho)+\Delta^\infty\G_\rho={\cal D}(\rho)
\circ\Gamma_\rho$. Putting
$\rho$ to zero, it follows from (\ref{qsm}) that the local functional 
${\cal D}(0)=0$, so that
${\cal D}(\rho)={\cal D}^\prime(\rho)\rho$. Differentiation of the
previous equation with respect to $\rho$ and putting $\rho$ to zero
then implies
$s^q \Xi^Q\circ\Gamma^\infty
= {\cal D}^\prime(0)\circ\G^\infty$, for some local functional
$\Xi^Q=\Xi+O(\hbar)$. 
At tree level, this equation implies that $\Delta^\prime(0)=\bar s \Xi
+O(\hbar)$, which gives the result. }

Let us now establish the quantum analog of the classical
equation (\ref{27}), i.e.,
\bea
\left\{\begin{array}{c}s^q B\circ\G^\infty+\lambda^\infty
\G^\infty=0,\cr
[\Delta^\infty,\lambda^\infty]=0,
\end{array}\right.\Longleftrightarrow
\left\{\begin{array}{c}
B\circ\G^\infty=s^q C\circ\G^\infty+\mu^\infty\G^\infty,\cr
\lambda^\infty=[\Delta^\infty,\mu^\infty].
\end{array}\right.\label{c27}
\eea
\proof{If the two equations on the right of the equivalence
  sign hold, the equations on the left just follow by applying $s^q$
  to the first equation on the right. Conversely, suppose that the
  equations on the left hold. From the first equation at order zero in
  $\hbar$, we deduce that $\bar s B_0+\lambda_0 S=0$ with
  $[\Delta_c,\lambda_0]=0$. According to (\ref{27}), this implies
  $B_0=\bar sC_0 +\mu_0 S$, with
  $\lambda_0=[\Delta_c,\mu_0]$. Applying the quantum action principles
  and lemma \ref{00}, we get $B\circ\G^\infty=s^q C_0^Q\circ\G^\infty
+\mu_0\circ\G^\infty+ \hbar B^\prime\circ\G^\infty$. Injecting into
the first equation gives $\hbar s^q
B^\prime\circ\G^\infty+(\lambda^\infty-[\Delta^\infty,\mu_0]) 
\G^\infty=0$. Because
$\lambda^\infty-[\Delta^\infty,\mu_0]=\hbar{\lambda^\prime}^\infty$,
with $[\Delta^\infty,{\lambda^\prime}^\infty]=0$, we can factorize
$\hbar$ and iterate the reasoning, which proves (\ref{c27}). 
}

If we put furthermore $\lambda^\infty$ to zero in (\ref{c27}), 
we see that the general solution to the $s^q$ cocycle condition is the
insertion $\mu^\infty\G^\infty$ with $[\Delta^\infty,\mu^\infty]=0$,
up to an $s^q$ coboundary. We have thus proved the following
theorem. 
\begin{theorem}
The cohomology of 
$s^q=(\Gamma^\infty,\cdot)+(\Delta^\infty)^L$ in the
space of local insertions is isomorphic to the cohomology of
$[\Delta^\infty,\cdot]$ in the space of graded right derivations in the
couplings $\xi^A$ that are formal power series in $\hbar$, the
isomorphism being given by 
$[\lambda^\infty\G^\infty]\longleftrightarrow[\lambda^\infty]$. 
\end{theorem}
These cohomology groups are defined to be the quantum BRST cohomology
groups of the extended antifield formalism. They can be considered to
be well defined versions of the formal quantum BRST cohomology
of the standard Batalin-Vilkovisky formalism, involving the
ill-defined second order operator
$\Delta=(-)^{a+1}\frac{\delta^R\cdot}{\delta\phi^a(x)}
\frac{\delta^R\cdot}{\delta\phi^*_a(x)}$ obtained through formal path
integral manipulations. 

\newpage

\mysection{Gauge parameter dependence}\label{s7}

The problem of the gauge dependence of the effective action and of 
the renormalization group functions has been extensively studied
in the mid seventies in the context of Yang-Mills theories 
\cite{Caswell:1974cj,Kluberg-Stern:1975rs,Kluberg-Stern:1975xv,
Kluberg-Stern:1975hc,Breitenlohner:1975qe}. An algebraic approach to
the problem, independent of the renormalization scheme, has been
proposed in \cite{Piguet:1985js}. On the assumption of the
existence of an invariant renormalization scheme, extensions to
generic, not necessarily power counting renormalizable theories have
been considered in 
\cite{Voronov:1982cp,Voronov:1982ur,Voronov:1982ph,Lavrov:1985hr}
and more recently in \cite{Anselmi:1994ry,Anselmi:1995zx}.

In this section, we combine the ideas of the above cited works and 
reinvestigate the problem in the general setting of the 
extended antifield formalism. 
 

We assume that the gauge fixing fermion $\Psi$ does not
depend on any essential coupling, whereas, as stated before, that the
minimal solution of the master equation only depends on the essential
couplings.  The gauge fixed action $S_{\rm gf}$
to be used as a starting point for perturbative quantization depends
on the gauge parameters $\alpha^i$, which we assume for simplicity to 
be bosonic, 
according to (\ref{gaugedep}). This
means that we have also to allow for a dependence of the effective
action, the local insertions
and the right derivations discussed so far on these gauge
parameters. 

\subsection{Gauge parameter dependence of effective action and anomalies}

According to the quantum action principle, $\partial^R_{\alpha^i}
\Gamma^\infty=K_i\circ\Gamma_{\rm gf}$, where $K_i=
-(S_{\rm gf},\partial_{\alpha^i}\Psi)_{\phi_c,\tilde\phi^*}
+O(\hbar)=-\bar s \partial^R_{\alpha^i}\Psi+O(\hbar)$. 
It follows from lemma \ref{00} that
this implies in a first step 
\beq
\partial^R_{\alpha^i}
\Gamma^\infty=s^q[-\partial^R_{\alpha^i}\Psi]^Q\circ
\Gamma^\infty
+\hbar K^\prime_i\circ\Gamma^\infty.\label{-6}
\eeq
Applying $s^q$ and
using (\ref{qsm}), we get
to lowest order in $\hbar$ the consistency condition
$[\Delta_1,\partial^R_{\alpha^i}] S_{\rm gf}+\bar s {K^\prime_i
  }_0=0$. Using $[\Delta_c,[\Delta_1,
\partial^R_{\alpha^i}]]=0$, equation
(\ref{27}) implies ${K^\prime_{i} }_0=-{\rho_i}_1
S_{\rm gf}+\bar s {N_i}_1$, with $[\Delta_1,
\partial^R_{\alpha^i}]+[\Delta_c,{\rho_i}_1]=0$. 
The quantum
action principle under the form $[{\rho_i}_1
S_{\rm gf}]\circ\G^\infty
={\rho_i}_1\G^\infty+\hbar I_i\circ\G^\infty$, for a local
insertion $I_i\circ\G^\infty$, together with lemma \ref{00} give
$(\partial^R_{\alpha^i}+\hbar{\rho_i}_1)\G^\infty=s^q([-
\partial_{\alpha^i}\Psi^Q+\hbar {N_i}_1^Q]\circ
\Gamma^\infty)+\hbar^2
K^{\prime\prime}_i\circ\G^\infty$. Defining $D_i^1=
\partial^R_{\alpha^i}+\hbar{\rho_i}_1$, we have
$[\Delta^\infty,D_i^1]=\hbar^2 \lambda_i +O(\hbar^3)$, so that 
$[\Delta_c,\lambda_i]=0$, and the reasoning can be pushed to higher
orders, with the result:
\bea
D^\infty_i \Gamma^\infty=s^q (L_{i}\circ
\Gamma^\infty),\label{e72}\\
D^\infty_i=\partial^R_{\alpha^i}+\sum_{n=1} \hbar^n{\rho_i}_n, 
\ [\Delta^\infty,D^\infty_i]=0.\label{e73}
\eea
where $L_{i}=-\partial^R_{\alpha^i}\Psi+O(\hbar)$ and the
coefficients ${\rho^A_i}_n$ of the right derivations
$\frac{\partial^R\cdot}{\partial \xi^A} {\rho^A_i}_n$
depend on the essential couplings $\xi^A$ and the gauge couplings
$\alpha^i$. Note that the relation (\ref{c27}) cannot be used in this
case, because $D^\infty_i$ does not belong to the space of graded
right derivations in the essential couplings alone (with a possible
gauge parameter dependence).

Let us now analyze the equations (\ref{e72}) and (\ref{e73}) for each
$i$ separately and drop the subscript $i$. Let us define the functions 
$\xi_\alpha^A=\xi_\alpha^A(\xi^B,\alpha)$ through system of 
differential equations
\beq
\frac{d}{d\alpha}\xi_\alpha^A=\rho^A(\xi_\alpha^A,\alpha),
\eeq
where $\rho^A=\sum_{n=1}\hbar^n \rho^A_n$.
If we introduce a subscript $\alpha$ for all quantities where
$\xi^A$ has been replaced by $\xi_\alpha^A$, equation (\ref{e72})
becomes 
\beq
\frac{d}{d\alpha}\Gamma^\infty_\alpha=s^q_\alpha (L_\alpha\circ
\Gamma^\infty_\alpha),
\eeq 
where $s^q_\alpha=(\Gamma^\infty_\alpha,\cdot)+\D^\infty_\alpha$,
with $\D^\infty_\alpha=\frac{\partial^R\cdot}{\partial\xi^A_\alpha}
{\Delta^\infty}^A_\alpha\equiv\frac{\partial^R\cdot}{\partial\xi^B}
\frac{\partial^R\xi^B}{\partial\xi^A_\alpha}{\Delta^\infty}^A_\alpha$.
Equation (\ref{e73}) then becomes
\beq
\frac{d}{d\alpha}\D^\infty_\alpha=0.
\eeq
We have thus proved:
\begin{theorem}\label{theor11}
For each gauge parameter separately, there exists redefinitions 
of the essential couplings by gauge
parameter dependent terms of higher order in $\hbar$ such that the
variation of the effective action with respect to the gauge parameter
is given by a quantum BRST coboundary, while the anomaly operator in 
terms of the redefined couplings $\D^\infty_\alpha$ is independent of
the gauge parameter.  
\end{theorem}

\subsection{Integrability condition
}

By adapting the extended BRST technique of \cite{Piguet:1985js} to the
present context, one can show 
\begin{lemma}\label{llem}
There exist 
local functionals $K_{[ij]}$ such that 
\beq
[D^\infty_{i},D^\infty_{j}]\Gamma^\infty+s^q
[K_{[ij]}\circ\Gamma^\infty]=0.\label{2.2}
\eeq
\end{lemma}
\proof{If we introduce parameters $\lambda^i$ of opposite 
Grassman parity to the $\alpha^i$ and define $S^e=S^\infty +
\partial^R_{\alpha^i}\Psi\lambda^i$. Using $\partial^R_{\alpha^i}S_{\rm gf}=
\bar s (-\partial^R_{\alpha^i}\Psi)$ and $\frac{{\partial^R}^2\Psi}
{\partial\alpha^i\alpha^j}\lambda^i\lambda^j=0$, it follows that 
\beq
\frac{1}{2}(S^e,S^e)+\Delta^\infty S^e+
D^\infty_i S^e\lambda^i=\frac{1}{2}(\partial^R_{\alpha^i}\Psi\lambda^i,
\partial^R_{\alpha^j}\Psi\lambda^j)+O(\hbar),\label{Aex}
\eeq
where $O(\hbar)$ is
a local functional of order at least $\hbar$. 
Applying the quantum action principle, it follows that 
$\frac{1}{2}(\G^e,\G^e)+\Delta^\infty\G^e+
D^\infty_i\G^e\lambda^i= \frac{1}{2}(\partial^R_{\alpha^i}\Psi\lambda^i,
\partial^R_{\alpha^b}\Psi\lambda^b)\circ\G^e+\hbar A\circ\G^e$. 
Putting $\lambda^i$ to zero and using (\ref{qsm}), it follows that 
$A=A_i\lambda^i$.
Differentiation with respect to $\lambda^i$ and putting
$\lambda^i$ to zero gives 
$s^q (\partial^R_{\alpha^i}\Psi^Q\circ \G^\infty)
+D^\infty_i\G^\infty=\hbar
A_{i}(0)\circ\G^\infty$. Using (\ref{e72}), we deduce that 
$A_{i}(0)\circ\G^\infty=s^q (L^\prime_{i}\circ \G^\infty)$, where 
$\hbar L^\prime_{i}\circ \G^\infty=\partial^R_{\alpha^i}\Psi^Q\circ 
\G^\infty+L_{i}\circ
\G^\infty$. If we now add to $S^e$ the counterterm $-
\hbar {L^\prime_i}_0\lambda^i$, we can absorb the lowest order contribution 
$A_{i}(0)$ up to terms of second order in $\hbar$ or of
first order in $\hbar$ and of second order in $\lambda^i$. For
the new $\G^e$, we end up with $\frac{1}{2}(\G^e,\G^e)+\Delta^\infty\G^e+
D^\infty_i\G^e\lambda^i=[\frac{1}{2}
B_{[ij]}(\lambda)\lambda^i\lambda^j +
\hbar^2 
A^\prime_i(0)\lambda^i]\circ\G^e$, where $B_{[ij]}(\lambda)
= (\partial^R_{\alpha^i}\Psi\lambda^i,
\partial^R_{\alpha^j}\Psi\lambda^j)+O(\hbar)$.
Differentiation with respect to $\lambda^i$ and
putting $\lambda^i$ to zero now gives $s^q(K_{i}\circ\G^\infty)=
A^\prime_i(0)\circ \G^\infty$, which implies that 
the lowest order contribution to $A^\prime_i(0)$ can be absorbed
by adding suitable counterterm proportional to $\lambda^i$ and of
order $\hbar^2$. Going on in the same way, one can achieve:
\beq
\frac{1}{2}(\G^e,\G^e)+\Delta^\infty\G^e+
D^\infty_i\G^e\lambda^i
=\frac{1}{2}K_{[ij]}(\lambda)
\circ\G^e\lambda^i\lambda^j,
\eeq
where $K_{[ij]}(\lambda)=(\partial^R_{\alpha^i}\Psi\lambda^j,
\partial^R_{\alpha^i}\Psi\lambda^j)+O(\hbar)$.

Acting with $
D^\infty_k\lambda^k$ on this equation, and using the same equation 
again, together with (\ref{e73}),
$((\G^e,\G^e),\G^e)=0$, and ${\Delta^\infty}^2=0$, we find
$(\G^e,\cdot)+{\Delta^\infty}^L
[\frac{1}{2}K_{[ij]}(\lambda)\lambda^i\lambda^j
\circ\G^e]+\frac{1}{2}[D_i,D_j]
\G^e\lambda^i\lambda^j=\frac{1}{2}D_k
[K_{[ij]}(\lambda)
\circ\G^e]\lambda^i\lambda^j \lambda^k$. 
Differentiating with respect to $\lambda^i$ and 
$\lambda^j$ and putting $\lambda$ to zero gives (\ref{2.2}).}


Note that $\partial^R_{\alpha^i} S_{\rm gf}\lambda^i=-(S_{\rm gf},
\partial_{\alpha^i}\Psi\lambda^i)$, and $\Delta_c\Psi=0$ imply in
 particular that
$\bar s (\partial^R_{\alpha^i}\Psi\lambda^i,
\partial^R_{\alpha^j}\Psi\lambda^j)
=0$, so that both terms of (\ref{2.2}) start
indeed at order $\hbar$.

Because $[\partial^R_{\alpha^i},\partial^R_{\alpha^j}]=0$, 
$[D^\infty_{i},D^\infty_{j}]$ belongs to the space of graded
derivations in the essential couplings alone (with a possible gauge
parameter dependence). Since furthermore, 
$[\Delta^\infty,[D^\infty_{i},D^\infty_{j}]]=0$ due to the graded
Jacobi identity, (\ref{c27}) can now be applied with the
result that 
\beq
[D^\infty_{i},D^\infty_{j}]=[\Delta^\infty,\rho_{ij}^\infty].
\label{incon}
\eeq
In the case of the standard antifield formalism, supposed to be stable
and anomaly free, so that in particular $\Delta^\infty=0$, 
equation (\ref{incon}), is the integrability
condition 
that states that the various redefinitions of the essential couplings 
can be made simultaneously \cite{Barnich:2000yx}. The same holds in the
general case, if one 
can prove that by suitable redefinitions of the $D^\infty_{i}$, the 
coefficients $\rho_{ij}^\infty$ can be made to vanish. In this case, 
the dependence of the effective
action on the all gauge parameters simultaneously is a quantum BRST
coboundary, while the redefined anomaly operator is independent of all
the gauge parameters. 

In the anomaly free stable case, a
quantum BRST coboundary, $(\G^\infty,I\circ \G^\infty)$
reduces to zero upon projection onto the physical
states, so that the gauge parameter dependence of the effective action
expressed in terms of the essential couplings on the quantum level
is also trivial in this physical sense and not only in the 
cohomological sense. 
It would be of interest to analyze (i) in what sense
a quantum BRST coboundary $s^q I\circ G^\infty=(\G^\infty,I\circ
G^\infty)+\Delta^\infty I\circ
\G^\infty$ is physically trivial and (ii) 
what can be concluded from (\ref{incon}) in the general case. 

\newpage

\mysection{Renormalization group and Callan-Symanzik equations}\label{s8}

\subsection{Renormalization group equation}

In addition to the gauge parameters, the effective action, the
local insertions and the right derivations also contain a dependence
on the renormalization scale $\mu$. 

Because $\mu\partial_\mu S_{\rm
  gf}= 0$, the same derivation that has allowed to prove (\ref{e72})
now gives 
\bea
D^\infty\G^\infty=\hbar s^q(C\circ\G^\infty),\label{e81}\\
D^\infty=\mu\partial_\mu+\sum_{n=1}\hbar^n\beta_n,\
[\Delta^\infty,D^\infty]=0. \label{e82}
\eea
The derivations $\beta_n=\frac{\partial^R\cdot}{\partial
  \xi^A}\beta^A_n$ only involve derivatives with respect to the
essential couplings. If in another version of the renormalization
group equation, there appear terms of the form 
$\frac{\partial^R\G^\infty}{\partial \alpha^i}\rho^i_n$,  
they can always be absorbed by using (\ref{e72}) at the expense of a
modification of the beta functions (of at least second order in
$\hbar$) and of the quantum BRST exact term 
$s^q(C\circ\G^\infty)$. This was first noted in the context of Yang-Mills
theory in \cite{Caswell:1974cj}. The same holds for any other
redundant coupling. We will then make the following definition: 

{\em The physical beta functions $\beta^A_n$ of the renormalization
group equation
are the coefficients of the derivatives $\frac{\partial^R\cdot}{\partial
  \xi^A}$ with respect to the essential couplings $\xi^A$ 
in the renormalization group equation where
the derivatives with respect to the redundant couplings have been
eliminated.}

Note that equation (\ref{e82}) can be written as
$\mu\partial_\mu\Delta^\infty=\mu\partial_\mu(\sum_{n=1}\hbar^n \Delta_n)
=-[\Delta^\infty,\sum_{n=1}\hbar^n\beta_n]$. It fixes the dependence
of the anomaly coefficients $(\sum_{n=1}\hbar^{n-1} \Delta_n$ on the
renormalization point $\mu$ to be a quantum BRST coboundary, with the
boundary term determined by the beta functions
$\beta^A=\sum_{n=1}\hbar^{n}\beta^A_n$. 

After integration of the renormalization group equations, i.e., after
the determination of functions $\xi^A_\mu=\xi^A_\mu(\xi^B,\mu)$
defined through the differential equation 
\bea
\mu\frac{d}{d\mu}\xi^A_\mu=\beta^A(\xi^A_\mu,\mu),
 \label{renormcoup},
\eea
equations
(\ref{e81}) and (\ref{e82}) in terms of the running couplings
$\xi^A_\mu$ become
\bea
\mu\frac{d}{d\mu}\G^\infty_\mu=\hbar s^q_\mu(C_\mu\circ\G^\infty_\mu),\\
\mu\frac{d}{d\mu}\Delta^\infty_\mu=0. 
\eea
We have thus shown: 
\begin{theorem}
If the theory is expressed in terms of the running (essential) couplings
$\xi^A_\mu$, the variation of the effective action with respect to the 
renormalization scale is a quantum BRST coboundary, while the
redefined anomaly operator $\Delta^\infty_\mu$
does not depend on the renormalization scale. 
\end{theorem}


\subsection{Gauge parameter dependence of renormalization group 
beta functions}

If one follows \cite{Caswell:1974cj} and
commutes the functional operators
of equations (\ref{e72}) and (\ref{e81}), one gets
\beq
[D^\infty,D^\infty_i]\G^\infty=s^q E_i
\eeq
where
$E_i=D^\infty[L_i\circ\Gamma^\infty]-D^\infty_i[\hbar
C\circ\Gamma^\infty]+(\hbar C\circ\Gamma^\infty,
L_i\circ\Gamma^\infty)
$.

Consider now the parameters $\alpha^{\bar i}=(\alpha^i,\mu)$,
the constants ghosts $\lambda^{\bar i}=(\lambda^i,\Lambda)$ 
and the differentials
$D^\infty_{\bar i}=(D_{i},D)$. It then follows that (\ref{Aex})
holds for the same $S^e$ but with $D^\infty_{i}\lambda^i$ replaced
by $D^\infty_{\bar i}\lambda^{\bar i}$. The proof that $[ 
D^\infty_{\bar i}, D^\infty_{\bar
  j}]=[\Delta^\infty,\sigma^\infty_{\bar i\bar j}]$ 
then proceeds exactly as the proof
of lemma \ref{llem} before
and includes in particular the result 
\beq
[D^\infty,D^\infty_{i}]=[\Delta^\infty,\sigma^\infty_i].
\eeq
Again, in the anomaly free standard formalism, one can deduce 
\cite{Barnich:2000yx} from
this equation (i) that the physical beta functions are gauge parameter
independent if the redefined couplings $\xi^A_\alpha$ have been used,
(ii) that the $\rho^A$ are renormalization scale independent if the
running couplings $\xi^A_\mu$ have been used, and (iii) that the
absorption of the gauge parameter dependence and of the renormalization 
scale can be made simultaneously.


\subsection{Power counting in the antifield formalism}

In order to get the analogous results for the Callan-Symanzik
equation, we have to replace the operator $\mu\partial_\mu$ by the
generator of (broken) dilatation invariance. 
In the antifield formalism, power counting can be implemented canonically
through the operator 
\bea
S_\eta=\int d^nx\ L_\eta=\int d^nx\ \phi^*_a(d^{(a)}+x^\mu\partial_\mu)
\phi^a,
\eea
where $\phi^a$ is a collective notation for the original fields and
the local ghosts associated to the gauge symmetries, while $d^{(a)}$ is
the canonical dimension of $\phi^a$ in units of inverse length. The  
bracket around the index $a$ means that there is no additional summation. 
We have
$(\phi^a(x),S_\eta)=(d^{(a)}+x^\mu\partial_\mu)\phi^a(x)$ 
and
$(\phi^*_a(x),S_\eta)=(n-d^{(a)}+x^\mu\partial_\mu)\phi^*_a(x)$,
so that the canonical dimension of the antifields is chosen to be
$n-d^{(A)}$. It is then straightforward to verify that for any
monomial $M(x)$ in the fields, the antifields and their derivatives of
homogeneous dimension $d^M$,
\bea
(M(x),S_\eta)=(d^M+x^\mu\partial_\mu)M(x).
\eea

\subsection{Callan-Symanzik equation}
\subsubsection{No dimensionful coupling constants}

For simplicity, we assume in a first stage that the 
coupling constants $\xi^A$ as well as the redundant coupling
constants all have dimension $0$.

Because there are non dimensionful parameters, 
all the terms of the Lagrangian 
$L$ of the (gauge fixed) solution of the extended master equation
have dimension $n$. Hence,
\bea
({L},S_\eta)=(n+x^\mu\partial_\mu){L}
=\partial_\mu(x^\mu{L}).\label{4}
\eea
Upon integration, we get 
\bea
(S,S_\eta)=0.
\label{bl}
\eea
Furthermore, 
$\Delta_c S_\eta=0=\Delta^\infty S_\eta$ because 
$S_\eta$ does not depend on $\xi^A$, which means $\bar s S_\eta=0$. 
We have
$(S^\infty,S_\eta)=O(\hbar)=(S^\infty,S_\eta)+{\Delta^\infty}^LS_\eta$,
so that the 
quantum action principle gives
\bea
(\G^\infty,S_\eta)=\hbar{\cal B}\circ\G^\infty=s^q S_\eta,\label{zin}
\eea
with ${\cal B}$ a local functional of ghost number $0$. 

Applying $s^q$, we get the consistency condition 
$s^q ({\cal B}\circ\G^\infty)=0$, 
which implies, to lowest
order in $\hbar$, $\bar s {\cal B}_0=0$ and hence 
\bea
{\cal B}_0
=-\beta_1 S-\bar s\Xi_1, \ [\Delta_c,\beta_1]=0,
\eea
with $ \beta_1=\frac{\partial^R \cdot}{\partial \xi^A}\beta^A_1$. 
According to the quantum action principle, we can replace 
$\beta_1 S\circ\G^\infty$ by 
$\beta_1 \G^\infty$ and the
difference will be the insertion of a local functional of 
order $\hbar$. Applying lemma \ref{00}, the local insertion
$[\bar s \Xi_1]\circ\G^\infty$ can be replaced by  
$s^q ({\Xi_1}^Q\circ\G^\infty)$, and
the difference will again be the insertion of a local functional of 
order $\hbar$.
We thus get 
\bea
s^q [S_\eta+\hbar\Xi^Q_1\circ\G^\infty]
+\hbar\beta_1\G^\infty
=\hbar^2{\cal B}^\prime\circ\G^\infty.\label{5}
 \eea
Acting with  $s^q$ on
(\ref{5}), using $[\Delta_c,\beta_1]=0$, we get the consistency
condition $\hbar[\Delta^\infty\beta_1]\G^\infty+\hbar^2
s^q({\cal B}^\prime\circ\G^\infty)=0$, giving to lowest order
$[\Delta_1,\beta_1]S+\bar s {\cal B}_0^\prime=0$. As in the previous
section, this means that the reasoning can be
pushed to all orders:
\bea
s^q [S_\eta-\hbar \Xi^\infty \circ\G^\infty]
+\hbar \beta^\infty \G^\infty=0,\label{68}
\eea
\bea
[\Delta^\infty,\beta^\infty]=0\label{69}.
\eea
with $\Xi^\infty =\Sigma_{n=1}\hbar^{n-1}\Xi^Q_n$ and
$\beta^\infty=\Sigma_{n=1}\hbar^{n-1}\beta_n$. 
Note that if $\beta^\infty=[\Delta^\infty,\sigma^\infty]$, the term
$\beta^\infty \G^\infty$ in (\ref{68}) can be absorbed by the
redefinition $\Xi^\infty \circ\G^\infty\longrightarrow 
\Xi^\infty\circ\G^\infty-\sigma^\infty \G^\infty$. We then get from
(\ref{69}):  
\begin{theorem}\label{t11}
The right derivation built out of the beta functions
of the Callan-Symanzik equation defines a 
non trivial quantum BRST cocycle in ghost number $0$. 
\end{theorem}

\subsubsection{Digression}
Let us consider for a moment the following particular 
case. 

(i) All the antibracket maps encoded in $f^A$ are zero so
that $\Delta_c=0$, $\bar s =s=(S(\xi),\cdot)$. 
This happens for instance if one couples only the BRST cohomological
classes in ghost number $0$ and if the Kluberg-Stern and
Zuber conjecture \cite{Kluberg-Stern:1975hc}, stating that these cohomological
classes can be described independently of the antifields, is valid. This
guarantees stability of the standard antifield formalism.  

(ii) The theory is anomaly free,
$\frac{1}{2}(\Gamma^\infty,\Gamma^\infty)=0$. 

(iii) The only possibility (for instance for power counting reasons)
for $\Xi_n$ is $\Xi_n=-\gamma_n\int
d^nx\ \phi^*_a\phi^a$, so that $\Xi_n$ is linear in the quantum fields 
and $[s \Xi_n]\circ\ \G^\infty$ can be replaced, according to the
quantum action principle, at each stage in $\hbar$ 
by $(\G^\infty,\Xi_n)$ up to the insertion of a local polynomial 
of higher order in $\hbar$.

Equation (\ref{68}) then reduces to 
\bea
(\G^\infty,S^\infty_\eta)+\hbar
{\partial^R\G^\infty\over\partial \xi^A}\beta^A=0,
\eea
with $S^\infty_\eta=\int d^nx\ \phi^*_a(d^{(a)}+\hbar \gamma
+x\cdot\partial)\phi^a$,
or explicitly 
\bea
\int d^nx\Big[\ {\delta^R \G^\infty\over\delta \phi^a(x)}(d^{(a)}
+\hbar\gamma+
x\cdot\partial)\phi^a(x)\nonumber\\+{\delta^R 
\G^\infty\over\delta \phi^*_a(x)}
(n-d^{(a)}-\hbar \gamma+x\cdot\partial)\phi^*_a(x)\Big]
+\hbar{\partial^R\G^\infty\over\partial \xi^A}\beta^A=0.\label{67}
\eea
After putting to zero the antifields, the second part of the first
integral vanishes and this equation is a familiar 
form of the Callan-Symanzik equation (see eg. \cite{ItZu}) 
in the massless case, 
with anomalous dimension $\gamma
=\Sigma_{k=1}\hbar^{k-1}\gamma_k$
for the fields.

\subsubsection{Remarks on explicit $x$ dependence.}

Note that $S_\eta$ is the generator of the dilatation symmetry 
of the theory. If it corresponds to a non trivial element
of $H^{-1,n}(s|d)$, the question arises whether it should be coupled with a
constant ghost in the extended solution $S(\xi)$ as in 
\cite{Tonin:1992wf,White:1992pa}. 
This depends on
whether or not we allow for explicit $x$ dependence in the local
functionals and the cohomology
classes of $s$ we are initially computing and then coupling 
to the solution of the master equation. 

In the previous section, we have supposed that
there is no explicit $x$ dependence in these functionals and  
cohomology classes, because if we 
assume the absence of dimensionful couplings, we cannot control
translation invariance through a corresponding cohomology class, its
generator $S_\mu=\int d^nx\ \phi^*\partial_\mu\phi$ being of dimension
$1$. 

We will assume here that one can apply the 
quantum action principles in the case of an explicit $x$ dependence of
the variation as in (\ref{zin}), at the price of allowing a priori for an 
explicit $x$ dependence of the inserted local functional ${\cal B}$. 
This assumption needs to be
checked by a more careful analysis of the renormalization properties
of the model which is beyond the scope of this review.

In order to prove 
then that ${\cal B}$ in eq (\ref{zin}) does 
not depend explicitly on $x$, we use 
translation invariance: classical translation invariance 
is expressed through $(S,S_\mu) =0$ with quantum version $(\G^\infty,S_\mu)
=\hbar{\cal D}_\mu\circ\G^\infty$, where the dimension of ${\cal D}_\mu$
is $1$, because there are no dimensionful parameters in the theory. 
Applying $(\cdot,S_\mu)$ to (\ref{zin}), using the graded Jacobi identity 
for the antibracket, the commutation relation $(S_\eta,S_\mu)=-S_\mu$ 
and the result on the dimension of ${\cal D}_\mu$, i.e., the relation
$({\cal D}_\mu,S_\eta)={\cal D}_\mu$, one finds to lowest order 
$({\cal B},S_\mu)=0$. This means that $(\partial_\mu-\partial/\partial x^\mu)
{\cal B}=0$, and since $\partial_\mu {\cal B}=0$, it shows 
that ${\cal B}$ does not depend explicitly on $x$.

In the general case where we allow for dimensionful couplings
considered below, we will assume that the theory is translation
invariant and that the generator $S_\mu$ is coupled through the
constant translation ghosts $\xi^\mu$. One can then show that the
local cohomology of the BRST operator in form degree $n$ 
for the extended theory can be chosen
to be independent of both $x^\mu$ and $\xi^\mu$ \cite{White:1992wu}.
In the same way, Lorentz invariance can then be controlled 
inside the formalism 
by coupling the appropriate generator. 

\subsubsection{General broken case}

We will now allow for coupling constants $\xi^A,\alpha^i$ of 
all possible dimensions $d^{(A)},d^{(i)}$ in the theory, 
which could be negative in the case of effective
field theories.
We have 
\bea
(L,S_\eta)+{\partial^R L\over\partial \xi^A}
d^{(A)}\xi^A+{\partial^R L\over\partial \alpha^i}
d^{(i)}\alpha^i=\partial_\mu (x^\mu L).\label{pc}
\eea
Integrating, one gets 
\bea
{\cal C}S=0,\label{77}
\eea
with ${\cal C}=(\cdot,S_\eta)+{\partial^R \cdot\over\partial \xi^A}
d^{(A)}\xi^A+{\partial^R \cdot\over\partial \alpha^i}
d^{(i)}\alpha^i$.


Using
(\ref{gaugedep}) and  $\D_c S_\eta=0=\Delta_c \partial^R_{\alpha_i}
\Psi$, (\ref{77}) becomes
\bea
\bar s (S_\eta-\partial^R_{\alpha_i}
\Psi d^{(i)}\alpha^i)
+\frac{\partial^R S}{\partial
\xi^A}d^{(A)}\xi^A=0.\label{8.21}
\eea 
Applying $\bar s$, we find $[\Delta_c,\beta_0]=0$, with $\beta_0=
\frac{\partial^R\cdot}{\partial
\xi^A}d^{(A)}\xi^A$.
%
The quantum version of this equation is 
\bea
s^q[S_\eta-(\partial^R_{\alpha_i}
\Psi)^Q d^{(i)}\alpha^i\circ \G^\infty]+\beta_0\G^\infty
=\hbar{\cal B}\circ\G^\infty.
\eea
Applying now $s^q$, the consistency condition to
lowest order implies $\bar s{\cal B}_0=0$. We can then get as in the
previous section the general form of the integrated 
Callan-Symanzik equation:
\bea
s^q[S_\eta-\Xi\circ\G^\infty]+\beta^\infty\G^\infty=0,\label{qcs}
\eea
\bea
[\Delta^\infty,\beta^\infty]=0,
\eea
with $\Xi=-\partial^R_{\alpha_i}
\Psi d^{(i)}\alpha^i+O(\hbar)$ and
$\beta^\infty=\Sigma_{n=0}\hbar^n\beta_n$. Hence, by defining 
$d^{(A)}\xi^A$ to be the tree level contribution of the beta function,
theorem \ref{t11} extends to the general case. 

\newpage

\mysection{Refined anomaly consistency condition}\label{s9}

We have seen in section \ref{s6} that the non trivial breakings of the
extended master equation are constrained by the cohomology of $\bar s$
in ghost number $1$ in the space of local functionals. 
In the same way,
one can study the constraints on the
anomalies to an invariant renormalization of symmetric integrated and non
integrated operators \cite{Dixon:1980zp}. 
In the cohomological reformulation, symmetric
integrated or non integrated operators correspond to BRST cohomological
classes in ghost number $0$ in the space of local functionals,
respectively in the space of local functions. More generally, one can
be interested in the anomalies appearing during the renormalization of
a BRST cohomological class in ghost number $g$ in these spaces. The
non trivial anomalies that can appear can be
shown to belong to the corresponding BRST cohomological classes in 
ghost number $g+1$. Some aspects of the renormalization of integrated
BRST cohomological classes in the extended antifield
formalism have already been discussed in section
\ref{s623}. 
 
The computation in the space of local functionals 
of the cohomology of the standard BRST differential $s$ with antifields 
can be reduced to the computation of
a relative cohomological group in the space of local $n$-forms by
introducing the space-time exterior derivative $d$. It is then related to
the cohomology of $s$ in the space of form 
valued local functions through descent equations
\cite{Stora:1976kd,Stora:1983ct,Zumino:1983ew,Zumino:1984rz}. The same 
holds for the BRST differential $\bar s$ associated to the 
extended antifield formalism. 

More generally, we will call the relative cohomological groups
$H^{g,p}(s|d)$ and $H^{g,p}(\bar s|d)$ in ghost number $g$ and 
form degree $p$ local BRST
cohomological groups.
A more detailed analysis 
of the descent equations \cite{Dubois-Violette:1985jb} shows 
that these groups
are characterized by two integers, the length $d$ of their descents and the 
length $l$ of their lifts. 

\subsection{Characterization of local BRST cohomological classes}
\label{char}

In this and the following section, we decompose the space $\Omega$ of
Lorentz-invariant polynomials or formal powers series in the $dx^\mu$,
the couplings $\xi^A$, the fields, antifields and their derivatives
into the direct sum of the constants and the remaining part,
$\Omega={\bf R}\oplus \Omega_+$. We have $\bar s\alpha=0=d\alpha$,
for a constant $\alpha$ and
$d\Omega_+ \subset
\Omega_+$. We furthermore assume that if $\bar s \omega=\alpha$ 
for a constant $\alpha$, then $\alpha=0$, which amounts to assuming
that the equations of motions are consistent (see the discussion in
chapter 9 of \cite{Barnich:2000zw}).  This means that 
$\bar s \Omega_+ \subset \Omega_+$. Hence, we can consider
the cohomological groups $H(\bar s,\Omega_+)$, $H(\bar
s|d,\Omega_+)$ and $H(d,\Omega_+)$.

By analyzing the cohomological groups $H(\bar s|d)$ (the space
$\Omega_+$ being always understood in the following) using descent 
equations, one can prove \cite{Dubois-Violette:1985jb}
that the elements
of these groups can be classified into chains of length $r$ with an
obstruction to further lifts and chains of length $s$ 
whose lifts are unobstructed,
i.e., chains with a non trivial element in degree $n$ 
(see also \cite{Talon:1985dz,Brandt,Henneaux:1999rp};
we follow here the notations of the review \cite{Barnich:2000zw}, 
where explicit proofs of the statements below can be found).
More precisely, we have 
$H^p(d)=0$, $p\leq n-1$, and there exists a basis 
\begin{eqnarray} 
\{[h^0_{i_r}],[\hat
h_{i_r}],[e^0_{\alpha_s}]\}\label{bass} 
\end{eqnarray} 
of $H(\bar s)$, for
$r=0,\dots,n-1$, $s=0,\dots,n$, such that a corresponding basis of
$H(\bar s|d)$ is
given by 
\begin{eqnarray}
\{[h^q_{i_r}],[e^p_{\alpha_s}]\}\label{bassd}
\end{eqnarray} 
for $q=0,\dots r$ and $p=0,\dots,s$, with 
\begin{eqnarray} 
\begin{array}{c}
\bar s h^{r+1}_{i_r}+d h^{r}_{i_r}=\hat h_{i_r},\\ 
\bar s h^r_{i_r}+ d
h^{r-1}_{i_r}=0,\\ 
{\vdots}\label{prop1}\\ 
\bar s h^1_{i_r}+ dh^0_{i_r}=0,\\ 
\bar s h^0_{i_r}=0, 
\end{array}
\end{eqnarray} 
and 
\begin{eqnarray} 
\begin{array}{c}
{\rm form\ degree}\ e^s_{\alpha_s}=n,\ de^s_{\alpha_s}=0, \\ 
\bar s e^{s}_{\alpha_s}+ d e^{s-1}_{\alpha_s} =0,\\ 
{\vdots}\label{prop2}\\
\bar s e^1_{\alpha_s}+ d e^0_{\alpha_s}=0,\\ 
\bar s
e^0_{\alpha_s}=0. 
\end{array}
\end{eqnarray} 
The cohomological group $H(\bar s)$ can thus be decomposed into elements 
$[e^0_{\alpha_s}]$ that are bottoms of unobstructed chains of length
$s$,
elements $[h^0_{i_r}]$ that are 
bottoms of obstructed chains of length $r$ and obstructions $[\hat
h_{i_r}]$ to chains of length $r$.
  
For the cohomological group $H(\bar s|d)$, 
the element $[h^l_{i_r}]$ is said to be the element of level $l$ of a
chain of length $r$ with obstruction; it has $l$ non trivial descents and 
$r-l$ non trivial lifts~; while the element
$[e^l_{\alpha_s}]$ is said to be the element of level $l$ of a chain of
length $s$ without obstructions, it has $l$ non trivial descents and
$s-l$ non trivial lifts.  

One can furthermore show that 
the general solution to a set of descent equations involving 
at most $l$ steps, 
$\bar s\omega^l+d\omega^{l-1}=0,$
$\bar s\omega^{l-1}+d\omega^{l-2}=0,\dots,$ $\bar s\omega^0=0$, 
can be written in terms of the above elements as   
\bea
\omega^l=
\sum_{q=0}^l\sum_{r=l-q}^{n-1} \lambda^{i_r}_{q}h^{l-q}_{i_r}
+\sum_{p=0}^l\sum_{s=l-p}^n \mu^{\alpha_s}_{p}e^{l-p}_{\alpha_s}
+\bar s\eta^l +d[\eta^{l-1}+\sum_{r=0}^{n-1}\nu^{(l)i_r}h^r_{i_r}],
\label{lenl} 
\eea
with $\eta^{-1}=0$. This means that a $\bar s$ modulo $d$ cocycle
which has $l$ non trivial descents, is a linear combination
of all elements of the chains (\ref{prop1}) and (\ref{prop2}) which have 
$l$ or less non trivial descents. 
If such a linear combination is $\bar s$ modulo $d$ trivial,
the coefficients of the linear combination must vanish, i.e.,   
\bea
\sum_{q=0}^{l}\sum_{r=l-q}^{n-1}
\lambda^{i_r}_{q}h^{l-q}_{i_r} +\sum_{p=0}^l\sum_{s= l-p}^n
\mu^{\alpha_s}_{p}e^{l-p}_{\alpha_s}=\bar s(\ )+d(\ ).
\label{cotr} 
\eea
implies that $\lambda^{i_r}_{q}=0=\mu^{\alpha_s}_{p}$.

\subsection{Lengths of descent and lifts of lowest order anomalies}
\label{sec4}
Let us now investigate the anomalies in the BRST invariant 
renormalization of a chain (\ref{prop2}) of length $s$ without 
obstructions. We follow the approach of 
\cite{Lucchesi:1987sp}, which consists in considering simultaneously the
anomalies for a whole chain of descent equations
(for a review, see \cite{Piguet:1995er}). 
Using the quantum action principles, one can show the analog of lemma
\ref{00} for a local form $a$ : 
\begin{eqnarray}
\bar s a\circ\G^\infty
=s^q a^Q\circ\G^\infty+\hbar b\circ\G^\infty,
\end{eqnarray} 
for some local form $b$.
When applied to the chain (\ref{prop2}), we find that 
the quantum version of this chain is 
\begin{eqnarray}
\begin{array}{c}
{\rm form\ degree}\ e^s_{\alpha_s}\circ\G^\infty =n,\ 
de^s_{\alpha_s}\circ\G^\infty =0,\\ 
s^q e^{s}_{\alpha_s}\circ\G^\infty+ d e^{s-1}_{\alpha_s}\circ\G^\infty  =
\hbar a^s_{\alpha_s}\circ\G^\infty,\\ 
{\vdots}\label{q1}\\
s^q e^1_{\alpha_s}\circ\G^\infty+ d e^0_{\alpha_s}\circ\G^\infty  =
\hbar a^1_{\alpha_s}\circ\G^\infty,\\ 
s^q e^0_{\alpha_s}\circ\G^\infty  =\hbar a^0_{\alpha_s}\circ\G^\infty
, 
\end{array}
\end{eqnarray} 
where $a^l_{\alpha_s}\circ\G^\infty=a^s_{\alpha_s}+O(\hbar)$ for a local 
function $a^l_{\alpha_s}$. 
Applying $s^q$, we get the consistency condition,
\begin{eqnarray} 
\begin{array}{c}
{\rm form\ degree}\ a^s_{\alpha_s}\circ\G^\infty =n,\ 
da^s_{\alpha_s}\circ\G^\infty=0 \\ 
s^q a^{s}_{\alpha_s}\circ\G^\infty+ d a^{s-1}_{\alpha_s}\circ\G^\infty =
0,\\ 
{\vdots}
\\
s^q a^1_{\alpha_s}\circ\G^\infty+ d a^0_{\alpha_s}\circ\G^\infty=
0,\\ 
s^q a^0_{\alpha_s}\circ\G^\infty=0.
\end{array}
\end{eqnarray} 
At lowest order in $\hbar$, we get 
\begin{eqnarray} 
\begin{array}{c}
{\rm form\ degree}\ a^s_{\alpha_s}=n,\ 
da^s_{\alpha_s}=0 \\ 
\bar s a^{s}_{\alpha_s}+ d a^{s-1}_{\alpha_s} =
0,\\ 
{\vdots}\label{locon1}\\
\bar s a^1_{\alpha_s}+ d a^0_{\alpha_s}=
0,\\ 
\bar s a^0_{\alpha_s}=0
.
\end{array} 
\end{eqnarray}
Using equation (\ref{lenl}) and the fact that the form
degree of $a^s_{\alpha_s}$ is $n$, it follows that 
\bea
a^l_{\alpha_s}=
\sum_{p=0}^l
{\mu^{\beta_{s-p}}_{p}}_{\alpha_s}
e^{l-p}_{\beta_{s-p}}
+\bar s\eta^l_{\alpha_s} +d[\eta^{l-1}_{\alpha_s}+
\sum_{r\geq 0}\nu^{(l)i_r}_{\alpha_s}h^r_{i_r}],\label{res1}
\eea
for $l=0,\dots,s$.
This gives our first result:

{\em The anomaly in the renormalization of an element of level $l$ 
of a chain of length $s$ without obstructions 
involves at most elements of chains of the same type with less non
trivial descents and the same number of non trivial lifts.}

For the anomalies for a chain with obstruction (\ref{prop1}), we get, 
\begin{eqnarray} 
\begin{array}{c}
s^q h^{r+1}_{i_r}\circ\G^\infty+ 
dh^r_{i_r}\circ\G^\infty 
=\hat h_{i_r}\circ\G^\infty+\hbar a^{r+1}_{i_r}\circ\G^\infty,\\ 
s^q h^{r}_{i_r}\circ\G^\infty+ d h^{r-1}_{i_r}\circ\G^\infty  =
\hbar a^r_{i_r}\circ\G^\infty,\\ 
{\vdots} \label{q2}\\
s^q h^1_{i_r}\circ\G^\infty+ d h^0_{i_r}\circ\G^\infty  =
\hbar a^1_{i_r}\circ\G^\infty,\\ 
s^q h^0_{i_r}\circ\G^\infty  =\hbar a^0_{i_r}\circ\G^\infty
, 
\end{array}
\end{eqnarray} 
We also have 
\bea
s^q \hat h_{i_r}\circ\G^\infty=-\hbar
\hat a_{i_r}\circ\G^\infty.\label{qo}
\eea
Applying $s^q$ gives 
$s^q \hat a_{i_r}\circ\G^\infty=0$ and then to lowest order, $\bar s
\hat a_{i_r}=0$. Applying now $s^q$ to the chain (\ref{q2}) gives
\bea
\begin{array}{c}
s^q a^{r+1}_{i_r}\circ\G^\infty+ da^r_{i_r}\circ\G^\infty 
=\hat a_{i_r}\circ\G^\infty,\\ 
s^q a^{r}_{i_r}\circ\G^\infty+ d a^{r-1}_{i_r}\circ\G^\infty  =
0,\\ 
{\vdots} 
\\
s^q a^1_{i_r}\circ\G^\infty+ d a^1_{i_r}\circ\G^\infty  =
0,\\ 
s^q  a^0_{i_r}\circ\G^\infty  =0
,
\end{array} 
\eea
and to lowest order,
\bea
\begin{array}{c}
\bar s a^{r+1}_{i_r}+ da^r_{i_r}
=\hat a_{i_r},\\ 
\bar s a^{r}_{i_r}+ d a^{r-1}_{i_r}  =
0,\\ 
{\vdots} \label{locon2}\\
\bar s a^1_{i_r}+ d a^1_{i_r}  =
0,\\ 
\bar s  a^0_{i_r}  =0
,
\end{array} 
\eea
On the one hand, it follows from (\ref{lenl}) that 
\bea
a^l_{i_r}=\sum_{q=0}^l\sum_{r^\prime= r-q}^{n-1} 
{\lambda^{j_{r^\prime}}_{q}}_{i_{r}}h^{l-q}_{j_{r^\prime}}
+\sum_{p=0}^l\sum_{s = r-p+1}^n
{\mu^{\beta_{s}}_{p}}_{i_r}
e^{l-p}_{\beta_{s}}
+\bar s\eta^l_{i_r}+d[\eta^{l-1}_{i_r}+
\sum_{r^\prime= 0}^{n-1}\nu^{(l)j_{r^\prime}}_{i_r}
h^{r^\prime}_{j_{r^\prime}}],\label{res2}
\eea
for $l=0,\dots,r$,
while on the other hand, the cohomology of $\bar s$ implies
\bea
\hat a_{i_r}=\sum_{r^\prime= 0}^{n-1}\alpha^{j_{r^\prime}}_{i_r}
\hat h_{j_{r^\prime}}+\sum_{r^\prime= 0}^{n-1}\beta^{j_{r^\prime}}_{i_r}
h^0_{i_r}+\sum_{s=0}^n\gamma^{\alpha_s}_{i_r}e^0_{\alpha_s}+\bar s
\hat o_{i_r}.\label{4.10}
\eea
Applying $d$ to (\ref{res2}) at $l=r$ gives
\bea
d a^r_{i_r}=-\sum_{q=0}^{r}\sum_{r^\prime=r-q+1}^{n-1}
{\lambda^{i_r^\prime}_{q}}_{i_{r}}\bar s h^{r-q+1}_{i_{r^\prime}}
-\sum_{p=0}^{r}\sum_{s=r-p+1}^n 
{\mu^{\beta_{s}}_{p}}_{i_r}
\bar s e^{r-p+1}_{\beta_{s}}
-\bar sd \eta^r_{i_r}\nonumber\\+
\sum_{q=0}^{r}{\lambda^{i_{r-q}}_{q}}_{i_{r}}(
-\bar s h^{r-q+1}_{i_{r-q}}+\hat h_{i_{r-q}}).\label{4.11}
\eea
Injecting now (\ref{4.10}) and (\ref{4.11}) into the first equation of
(\ref{locon2})
gives first of all
$\beta^{j_{r^\prime}}_{i_r}=0=\gamma^{\alpha_s}_{i_r}$ and also
\bea
\hat a_{i_r}=\sum_{q=0}^r{\lambda^{j_{r-q}}_{q}}_{i_r}
\hat h_{j_{r-q}}+\bar s
\hat o_{i_r},\label{res0}
\eea
and then,
\bea
\bar s 
(a^{r+1}_{i_r}-\sum_{q=0}^{r}\sum_{r^\prime=r-q+1}^{n-1} 
{\lambda^{i_{r^\prime}}_{q}}_{i_{r}}h^{r-q+1}_{i_{r^\prime}}
-\sum_{p=0}^{r}\sum_{s=r-p+1}^n  
{\mu^{\beta_{s}}_{p}}_{i_r}
e^{r-p+1}_{\beta_{s}}\nonumber\\
-\sum_{q=0}^{r}{\lambda^{i_{r-q}}_{q}}_{i_r}
h^{r-q+1}_{i_{r-q}}-d \eta^r_{i_r}-\hat o_{i_r})=0,
\eea
so that, using the cohomology of $\bar s$,
\bea
a^{r+1}_{i_r}=\sum_{q=0}^{r+1}\sum_{r^\prime=r-q}^{n-1} 
{\lambda^{i_{r^\prime}}_{q}}_{i_r}h^{r-q+1}_{i_{r^\prime}}
+\sum_{p=0}^{r+1}\sum_{s=r-p+1}^n 
{\mu^{\beta_{s}}_{p}}_{i_r}
e^{r-p+1}_{\beta_{s}}\nonumber\\
+d [\eta^r_{i_r}+\sum_{r^\prime=0}^{n-1}
\nu^{(r+1)j_{r^\prime}}_{i_r}h^{r^\prime}_{j_{r^\prime}}]
+\hat o_{i_r}+\bar s
\eta^{r+1}_{i_r}.\label{c5}
\eea
Our second result is then: 

{\em The anomaly in the renormalization of an element of level $l$ 
of a chain of length $r$ with obstructions 
involves at most elements of chains with obstructions with less non
trivial descents and more non trivial lifts and elements of 
chains without obstructions with less non trivial descents and 
strictly more non trivial lifts.}

Let us now rewrite (\ref{res1}) and (\ref{res2}) at $l$=0 as
\bea
a^0_{\alpha_s}=
{\mu^{\beta_{s}}_{0}}_{\alpha_s}
e^{0}_{\beta_{s}}
+\bar s[\eta^0_{\alpha_s} -\sum_{r= 0}^{n-1}\nu^{(0)i_r}_{\alpha_s}
h^{r+1}_{i_r}]+
\sum_{r= 0}^{n-1}\nu^{(0)i_r}_{\alpha_s}\hat h_{i_r},\label{res7}\\
a^0_{i_r}=\sum_{r^\prime= r}^{n-1}
{\lambda^{j_{r^\prime}}_{0}}_{i_{r}}h^{0}_{j_{r^\prime}}
+\sum_{s = r+1}^{n}
{\mu^{\beta_{s}}_{0}}_{i_r}
e^{0}_{\beta_{s}}
+\bar s[\eta^0_{i_r}-\sum_{r^\prime= 0}^{n-1}\nu^{(0)j_{r^\prime}}_{i_r}
h^{r^\prime+1}_{j_{r^\prime}}] +
\sum_{r^\prime=0}^{n-1}\nu^{(0)j_{r^\prime}}_{i_r}
\hat h_{j_{r^\prime}}.\label{res8}
\eea
Combined with (\ref{res0}), our third result on anomalies 
in the renormalization of elements of $H(\bar s)$ is accordingly:

{\em The anomaly in the renormalization of obstructions to chains of
  length $r$ 
  involves at most obstructions to shorter chains; the anomaly in 
  the renormalization of bottoms of unobstructed chains of length $s$
  involves at most bottoms of unobstructed chains of the same length
  and obstructions to chains of all possible lengths; the anomaly in
  the renormalization of bottoms of obstructed chains of length $r$
  involves at most bottoms of obstructed chains of 
greater length, bottoms of unobstructed chains of strictly 
greater length and obstructions to chains of all possible lengths.}

\subsection{Differentials associated to one loop anomalies}
\label{alt}

A different, more compact, way to formulate and prove
the results of section \ref{char} and \ref{sec4} is to use the 
exact couple describing the descent 
and the associated spectral sequence \cite{Dubois-Violette:1985jb}. 

Indeed, the diagram  
\begin{eqnarray}
\begin{array}{ccc}
H(\bar s|d )\stackrel{{\cal D}}{\longrightarrow}
H(\bar s|d)\\
i_0\nwarrow\ \swarrow l_0\\
H(\bar s)
\end{array}\label{ec}
\end{eqnarray}
can be shown to be exact at all corners. 
The various maps are defined as follows:  $i_0$ is the map which
consists in regarding an element of $H(\bar s)$ as an element of
$H(\bar s|d)$,  
$i_0:H(\bar s)\longrightarrow
H(\bar s|d)$, with $i_0[a]=[a]$. It is well defined  
because every $\bar s$ cocycle is a 
$\bar s$ cocycle
modulo $d$ and every $\bar s$ coboundary is a $\bar s$ 
coboundary modulo $d$. 
The descent homomorphism 
${\cal D}:
H^{k,l}(\bar s|d)\longrightarrow
H^{k+1,l-1}(\bar s|d)$ with 
${\cal D}[a]=[b]$, if $\bar s a + db=0$ is 
well defined because of the triviality of the cohomology of $d$ in
form degree $p\leq n-1$.   
Finally, the map $l_0:H^{k+1,l-1}(\bar s|d)\longrightarrow
H^{k+1,l}(\bar s)$ is defined by $l_0[a]=[d a]$. It is well defined
because the relation $\{\bar s ,d\}=0$ implies that
it maps
cocycles to cocycles and coboundaries to coboundaries. 

To such an exact
couple $(H(\bar s|d),K_0=H(\bar s))$, 
one can associate in a standard
way derived exact couples $({\cal D}^rH(\bar s|d), K_r)$,
\begin{eqnarray}
\begin{array}{ccc}
{\cal D}^r H(\bar s|d )\stackrel{{\cal D}}{\longrightarrow}
{\cal D}^r H(\bar s|d)\\
i_r\nwarrow\ \swarrow l_r\\
K_r,
\end{array}\label{ecr}
\end{eqnarray}
and a spectral sequence $K_{r+1}=H(d_r,K_r)$, with 
$K_0\equiv H(\bar s)$. The maps of these exact couples are
defined recursively as follows: the map $d_{r-1}=l_{r-1}\circ i_{r-1}$ 
can be shown to be a differential, the map 
$i_r$ is the map induced by $i_{r-1}$ in $K_r$, while $l_r {\cal D} 
[a]=l_{r-1}[a]$.

Explicitly, 
the differential $d_0: K_0\longrightarrow K_0$ is defined by
$d_0[a]=[da]$, where $\bar s a=0$.  
An element $k_r\in K_r$ is
identified with the equivalence class $[a]_r$ of an element $[a]\in 
H(\bar s)$, 
where
$[a]\sim_r [a^\prime]$ if $[a]-[a^\prime]\in \oplus_{q=0}^{r-1}{\rm
  im}
\ d_q$. 
The relations $d_q k_{r}=0$, $q=0,\dots,r-1$ mean
that $k_r$ is a bottom that can be lifted at least $r$ times, i.e., 
there exist
$c_1,\dots,c_{r+1}$ such that $\bar s a=0, da+\bar s c_1=0,\dots, 
dc_{r-1}+\bar s c_{r}=0$. Then, the differential $d_r$ is defined by 
$d_r k_r=[dc_r]_r$.

Because there are no forms
of form degree higher than $n$, ${\cal D}^{n+1} H(\bar
s|d)=0$ and $d_n\equiv 0$ so that the construction stops at $r=n$. 

The space of local forms $\Omega$ is decomposed as 
$\Omega=E_0\oplus G\oplus \bar s G \oplus {\bf R}$, with  
$E_0\simeq K_0=H(\bar s)$. 
If we define $E_r,F_{r-1}\subset E_{r-1}$ through 
$E_{r-1}=E_{r-1}\oplus E_r\oplus 
d_{r-1} F_{r-1}$ with $E_r\simeq K_r$, we get the decomposition
\bea
E_0=F_0\oplus \dots\oplus F_{n-1}\oplus E_n\oplus
d_{r-1} F_{r-1}\oplus\dots\oplus d_0 F_0. \label{decomp}
\eea
The $e^0_{\alpha_s}$ are elements of a basis of $E_n$
that can be lifted $s$ times before hitting form degree $n$, i.e., 
that are of form degree $n-s$, while 
$\hat h_{i_r}$ and $h^0_{i_r}$ are elements of a basis
of $d_r F_r $ and $F_r$ respectively. This sums up the results of 
section \ref{char}. 

Let us now define the linear map 
\bea
\delta_0: H^g(\bar s)\longrightarrow
H^{g+1}(\bar s),\\
\delta_0[a]=[b],\ \mbox{where} 
\ s^q a\circ
\G^\infty=\hbar b\circ\G^\infty. 
\eea
The map associates to a given BRST
cohomological class the non trivial order $\hbar$, i.e., 1 loop
contribution of its anomaly.

The map is well defined, because the 
consistency condition implies that $\bar s b=0$. Furthermore, 
if $a=\bar s c$,
$a\circ \G^\infty =s^q c\circ \G^\infty+\hbar d\circ \G^\infty$, so that 
$s^q a\circ \G^\infty=\hbar s^q d\circ \G^\infty$, meaning that $b=\bar
s d$. This implies that the map does not depend on the choice of the
representative. In addition, this map is a differential
\bea
\delta_0^2=0. 
\eea 
Indeed, if $[a]=\delta_0[c]$, we have $a\circ\G^\infty 
=\frac{1}{\hbar}s^q c\circ \G^\infty$. It follows that $s^q
a\circ\G^\infty=0$. 
A BRST cohomological class which is a  $\delta_0$-cocycle has no
1-loop anomaly, while a BRST cohomological class which is a 
 $\delta_0$-coboundary is the 1-loop anomaly of some other 
BRST cohomological class. 

We thus have two differentials in $K_0=H(\bar
s)$, $d_0$ introduced above and $\delta_0$. These 
differentials anticommute, 
\bea
\{d_0,\delta_0\}=0. 
\eea
Indeed, if $s^q a\circ
\G^\infty=\hbar b\circ\G^\infty$, $d_0\delta_0 [a]=d_0[b]=[db]$, while 
$\delta_0 d_0[a]=\delta_0 [da]$, and $s^q(da\circ\G^\infty)=s^q 
d(a\circ \G^\infty +\hbar c\circ\G^\infty)=-d s^q (a
\circ \G^\infty +\hbar c\circ\G^\infty)=-\hbar d(b\circ \G^\infty+
s^q c\circ\G^\infty)$, so that $\delta_0 [da]=[-d(b+\bar s
c)]=-[db]$. 

The relation $d_0 \delta_0 [a]= -\delta_0 d_0[a]$ means:
\begin{itemize}
\item if $[a]$ belongs to ${\rm im}\ d_0$, $[a]=d_0[b]$, then  
$\delta_0[a]=-d_0\delta_0[b]$,
i.e., if $[a]$ represents an obstruction to the lift of an element
$[b]$, its anomaly represents minus the obstruction to the lift of the
anomaly of $[b]$, 
\item if $d_0 [a]=0$, then $d_0\delta_0[a]=0$, i.e., if $[a]$ can be
  lifted, then so does its anomaly $\delta_0[a]$,
\item the
  anomaly in a bottom $[a]$ of $K_0$ that cannot be lifted is minus 
the anomaly of the corresponding obstruction, up to elements that can
  be lifted.
\end{itemize}
If we organize the space $E_0\simeq K_0=H(\bar s)$ as $E_0=F_0\oplus 
E_1\oplus d_0 F_0$, with $E_1\simeq K_1$, 
we have shown that 1-loop contribution
to the anomaly of an element in one of these subspaces belongs to the
same subspace or to a subspace that stands to the right. Together with
the last point of the previous list, this sums up the results for the 
elements of $H(\bar s)$, i.e., for the obstructions and the 
bottoms contained in (\ref{res0}), (\ref{res7}) and (\ref{res8}) 
restricted to $r=0$. 

In the same way, these results for all $r$ and $s$ 
follow from the fact that $\delta_0$ induces a well-defined
differential (also called $\delta_0$ in the following) in the spaces
$K_r$, anticommuting with $d_r$, $\delta_0:
K_r\longrightarrow K_r$ with $\{\delta_0,d_r\}=0$. 

Indeed, suppose
this result to be true for $K_0,\dots, K_{r-1}$, $d_0,\dots,d_{r-1}$. 
An element $[a]_r\in K_r$ 
satisfies $\bar s a=0$, $da +\bar s c_1=0$, $\dots,$ $dc_{r-1}+\bar s
c_r=0$. This implies $s^q a\circ\G^\infty=\hbar b\circ\G^\infty$, 
$d a\circ\G^\infty +s^q
c_1\circ\G^\infty=\hbar f_1\circ\G^\infty$, $\dots,$ 
$dc_{r-1}\circ\G^\infty+s^q
c_r\circ\G^\infty=\hbar f_r\circ\G^\infty$. Applying $s^q$ gives to
lowest order 
 $\bar s b=0$, $db +\bar s f_1=0$, $\dots,$ $df_{r-1}+\bar s
f_r=0$, so that $[b]_r$ is well defined. Suppose now that 
$[a]_r=d_0[g_0]_0+\dots+d_{r-1}[g_{r-1}]_{r-1}$. Anticommutativity of
$\delta_0$ with $d_0,\dots, d_{r-1}$ then implies that
$\delta_0[a]_r=0$. Hence, $\delta_0$ does not depend on the
representative and is well defined in $K_r$. Finally,
$d_r\delta_0[a]_r=d_r[b]_r=[df_r]_r$,
while $\delta_0d_r[a]_r=\delta_0 [dc_r]_r$, and $s^q
(dc_r)\circ\G^\infty=s^qd(c_r\circ \G^\infty+\hbar c^\prime\circ\G^\infty)=
-d( s^q c_r\circ\G^\infty+\hbar s^q c^\prime\circ \G^\infty)$, so that 
$\delta_0 [dc_r]_r=-[d(f_r+\bar s c^\prime)]_r=-d_r\delta_0[a]_r$. 

The results (\ref{res0}), (\ref{res7}) and (\ref{res8}) can then be 
summarized by the statement that an anomaly in one of the subspaces
of the decomposition (\ref{decomp})
must belong to the same subspace or to one that stands to the right; 
furthermore, the part of the anomaly of an element of $F_i$ in $F_i$
is minus the part of the anomaly of
the corresponding element of $d_i F_i$ in $d_{i}F_{i}$.

In order to recover the results for elements of $H(\bar s|d)$, 
we define $\Delta_0$ to be the equivalent of $\delta_0$ for modulo $d$
BRST cohomological classes, 
$\Delta_0 [a]=[b]$, for $[a],[b]\in H(\bar s|d)$, 
where $\bar s a+dm=0$, $\bar s m+du=0$, 
$s^qa\circ \G^\infty+d(m\circ\G^\infty)=\hbar b\circ \G^\infty$, 
$s^qm\circ \G^\infty+d(u\circ\G^\infty)=\hbar n\circ \G^\infty$.
Indeed, the map is well defined because 
the consistency condition implies to lowest order
$\bar s b+ dn=0$, while if $a=\bar s c+ dg$, we have $m=\bar s g +d
u$, so that $s^q a\circ\G^\infty+dm\circ\G^\infty=\hbar b\circ
\G^\infty$ gives $s^q(s^q c\circ\G^\infty+d g\circ\G^\infty+\hbar
f\circ\G^\infty)+d(s^q g\circ\G^\infty+du\circ\G^\infty+\hbar
v\circ\G^\infty)=\hbar b\circ\G^\infty$, which implies $b=\bar s f+dv$ as it
should. 

The following properties are straightforward to check: 
$[\Delta_0,{\cal D}]=0$,
$l \Delta_0 =-\delta_0 l$, $i_0\delta_0=\Delta_0 i_0$. 
One says (see e.g. \cite{Hu}, Chapter VIII.9) that
$(\Delta_0,\delta_0)$ is a mapping of the exact couple $(H(\bar
s|d),H(\bar s))$. 

The previous result, that $\delta_0$ induces well
defined maps in the spaces of the spectral sequence anticommuting with 
the differentials $d_r$, follows directly from the way the spectral 
sequence is associated to an exact couple. The relation between
(\ref{res1}), (\ref{res2}) at $l=0$ and (\ref{res7}), (\ref{res8}) is
summarized by $i_0\delta_0=\Delta_0 i_0$~; the relations between 
(\ref{res1}), (\ref{res2}) at different values of $l$ are summarized
by $[\Delta_0,{\cal D}]=0$~; finally, the relation between
(\ref{res2}) at $l=r$ and (\ref{res0}) is summarized by 
$l \Delta_0 =-\delta_0 l$. 

Note that in this case, we have furthermore the property that
$\Delta_0$ is a differential, $\Delta_0^2=0$.

{\bf Remark:} It follows from the above analysis that the relevant
property of the differentials $d_r$ is $\{\delta_0,d_r\}=0$. 
This means that analogous results that constrain the anomalies to
belong to particular subspaces of $H(\bar s)$ or $H(\bar s|d)$ 
can be derived if one can find maps $\lambda_0:H(\bar s)\longrightarrow 
H(\bar s)$, respectively $\Lambda_0:H(\bar s|d)\longrightarrow 
H(\bar s|d)$ such that $[\delta_0,\lambda_0]=0$, respectively
$[\Delta_0,\Lambda_0]=0$. 

The discussion in section on the length of descents and 
lifts of BRST
cohomological classes and their anomalies in this and the previous
subsection does not rely on the use 
of the extended antifield formalism. It can be done along the same
lines in the 
standard set-up as long as one assumes the quantum theory to be  
anomaly free and stable, so that the standard Zinn-Justin equation
$\frac{1}{2}(\Gamma,\Gamma)=0$ holds.  
\newpage

\mysection{Application 2: non renormalization of the Y-M gauge anomaly}

\label{s10}

We now show that the anomalous master equation for
Yang-Mills theories discussed in 
\cite{Costa:1977pd,Aoyama:1981yw,Tonin:1992wf} can be viewed as a
particular case of the anomalous master equation of the extended
antifield formalism. Then, we discuss how the Adler-Bardeen theorem
for the non abelian gauge anomaly
\cite{Bardeen,Costa:1977pd,Bandelloni:1980sp,Piguet:1992yg,Piguet:1993ds} 
 can be understood as a direct consequence of the fact that the length
of the descent of the gauge anomaly is $4$, while the length of the
descent of all the other cohomological classes coupled to the action is $0$.
These considerations are purely cohomological, so that they do 
not depend on the way the gauge is fixed or on power counting
restrictions. Furthermore, this approach to the 
Adler-Bardeen theorem does not require the use of the Callan-Symanzik
equation or assumptions on the beta functions of the theory.  

In the case of standard Yang-Mills theory, 
it is sufficient for our purpose to
couple the
local BRST cohomology classes in ghost number $0$ and ghost number $1$,
because this will be enough, under some assumptions stated explicitly below,  
to guarantee stability and to control the
anomalies. The starting point action contains from the beginning 
a coupling to the Adler-Bardeen
anomaly as in 
\cite{Costa:1977pd,Aoyama:1981yw,Tonin:1992wf}, 
with additional couplings to possibly
higher dimensional gauge invariant operators, 
if one does not want to restrict oneself to the
power counting renormalizable case \cite{Gomis:1996jp}. 
More precisely, the starting point 
is the action 
\bea
S_\rho=\int d^4x\ [-\frac{1}{4g^2}F^{\mu\nu}_I F_{\mu\nu}^I+
L^{\rm kin}_{\rm matter}(y^i,D_\mu y^i)]\nonumber \\
+\int d^4x\ [-D_\mu C^IA^{*\mu}_I+ C^I
T^i_{Ij}y^j y^*_i
-\frac{1}{2}C^IC^J{f_{JI}}^KC^*_K]\nonumber  \\
+g^i \int d^4x\ {\cal O}_i +{\cal A}\rho,\label{6.1}
\eea
satisfying the master equation
\bea
\frac{1}{2}(S_\rho,S_\rho)=0.
\eea
The Lagrangian $L^{\rm kin}_{\rm matter}(y^i,D_\mu y^i)$ is the gauge 
invariant extension of the kinetic terms for the matter fields $y^i$. 
For simplicity, we assume the gauge group to be $SU(3)$. The 
${\cal O}_i$ are gauge invariant local functions built 
out of the field strengths $F_{\mu\nu}^I$, the matter fields $y^i$ 
and their covariant derivatives such 
that the $\int d^4x\ {\cal O}_i$ (which can, but need not, be assumed to 
be power counting renormalizable) and $\int d^4x\
-\frac{1}{4g^2}F^{\mu\nu}_I F_{\mu\nu}^I$ are linearly independent
even when the gauge covariant equations of motions hold. Finally,
${\cal A}=\int {\rm Tr}\ [Cd(AdA+\frac{1}{2}A^3)]$ 
is the Adler-Bardeen gauge anomaly, $g$ is the gauge coupling
constant, $g^i$ are coupling constants for the other gauge invariant
operators, while $\rho$ is a Grassmann odd coupling constant 
with ghost number $-1$ for the 
Adler-Bardeen anomaly.
In this particular case, 
$\Delta_c=0$. This can be traced back to the fact that all
representatives of the local BRST cohomological classes in ghost
number $0$ and $1$ can be chosen to be independent of the
antifields.  

The gauge is fixed by introducing the cohomologically 
trivial non minimal
sector consisting of the antighost $\bar C^I$ and the Lagrange
multiplier $B^I$ and their antifields. One adds to the action (\ref{6.1}) 
the term 
$\int d^4x\ \bar C_I^* B^I$ and chooses an appropriate gauge 
fixing fermion $\Psi$, 
which is used to  
generate an anticanonical transformation in the fields and
antifields such that the propagators of the
theory are well defined. The gauge fixing is irrelevant for the 
cohomological considerations below. 

For the question of stability and anomalies, we have to analyze the
cohomology $H^{0,4}(s_\rho|d)$ and $H^{1,4}(s_\rho|d)$ in the space of 
functions in the couplings 
$g,g^i,\rho$ with coefficients that are Lorentz invariant polynomials
in the $dx^\mu$, the fields, the antifields and their derivatives. 

In order to compute this cohomology, we decompose, as in 
\cite{Costa:1977pd},
 both the BRST differential $s_\rho$ and the local forms into 
parts independent of
$\rho$ and parts linear in $\rho$. Explicitly, $s_\rho=s_0+s_1$, where 
$s$ is the standard BRST differential associated to the solution
$S_{\rho=0}$ of the master equation, while $s_1=({\cal A}\rho,\cdot)$.
The $\rho$ independent part of the cocycle condition 
$s_\rho \omega(\rho)+d\eta(\rho)=0$ 
in form degree $4$ gives (see e.g. \cite{Barnich:2000zw} for a review) 
$\omega(0)=\alpha(g,g^i)d^4x\ F^{\mu\nu}_I
F_{\mu\nu}^I+\alpha^j(g,g^i)d^4x\ {\cal O}_i +s(\ ) +d(\ )$. This
implies $\omega(\rho)=\alpha(g,g^i)d^4x\ F^{\mu\nu}_I
F_{\mu\nu}^I+\alpha^j(g,g^i)d^4x\ {\cal
  O}_i+\omega^\prime_1\rho +s_\rho(\ )+d(\ ).$ Because $d^4x\ F^{\mu\nu}_I
F_{\mu\nu}^I$ and $d^4x\ {\cal O}_i$ are also $s_1$ closed and $\rho^2=0$, the
cocycle condition reduces to $s\omega^\prime_1+d(\ )=0$, where the 
ghost number of $\omega^\prime_1$ is $1$. It follows that  $\omega^\prime_1=
\lambda(g,g^i){\rm Tr}\ [Cd(AdA+\frac{1}{2}A^3)]+s(\ )+d(\
)$. Hence, the general solution of the consistency condition in the
space of local functionals in ghost number $0$ is given by 
\bea
\alpha(g,g^i)\frac{\partial S_\rho}{\partial g}+
\alpha^j(g,g^i)\frac{\partial S_\rho}{\partial g^i}+ 
\frac{\partial^R S_\rho}{\partial
  \rho}\rho\lambda(g,g^i)+(S_\rho,\Xi_\rho)
\label{6.4}
\eea
for some local functional $\Xi_\rho$ in ghost number $-1$. 
This implies that the theory is stable. 

Similarly, in ghost number $1$, the $\rho$ independent part of the
cohomology gives as only anomaly candidate the Adler-Bardeen anomaly.
There could however be a $\rho$ linear non trivial contribution to the anomaly,
because the cohomology of 
$s$ in form degree $4$ and ghost number $2$ is not necessarily empty 
(see \cite{Barnich:2000zw}, section 12.4), contrary to the 
claim in \cite{Costa:1977pd}. 
More precisely, to each $x^\mu$-independent, 
gauge and Lorentz invariant non trivial conserved 
current $j_\Delta=j^\mu_\Delta\epsilon_{\mu\nu_1\nu_2\nu_3}
dx^{\nu_1}dx^{\nu_2}dx^{\nu_3}$, there corresponds the cohomological class
$V^{2,4}_\Delta=j_\Delta [{\rm Tr} C^3]^1+{\it antifield\ dependent\ terms}$
in $H^{2,n}(s|d)$, with $s[{\rm Tr} C^3]^1+{\rm Tr} C^3=0$. (There 
could in principle be another type of antifield dependent cohomology classes 
in exceptional situations \cite{Barnich:2000zw}, which we exclude 
from the present considerations).

If there are such non trivial currents $j_\Delta$, we have to change our 
starting point and also couple the ``anomaly for anomaly candidates''
$V^{2,4}$ 
from the beginning with couplings in ghost number $-2$. But then the 
cohomology of $s$ in ghost number $3$ becomes relevant. 
There are plenty of such classes, for instance classes of the form 
$d^4x\ I\ {\rm Tr}C^3$, where $I$ are invariant functions built 
out of the field strengths, the matter fields and their covariant 
derivatives. In this way, one is led to use the full extended 
antifield formalism as described in section \ref{s5} 
with all BRST cohomology 
classes in positive ghost 
number coupled from the beginning. In the case where the algebra of the 
non trivial symmetries associated to the currents $j_\Delta$ is non abelian, 
the operator $\Delta_c$ will be non vanishing at the classical
level and involve in particular the structure constants of this algebra.

Another possibility is to try to show that the
anomaly candidates $V^{2,4}_\Delta$ do not effectively arise in the theory, 
by using higher order cohomological restrictions: as in the proof of the
absence of similar instabilities in the presence of abelian factors
for standard Yang-Mills theories in section \ref{s3}
one couples with external fields gauge invariant functions that break
the symmetries associated to the currents $j_\Delta$. In this way, one 
eliminates the currents $j_\Delta$ and the associated anomaly for
anomaly candidates $V^{2,4}_\Delta$ from the extended theory. At the
end of the computations, the external fields can be put to zero. 

Because this discussion is not central to the argument below, we will simply
assume here that 
the observables ${\cal O}_i$ are
such that there are no non trivial currents $j_\Delta$ and thus no
anomaly for anomaly candidates $V^{2,4}_{\Delta}$ in the
theory. 
The general solution of the consistency condition in ghost number $1$
is then given by 
\bea
\frac{\partial S_\rho}{\partial \rho}\sigma(g,g^i)
+(S_\rho,\Sigma_\rho), \label{6.5}
\eea
for some local functional $\Sigma_\rho$ in ghost number
$0$.

By standard arguments, using in addition the same reasoning as in 
section \ref{sec2}, it follows from (\ref{6.4}) and (\ref{6.5}) 
that the model is renormalizable and 
that the renormalized 
generating functional for 1 particle irreducible vertex functions
$\Gamma_\rho$ satisfies 
\bea
\frac{1}{2}(\Gamma_\rho,\Gamma_\rho)+\frac{\partial^R\Gamma_\rho}{\partial 
  \rho}\hbar \sigma(g,g^i)=0,\label{fun}
\eea
where $\sigma(g,g^i)$ is a formal power series in $\hbar$. Hence, in
this case $\Delta^\infty=\frac{\partial^R\cdot}{\partial 
  \rho}\hbar \sigma(g,g^i)$ and the quantum BRST differential is 
$s^q=(\Gamma_\rho,\cdot)-\hbar\sigma(g,g^i){\partial^L}{\partial 
  \rho}$. 

Let us now investigate the renormalization of the operators 
$d^4x\ F^{\mu\nu}_I F_{\mu\nu}^I$ and $d^4x\ {\cal O}_i$. According to 
the classification in section \ref{char}, they are both of the type 
$e^0_{\alpha_0}$, because they are non trivial
bottoms in maximal form degree $4$, while the Adler-Bardeen anomaly 
${\rm Tr}\ [Cd(AdA+\frac{1}{2}A^3)]$ is of the type
$e^4_{\alpha_4}$, as it descends to the non trivial bottom 
${\rm Tr}C^5$. Because there is no $e^0_{\alpha_0}$ in ghost
number $1$ (and form degree 4) 
and no $h^r_{i_r}$ in form degree $3$ and ghost number $1$,
equation (\ref{res1}) implies that the lowest order contribution to the 
anomaly in the renormalization of $d^4x\ F^{\mu\nu}_I F_{\mu\nu}^I$
and of $d^4x\ {\cal O}_i$ is $s_\rho$ exact and can thus be absorbed
through a BRST breaking counterterm added to these operators. This
reasoning can be pushed to all orders, with the result that one can
achieve, through the addition of suitable counterterms,
\bea
s^q [d^4x\ F^{\mu\nu}_I F_{\mu\nu}^I\circ\G_\rho]=0,\label{36}\\
s^q [d^4x\ {\cal O}_i\circ\G_\rho]=0\label{37}.
\eea
If we now apply $-\frac{g}{2}\frac{\partial}{\partial g}$,
respectively $\frac{\partial}{\partial g^j}$ to (\ref{fun}), we get 
on the one hand,
\bea
s^q [\int d^4x\ -\frac{1}{4g^2}F^{\mu\nu}_I F_{\mu\nu}^I\circ\G_\rho]+
\frac{\partial^R\Gamma_\rho}{\partial 
  \rho}\hbar [-\frac{g}{2}\frac{\partial\sigma(g,g^i)}{\partial g}]=0,\\
s^q [\int d^4x\ {\cal O}_i\circ\G_\rho]+\frac{\partial^R\Gamma_\rho}{\partial 
  \rho}\hbar [\frac{\partial\sigma(g,g^i)}{\partial g^j}]=0.
\eea 
Comparing on the other hand with the integrated versions of (\ref{36})
and (\ref{37}), we deduce that 
\bea
\frac{\partial\sigma(g,g^i)}{\partial g}=0,\ 
\frac{\partial\sigma(g,g^i)}{\partial g^j}=0,
\eea
which is crucial in the proof of the Adler-Bardeen theorem (see e.g. section
6.3.2 of \cite{Piguet:1995er}). 

\newpage

\mysection{Conclusion}

Starting from the standard antifield formalism, one first 
computes a basis for the representatives of the local BRST cohomology
of the theory. The representatives of those classes that are not
already contained in the standard solution of the master equation are
coupled with new coupling
constants. The extended formalism can then 
be constructed perturbatively and involves the antibracket 
algebra of these representatives. The extended master
equation $\frac{1}{2}(S,S)+\Delta_c S=0$ is stable by construction,
because the associated differential $\bar s = (S,\cdot)+\Delta_c^L$
takes into account higher order cohomological restrictions encoded in
$\Delta_c^L$. 

The extended master equation can be constructed
consistently in two particular cases. The first case is 
the extended 
antifield formalism of \cite{Brandt:1998cz}, where only the 
local BRST cohomological classes in strictly negative ghost numbers
have been coupled. This
is the relevant formalism to control the algebra of the non trivial
generalized global symmetries of the theory.
In the second case, only the classes in zero
and positive ghost numbers are coupled. 
The formalism then allows to control the renormalization,
where only the gauge symmetries and no global symmetries are taken
into account. 

The stability of the formalism guarantees renormalizability in the
modern sense for generic gauge theories. The renormalized effective
action $\Gamma^\infty$ satisfies the extended master equation 
$\frac{1}{2}(\G^\infty,\G^\infty)+\Delta^\infty \G^\infty=0$, where the
differential $\Delta^\infty$ is a deformation of the differential
$\Delta_c$ and encodes all the information about the
anomalies. The associated quantum BRST differential
$s^q=(\G^\infty,\cdot)+
(\Delta^\infty)^L$ is well defined and reduces to $\bar s$ in the
classical limit.  

The standard master equation, both on the classical and the quantum
level, expresses gauge
invariance of the original theory. The extended master equation allows for a
breaking of gauge invariance, even on the classical level. This
happens if $\Delta_c$ does not vanish. 
The important question of gauge invariance on the quantum level for the
original theory can be answered by putting to zero the additional couplings
after renormalization. If this can be done and $\Delta^\infty$ 
then vanishes, the quantum
theory is gauge invariant in the sense that the standard Zinn-Justin
equation for the effective action holds. 

Controlling the symmetry and anomaly structure 
of the theory, by first computing the complete classical 
cohomology, then constructing a
stable formalism and finally quantizing, i.e., computing $\G^\infty$
and $\D^\infty$, without any assumptions 
about anomalies
needed, could be called ``cohomological renormalization'', in
contrast to ``algebraic renormalization'', where for a given theory,
stability has to be checked and absence of anomalies is required. 

\newpage

\section*{Acknowledgements}

\addcontentsline{toc}{section}{Acknowledgements}

The author is Scientific Research Worker of the ``Fonds National Belge de la 
Recherche Scientific''. His work is supported in part by the ``Actions de
Recherche Concert{\'e}es" of the ``Direction de la Recherche
Scientifique - Communaut{\'e} Fran{\c c}aise de Belgique", by
IISN - Belgium (convention 4.4505.86) and by
Proyectos FONDECYT 1970151 and 7960001 (Chile).
He acknowledges useful discussions with 
J.A.~de~Azc\'arraga, F.~Brandt, P.A.~Grassi, M.~Henneaux,
J.M.~Izquierdo, D.~Maison, C.~Schomblond, M.~Tonin, J.~Stasheff and
A.~Wilch.

\end{document}